\documentclass{jfm}

\usepackage{graphicx}
\usepackage{amsmath}
\usepackage{mathtools}
\usepackage{relsize}
\usepackage{amssymb}  
\usepackage{dcolumn}
\usepackage{bm}
\usepackage{bigints}
\usepackage{enumitem}
\usepackage{float}
\usepackage{verbatim}
\usepackage{makecell}
\usepackage{enumitem}
\usepackage[export]{adjustbox}
\usepackage{scalerel}[2016-12-29]
\usepackage[dvipsnames]{xcolor}
\usepackage{multirow,bigdelim}
\usepackage{lscape}
\usepackage{tikz}
\usepackage{siunitx}
\usepackage{cancel}
\usepackage{lineno}
\usepackage{bigints}
\DeclareMathOperator\atanh{atanh}
\usepackage[normalem]{ulem}

\usepackage[colorlinks=true, linkcolor=blue, urlcolor=blue, citecolor=blue]{hyperref}

\newcount\ndots
\def\drwln#1#2{\raise 2.5pt\vbox{\hrule width #1pt height #2pt}}

\def\filsqr   {${\vcenter{\hrule height 2pt
                          \hbox{\vrule width 2.2pt height 0.2pt \kern 0.1pt
                                \vrule width 2.2pt}
                                \hrule height 2.2pt}}$\nobreak\ }

\newcommand*{\Scale}[2][4]{\scalebox{#1}{$#2$}}%

\begin{document}


\title{\textbf{Leveraging unstructured grids for direct numerical simulations of wall turbulence} 
}


\shortauthor{Rouhi, Kumar, Wu, Kozul and Lehmkuhl}

\shorttitle{leveraging unstructured grids for DNS}

\author{Amirreza Rouhi$^{1}$ \corresp{\email{amirreza.rouhi@ntu.ac.uk}}, Vishal Kumar$^{2}$, Wen Wu$^3$, Melissa Kozul$^4$, \and Oriol Lehmkuhl$^{2}$}
\affiliation{$^1$Department of Engineering, School of Science and Technology\\
                Nottingham Trent University, Nottingham, UK\\[5pt]                
             $^2$CASE, Barcelona Supercomputing Center (BSC), Barcelona, Spain\\[5pt]
             $^3$Department of Mechanical Engineering, University of Mississippi, Oxford, USA\\[5pt]
             $^4$Department of Mechanical Engineering, University of Melbourne, Victoria 3010, Australia}


\maketitle

\begin{abstract}
Towards computational cost saving for direct numerical simulations (DNSs) of wall turbulence, we formulate an unstructured grid-generation framework, termed $\eta$-grid, where the wall-normal ($y$) and spanwise ($z$) grid sizes are proportional to the local Kolmogorov scale $\eta$. The framework consists of an inner layer, with a thickness $\sim 50$ viscous units, with viscous-scaled grid sizes similar to a conventional DNS grid: $0.3 \lesssim \Delta y^+ \lesssim 4, \Delta z^+ \simeq 5$ over a smooth wall, and $\ell^+/30 \lesssim \Delta y^+, \Delta z^+ \lesssim 4$ over uneven surfaces, where $\ell^+$ is the smallest surface wavelength. Above the inner layer, $\Delta y^+ \simeq \Delta z^+ \simeq 2\eta^+$. We test $\eta$-grid with finite volume and spectral element solvers, and conduct DNSs of turbulent channel flows and boundary layers over smooth wall and various streamwise-aligned riblets, up to friction Reynolds number $\delta^+_0 = 1000$. We assess the accuracy of $\eta$-grid against the conventional Cartesian grids, through comparison with the reference DNS and experimental data. Results from $\eta$-grid and the Cartesian grids differ by less than $1\%$, in terms of turbulence statistics up to second-order, and the energy spectra. For turbulent channel flows with $10^3 \lesssim \delta^+_0 \lesssim 10^4$, the number of grid points with $\eta$-grid ($N_\eta$) scales $\propto {\delta^+_0}^{2.47}$ over a smooth wall, and $\propto {\delta^+_0}^{2.0-2.47}$ over riblets, whereas the number of grid points with a Cartesian grid and hyperbolic-tangent $y$-grid ($N_\mathrm{Tanh}$) scales $\propto {\delta^+_0}^{3.0}$. By $\delta^+_0 = 6000$, $N_\eta/N_\mathrm{Tanh} \simeq 0.1$ over a smooth wall, and $N_\eta/N_\mathrm{Tanh} \simeq 0.04$ over typical drag-reducing riblets, with viscous-scaled spacing $15$.  


\end{abstract}




\section{Introduction}
Wall turbulence is a prevalent flow configuration in nature and industry. The interaction of atmospheric boundary layer with complex terrains (urban areas or forest canopies) dictates the transport of species (moisture or pollution), hence impacts the environmental and health sectors~\citep{monin1970atmospheric,panofsky1974atmospheric,zeman1981progress}. This same atmospheric boundary layer can be exploited by wind farms to have a favourable impact on the energy sector~\citep{de1983theory,sorensen2011aerodynamic,stevens2017flow}. In the transport sector, the aerodynamics of ground vehicles or aircraft are tied to their interactions with turbulent boundary layers~\citep{landweber1979ship,hucho1993aerodynamics,schetz2001aerodynamics}. Owing to the complexity and vastness of wall-turbulence research, since 1969 the pioneers of this field have been publishing comprehensive review articles in \textit{Annual Review of Fluid Mechanics} on various aspects of this field, such as the physics of wall turbulence~\citep{kovasznay1970turbulent,smits2011high}, its phenomenological models~\citep{yaglom1979similarity,marusic2019attached}, or computational aspects~\citep{wu2017inflow,moin1998direct,piomelli2002wall,bose2018wall}, controlling wall turbulence~\citep{lumley1969drag,berman1978drag,smits1985response,bushnell1989turbulence,karniadakis2003mechanisms,fukagata2024turbulent}, turbulent flows over rough surfaces~\citep{raupach1981turbulence,jimenez2004turbulent,chung2021predicting}, and non-equilibrium effects in wall turbulence~\citep{adamson1980analysis,smith1986steady,clemens2014low}. 

\begin{figure}
  \centering
 \includegraphics[width=.99\linewidth,trim={{0.1\textwidth} {0.65\textwidth} {0.1\textwidth} {0.0\textwidth}},clip]{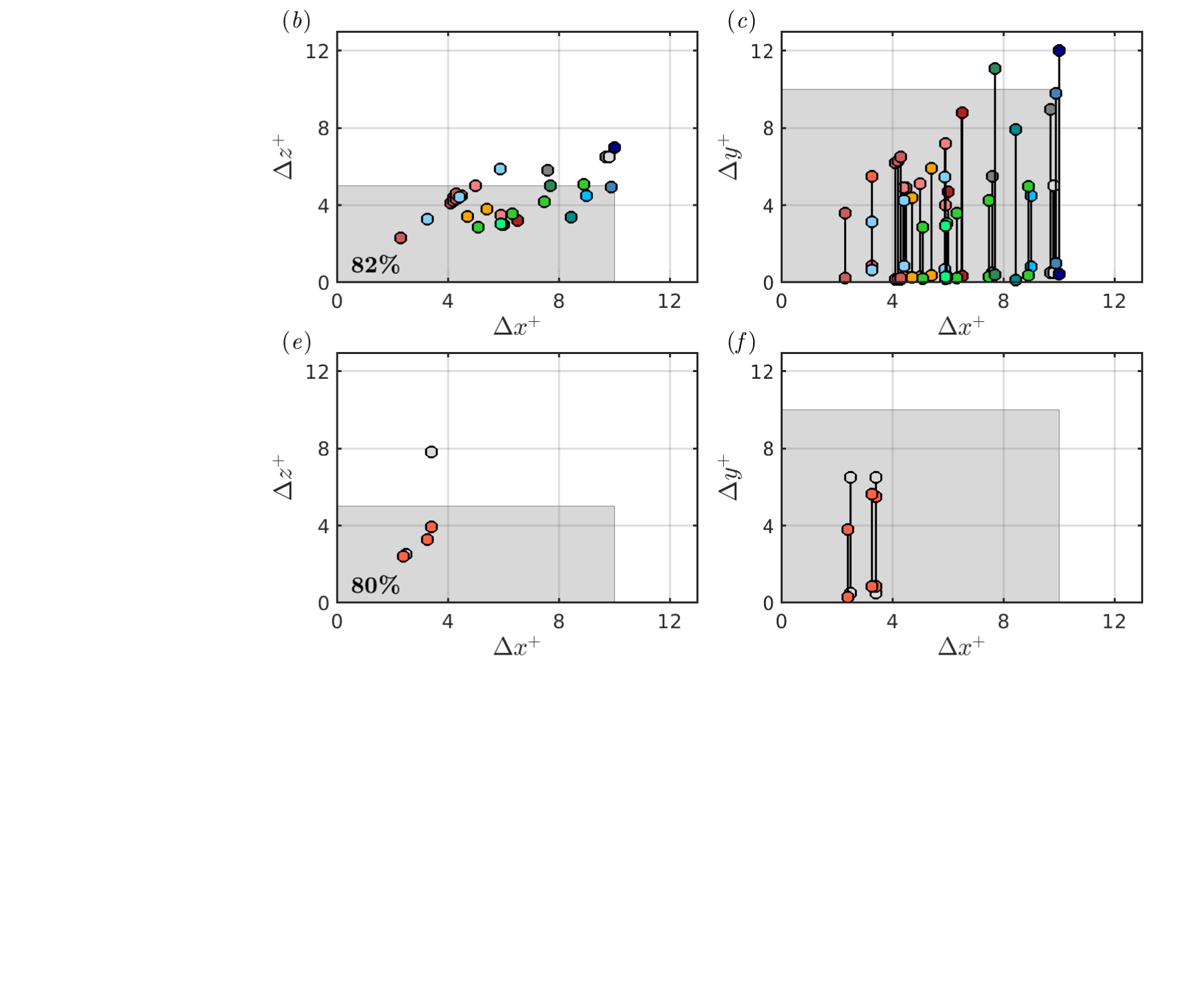}
    \makebox[0pt][r]{
    \raisebox{1em}{%
      \includegraphics[width=.19\linewidth]{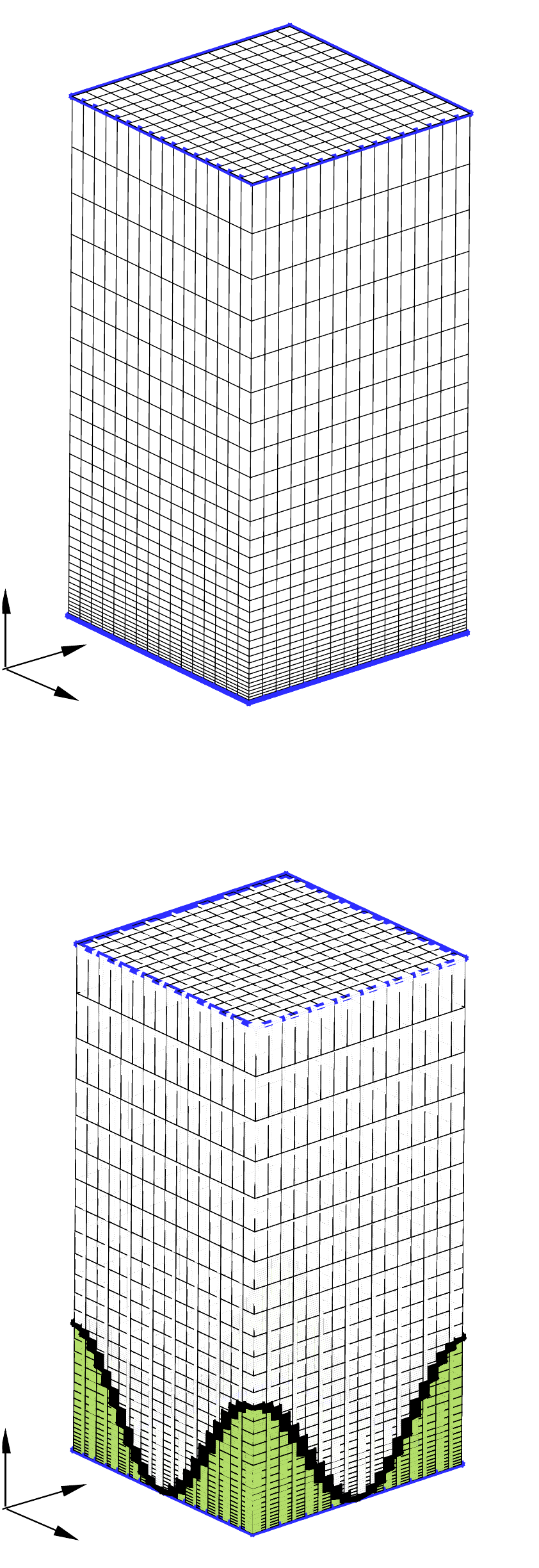}
    }\hspace*{34em}%
  }  
  \put(-380,220){\Scale[0.95]{(\textit{a})}}
  \put(-380,105){\Scale[0.95]{(\textit{d})}}
  \put(-380,121){$\Scale[0.75]{z}$}
  \put(-377,129){$\Scale[0.75]{x}$}
  \put(-385,140){$\Scale[0.75]{y}$} 
  \put(-380,8){$\Scale[0.75]{z}$}
  \put(-377,16){$\Scale[0.75]{x}$}
  \put(-385,27){$\Scale[0.75]{y}$}    
  \\
 \includegraphics[width=.99\linewidth,trim={{0.1\textwidth} {0.75\textwidth} {0.1\textwidth} {0.0\textwidth}},clip]{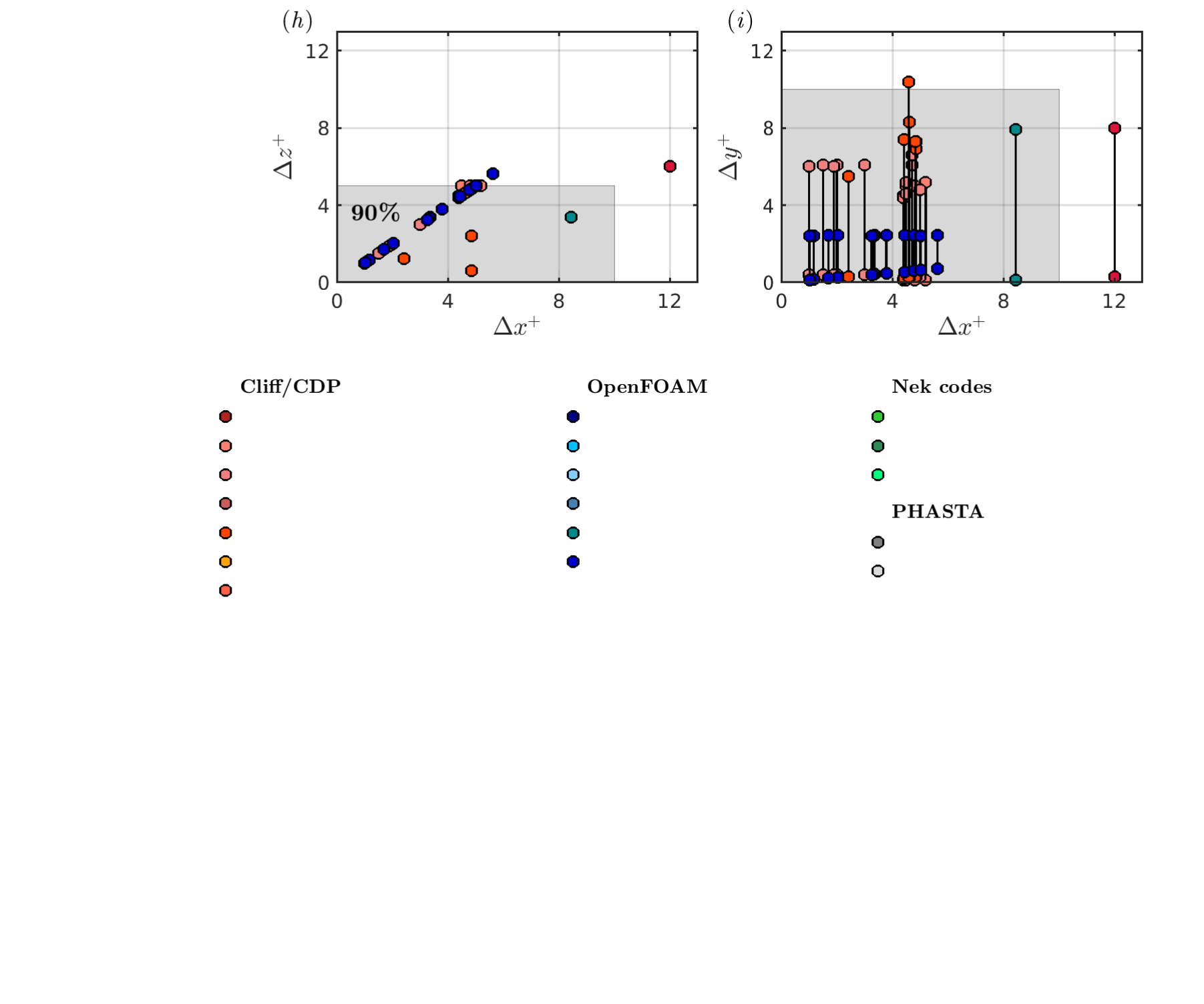}   
  \makebox[0pt][r]{
    \raisebox{11em}{%
      \includegraphics[width=.19\linewidth]{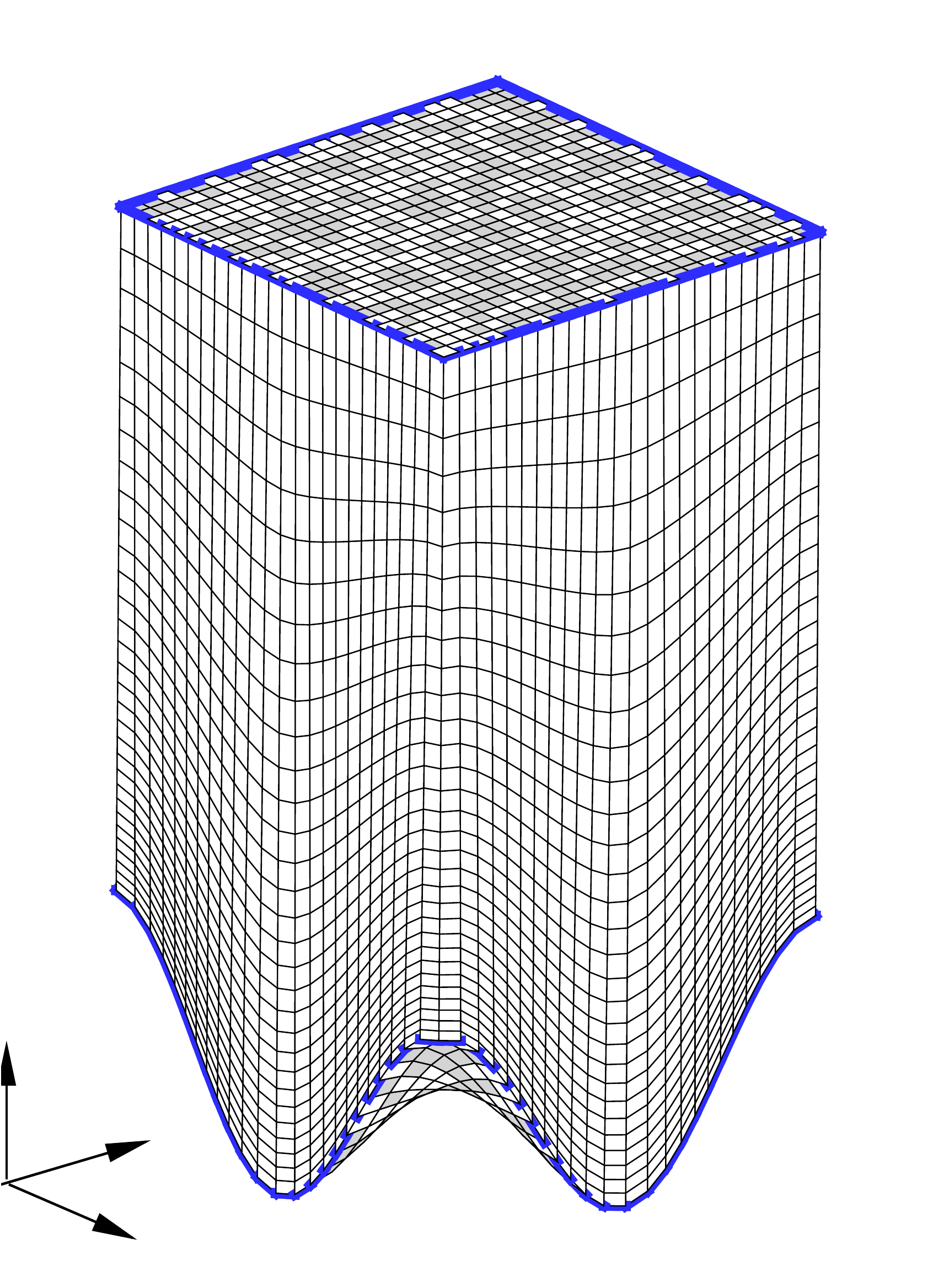}
    }\hspace*{34em}%
  }%
  \put(-380,194){\Scale[0.95]{(\textit{g})}}
  \put(-380,100){$\Scale[0.75]{z}$}
  \put(-377,108){$\Scale[0.75]{x}$}
  \put(-385,119){$\Scale[0.75]{y}$}  
  \put(-319,61){{\scriptsize \cite{endrikat2022reorganisation}}} 
  \put(-319,51){{\scriptsize \cite{chung2015fast}}}
  \put(-319,41){{\scriptsize \cite{macdonald2016turbulent}}}
  \put(-319,31){{\scriptsize \cite{macdonald2017minimal}}}
  \put(-319,21){{\scriptsize \cite{macdonald2019roughness}}}
  \put(-319,11){{\scriptsize \cite{anantharamu2020analysis}}}
  \put(-319,1){{\scriptsize \cite{ma2021direct}}}
  \put(-198,61){{\scriptsize \cite{rezaeiravesh2021numerical}}} 
  \put(-198,51){{\scriptsize \cite{komen2014quasi}}} 
  \put(-198,41){{\scriptsize \cite{xu2025temporal}}}
  \put(-198,31){{\scriptsize \cite{zhang2024direct}}}
  \put(-198,21){{\scriptsize \cite{lecrivain2016using}}}
  \put(-198,11){{\scriptsize \cite{zhang2024numerical}}}
  \put(-92,61){{\scriptsize \cite{zahtila2023systematic}}}
  \put(-92,51){{\scriptsize \cite{stanly2026influence}}}
  \put(-92,41){{\scriptsize \cite{chen2023backflow}}}
  \put(-92,18){{\scriptsize \cite{trofimova2009direct}}}
  \put(-92,8){{\scriptsize \cite{mishra2015dns}}}
  \caption{Compilation of the DNS grid sizes by the studies that used unstructured-grid solvers. Turbulent channel flows or TBLs over (\textit{a-c}) smooth surfaces with Cartesian grids, (\textit{d-f}) uneven surfaces with Cartesian grids and IBM, and (\textit{g-i}) uneven surfaces with curvilinear grids. (\textit{a,d,g}) Show a representative grid from the compiled cases in the plots on the right. (\textit{b,e,h}) $\Delta z^+$ vs.\ $\Delta x^+$, and (\textit{c,f,i}) $\Delta y^+$ vs.\ $\Delta x^+$. Each single bullet in (\textit{b,e,h}) represents a fixed $(\Delta x^+, \Delta z^+)$ of a DNS case, and each pair of bullets with a connected line in (\textit{c,f,i}) represents a fixed $\Delta x^+$ but varying $\Delta y^+$ of a DNS case. The bullets are colored based on the solver used: CDP/Cliff (red scale), OpenFOAM (blue scale), PHASTA (gray scale), and Nek5000, NekRS and Nektar++ (green scale). For the latter SEM solvers, we report the average spacing between the polynomial points. The gray regions shade the conventional range for DNS grid sizes ($\Delta x^+ \le 10, \Delta y^+ \le 10, \Delta z^+ \le 5$). In (\textit{b,e,f}), the numbers in bold are the percentage of the cases that fall into the gray zone. }
  \label{fig:grids_intro_p1}
\end{figure}

Direct Numerical Simulation (DNS) is a primary computational technique for studying wall turbulence~\citep{moin1998direct}. It provides accurate high-fidelity three-dimensional flow fields, that allow us to study the flow physics to unprecedented detail. We can deduce or improve scaling laws and models with DNS, or we can calculate quantities that are difficult to measure via laboratory experiments, e.g.\ wall shear-stress~\citep{hutchins2002accurate,baars2016wall,neuhauser2025predicting}. The pioneering DNS studies of wall turbulence focused on smooth-wall turbulent channel flow~\citep{kim1987turbulence,moser1999direct,hoyas2006scaling}, and zero pressure-gradient (ZPG) turbulent boundary layer (TBL)~\citep{spalart1988direct,simens2009high,wu2009direct,schlatter2009turbulent}. They developed efficient computational solvers with high-order finite difference or spectral methods for spatial discretisation; these solvers operate on Cartesian grids, as in figure~\ref{fig:grids_intro_p1}(\textit{a}). The mentioned pioneering DNSs, generated Cartesian grids with viscous-scaled streamwise and spanwise spacing $\Delta x^+ \in [4, 12]$ and $\Delta z^+ \in [3, 7]$, and the wall-normal grid size was stretched from $\Delta y^+_w \in [0.1, 0.3]$ at the wall to $\Delta y^+_{\delta} \in [7, 10]$ at the channel half-height (or TBL thickness). These grid size prescriptions have become the convention for DNS of smooth wall-bounded turbulent flows, as well as non-canonical turbulent flows, such as turbulent flow over periodic hills~\citep{krank2018direct}, bumps~\citep{okochi2025direct}, or separating TBL~\citep{wu2020spatio}.

 

\begin{figure}
  \centering
  \hspace{2cm}
 \includegraphics[width=0.99\linewidth,trim={{0.1\textwidth} {0.2\textwidth} {0.1\textwidth} {0.0\textwidth}},clip]{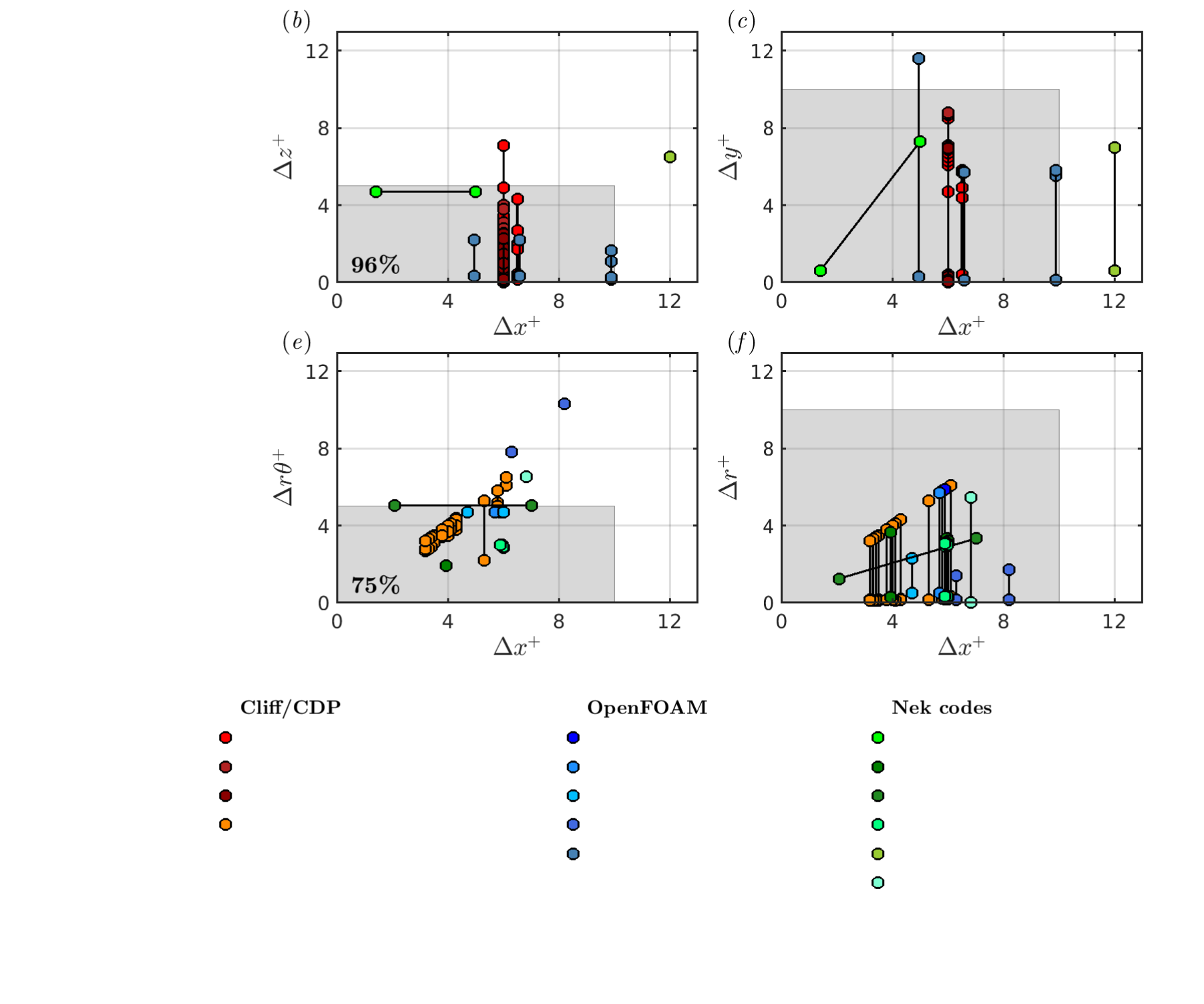}
    \makebox[0pt][r]{
    \raisebox{11em}{%
      \includegraphics[width=.20\linewidth]{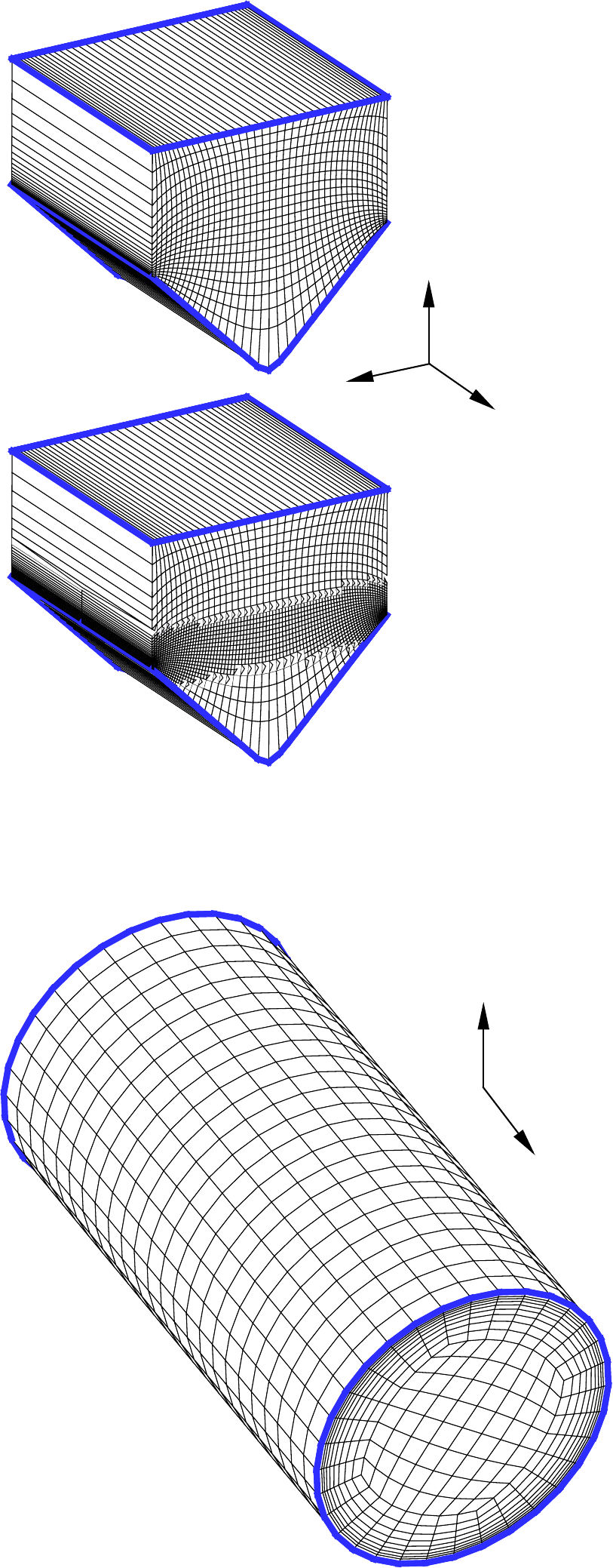}
    }\hspace*{34em}%
  }
  \put(-380,300){\Scale[0.95]{(\textit{a})}}
  \put(-348,246){$\Scale[0.75]{z}$}
  \put(-335,243){$\Scale[0.75]{x}$}
  \put(-336,263){$\Scale[0.75]{y}$}  
  \put(-380,190){\Scale[0.95]{(\textit{d})}}
  \put(-328,149){$\Scale[0.75]{x}$}
  \put(-335,170){$\Scale[0.75]{r}$}  
\put(-316,52){{\scriptsize \cite{endrikat2021influence}}} 
\put(-316,42){{\scriptsize \cite{endrikat2022reorganisation}}}
\put(-316,32){{\scriptsize \cite{wong2024viscous}}}
\put(-316,22){{\scriptsize \cite{chan2015systematic}}}
\put(-195,52){{\scriptsize \cite{cheng2020forcing}}}
\put(-195,42){{\scriptsize \cite{zheng2019direct}}}
\put(-195,32){{\scriptsize \cite{komen2014quasi}}}
\put(-195,22){{\scriptsize \cite{chu2016direct}}}
\put(-195,12){{\scriptsize \cite{zhang2024direct}}}
\put(-90,52){{\scriptsize \cite{deshpande2024pressure}}} 
\put(-90,42){{\scriptsize \cite{el2013direct}}}
\put(-90,32){{\scriptsize \cite{hufnagel2018three}}}
\put(-90,22){{\scriptsize \cite{chen2023backflow}}}
\put(-90,12){{\scriptsize \cite{chan2023effect}}}
\put(-90,2){{\scriptsize \cite{fei2025extreme}}}
  \caption{Same as figure~\ref{fig:grids_intro_p1}, but for (\textit{a-c}) turbulent channel flows over riblets or spanwise-aligned bars, and (\textit{d-f}) turbulent pipe flows over smooth or wavy walls. Unlike figure~\ref{fig:grids_intro_p1}, some studies have simultaneously varied the grid size in the wall-normal ($\Delta y^+/\Delta r^+$) and streamwise ($\Delta x^+$) directions, or wall-normal ($\Delta y^+/\Delta r^+$) and azimuthal ($\Delta z^+/\Delta r\theta^+$) directions.}
  \label{fig:grids_intro_p2}
\end{figure}

Cartesian-grid solvers are also widely used for DNSs of turbulent flows over non-smooth surfaces. Some studies mimic the uneven surface characteristics via a surrogate parametric forcing, e.g.\ roughness forcing techniques~\citep{busse2012parametric,varghese2020representing}, or slip boundary conditions for superhydrophobic surfaces~\citep{min2004effects,jelly2014turbulence} and porous media~\citep{ochoa1995momentum,rosti2015direct}. Alternatively, studies explicitly resolve the surface geometry via an Immersed-Boundary Method, IBM~\citep{yuan2014estimation,jelly2018reynolds,rouhi2019direct}; the surface is treated as a solid phase in the computational domain (figure~\ref{fig:grids_intro_p1}\textit{d}). Application of IBMs has gained an explosive popularity with over $2000$ annual publications on this subject~\citep{verzicco2023immersed}. However, a constraint of the Cartesian-grid solvers with IBM is their inability to increase $\Delta x^+, \Delta z^+$ with wall distance $y$. This constraint is a computational bottleneck for simulating microsized surfaces with viscous length-scales $\ell^+ \lesssim \mathcal{O}(10)$. A prime example is a turbulent flow over riblets (figures~\ref{fig:grids_intro_p2}\textit{a-c}). Riblets yield optimal drag-reducing performance when their viscous-scaled spanwise spacing $s^+ \simeq 15$~\citep{garcia2011hydrodynamic}; sufficiently resolving such fine-grained spacing requires $\Delta z^+ \simeq 0.5 - 1.0$ ($15 - 30$ grid points per $s^+$). With Cartesian grids and IBM, such stringent $\Delta z^+$ will be extruded across the domain, as was the case in \cite{kozul2023direct}, \cite{malathi2023riblet} and \cite{savino2025attached}. 

With unstructured-grid solvers, we are able to increase $\Delta x^+, \Delta z^+$ away from the wall. This feature has attracted some researchers to opt for unstructured-grid solvers for simulating turbulent flows over regular rough surfaces or riblets. In figures~\ref{fig:grids_intro_p1} and \ref{fig:grids_intro_p2}, we compile studies that have applied some popular unstructured-grid solvers for DNSs of wall turbulence: Finite-Volume Method (FVM) solvers OpenFOAM~\citep{weller1998tensorial}, CDP and its variants~\citep{ham2004energy,mahesh2004numerical,ham2006accurate}, Finite-Element Method (FEM) solver PHASTA~\citep{jansen1999stabilized,whiting2001stabilized}, and Spectral-Element Method (SEM) solvers Nek5000~\citep{fischer2008nek5000}, NekRS~\citep{fischer2022nekrs}, and Nektar++~\citep{cantwell2015nektar,moxey2020nektar}. Figure~\ref{fig:grids_intro_p1} compiles cases on turbulent channel flow or TBL over smooth surfaces (figures~\ref{fig:grids_intro_p1}\textit{a-c}), as well as uneven surfaces (figures~\ref{fig:grids_intro_p1}\textit{d-i}). Figure~\ref{fig:grids_intro_p2} compiles cases on turbulent channel flow over riblets or spanwise-aligned bars (figures~\ref{fig:grids_intro_p2}\textit{a-c}), and turbulent pipe flow over smooth or wavy surfaces (figures~\ref{fig:grids_intro_p2}\textit{d-f}). Cases in figure~\ref{fig:grids_intro_p1} generate structured grids (Cartesian or curvilinear), with fixed $\Delta x^+, \Delta z^+$ and stretched $\Delta y^+$; $80\%$ of these cases set the grid sizes within the range as set for a conventional DNS with a structured-grid solver, i.e.\ $\Delta x^+ \le 10, \Delta y^+ \le 10, \Delta z^+ \le 5$ (percentages in figures~\ref{fig:grids_intro_p1}\textit{b,e,h}). Some cases in figure~\ref{fig:grids_intro_p2} make use of the unstructured-grid solvers to locally refine the grid; e.g.\ for riblets, they refine the grid near the riblet crests (figure~\ref{fig:grids_intro_p2}\textit{a}, \citealt{endrikat2021influence}). Nevertheless, $96\%$ of the riblet studies, and $75\%$ of the pipe studies, restrict their maximum grid sizes to the prescriptions for the structured-grid codes (figures~\ref{fig:grids_intro_p2}\textit{b,e}). 



 \begin{figure}
  \centering
 \includegraphics[width=\textwidth,trim={{0.0\textwidth} {0.0\textwidth} {0.0\textwidth} {0.0\textwidth}},clip]{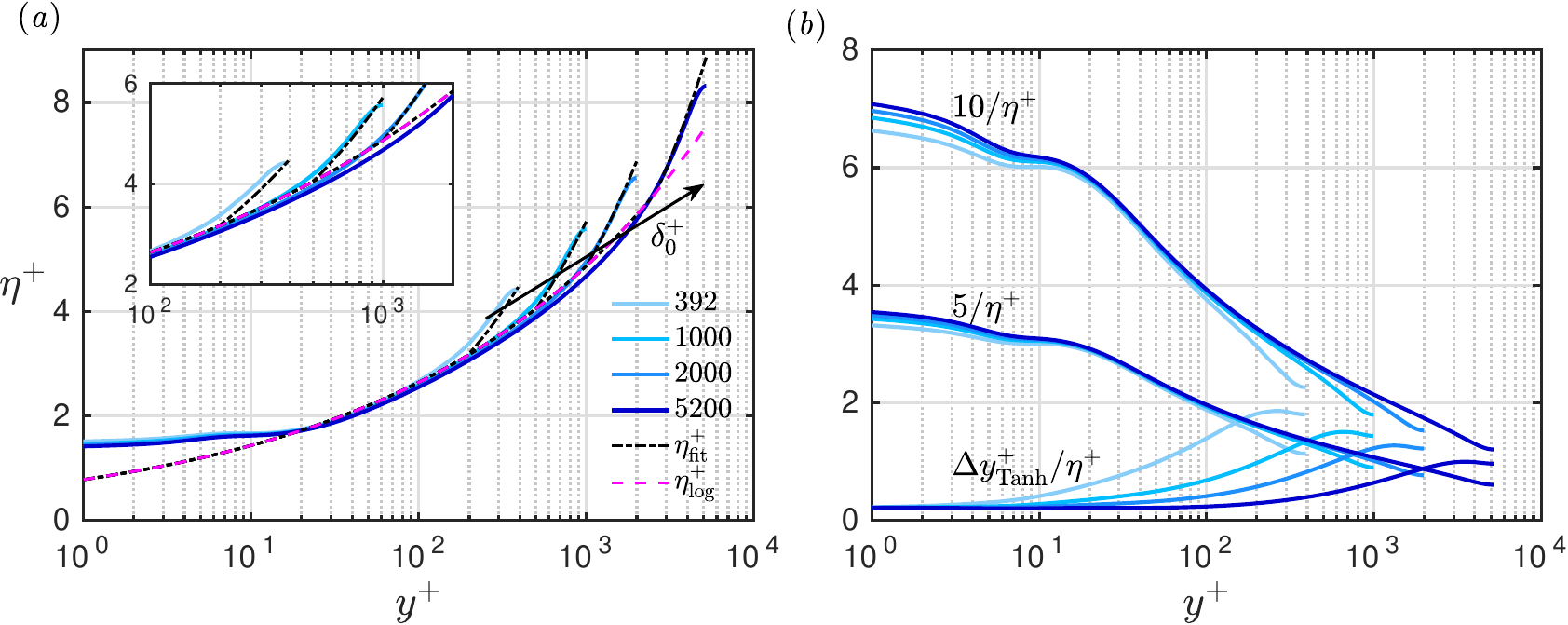}
  \includegraphics[width=\textwidth,trim={{0.0\textwidth} {0.0\textwidth} {0.0\textwidth} {0.0\textwidth}},clip]{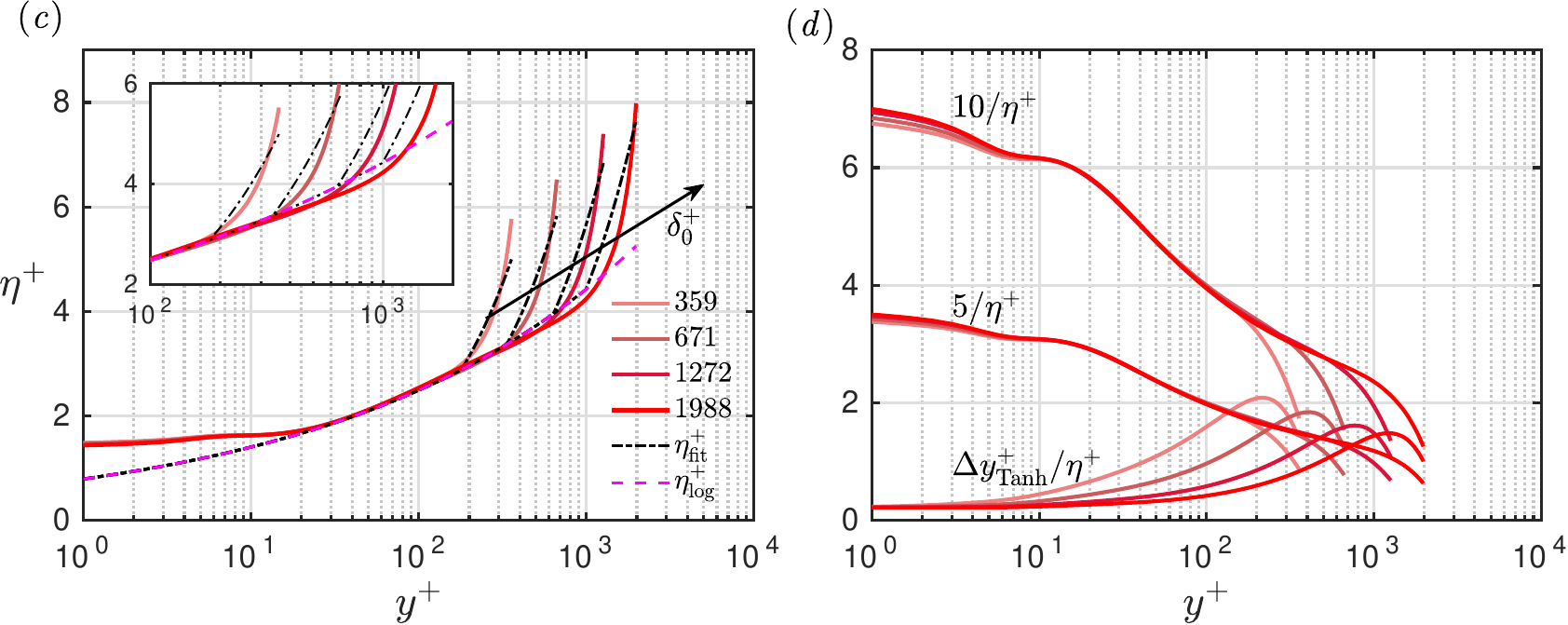}
  \caption{Profiles of $\eta^+$ versus $y^+$ for (\textit{a}) turbulent channel flow from \cite{moser1999direct} ($\delta^+_0 = 392$), \cite{hoyas2006scaling} ($\delta^+_0 = 2000$), and \cite{lee2015direct} ($\delta^+_0 = 1000, 5200$), and (\textit{c}) ZPG TBL from \cite{schlatter2010assessment} ($\delta^+_0 = 359, 671, 1272$) and \cite{sillero2013one} ($\delta^+_0 = 1988$); the dashed-dotted lines plot $\eta^+_\mathrm{fit}$ (\ref{eq:eta_fit}). (\textit{b,d}) Ratios of a conventional Cartesian DNS grid over $\eta^+$ for turbulent channel flow and ZPG TBL, respectively, where we assume $\Delta x^+ = 10, \Delta z^+ = 5$, and a hyperbolic tangent $y$-grid mapping $\Delta y^+_\mathrm{Tanh}$ (\ref{eq:tanh}), with $\Delta y^+_w = 0.3$ and $\Delta y^+_{\delta_0} = 8.0$.}
  \label{fig:eta_Dxp_Dzp}
\end{figure}



The conventional DNS grid sizes ($\Delta x^+ \lesssim 10, \Delta y^+_w \lesssim 1, \Delta z^+ \lesssim 5$) are based on the resolution requirements of the near-wall scales up the buffer region ($y^+ \lesssim 20$), with the smallest Kolmogorov scale ($\eta^+ \simeq 1.6$, figures~\ref{fig:eta_Dxp_Dzp}\textit{a,c}). For instance, DNS of \cite{kim1987turbulence} with $\Delta x^+ = 12, \Delta y^+_w = 0.05$ and $\Delta z^+  = 7$, yields $\Delta x^+/\eta^+ \simeq 7.5, \Delta y^+_w/\eta^+ \simeq 0.03$ and $\Delta z^+/\eta^+ \simeq 4.4$ near the wall~\citep{moin1998direct}. This resolution is shown to be sufficient to reliably resolve up to the second-order statistics with pseudo-spectral codes~\citep{moin1998direct,moser1999direct,lee2015direct}, as well as with fourth-order~\citep{chung2014idealised,rouhi2025spanwise} and second-order~\citep{bernardini2014velocity} finite-difference codes. Beyond the buffer region, turbulence scales grow with the wall distance; this is evident in the profiles of $\eta^+$ versus $y^+$ for turbulent channel flow (figure~\ref{fig:eta_Dxp_Dzp}\textit{a}), as well as zero pressure-gradient (ZPG) turbulent boundary layer (TBL) (figure~\ref{fig:eta_Dxp_Dzp}\textit{c}). We plot the profiles from the reference DNSs~\citep{moser1999direct,hoyas2006scaling,lee2015direct,schlatter2010assessment,sillero2013one}. In figures~\ref{fig:eta_Dxp_Dzp}(\textit{b,d}), we plot the ratios $\Delta x^+/\eta^+, \Delta y^+_\mathrm{Tanh}/\eta^+, \Delta z^+/\eta^+$ by assuming a Cartesian grid with constant $\Delta x^+ = 10, \Delta z^+ = 5$, and a hyperbolic-tangent $y$-grid mapping, as widely chosen for DNSs of wall turbulence~\citep{park1999effects,zhu2025influences,rouhi2025spanwise} 
\begin{align}
\Delta y^+_\mathrm{Tanh}(j) &= \frac{\delta^+_0}{N_{y_\mathrm{Tanh}}}\frac{\alpha}{\tanh(\alpha)} \left\{ 1 - \tanh^2\left[ \alpha \left( \frac{j}{N_{y_\mathrm{Tanh}}} -1 \right) \right] \right\} \tag{1.1\textit{a}} \label{eq:tanh} \\
\alpha &= \atanh \left[ \left( 1 - \frac{\Delta y^+_w}{\Delta y^+_{\delta_0}} \right)^{1/2} \right], \quad N_{y_\mathrm{Tanh}} = \frac{\alpha}{\tanh(\alpha)}\frac{\delta^+_0}{\Delta y^+_{\delta_0}}, \tag{1.1\textit{b,c}} \label{eq:alpha_N}
\end{align}
where the integer $j$ varies from $1$ to $N_{y_\mathrm{Tanh}}$. Throughout this manuscript, $\delta_0$ denotes full-channel half-height (walls at the top and bottom boundaries), open-channel height (wall at the bottom boundary only), or the TBL thickness at the target location. In other words, $\delta^+_0 \equiv \delta_0 u_{\tau_0}/\nu$ is the target friction Reynolds number with friction velocity $u_{\tau_0}$. In (\ref{eq:alpha_N}), $\Delta y^+_w$ and $\Delta y^+_{\delta_0}$ are respectively the grid sizes at the bottom wall, and at the full-channel centerline (open-channel top boundary or edge of the TBL). For $\Delta y^+_\mathrm{Tanh}/\eta^+$ in figures~\ref{fig:eta_Dxp_Dzp}(\textit{b,d}), we set $\Delta y^+_w = 0.3, \Delta y^+_{\delta_0} = 8.0$, being common choices~\citep{hoyas2006scaling,alcantara2021direct,rouhi2025spanwise}. Up to the buffer region $(y^+ \lesssim 20)$, $\Delta x^+/\eta^+ \simeq 6.0 - 7.0, \Delta y^+_\mathrm{Tanh}/\eta^+ \lesssim 0.6, \Delta z^+/\eta^+ \simeq 3.0 - 3.5$.  Beyond $y^+ \simeq 20$, $\eta^+$ increases with $y^+$ and the Cartesian grid over resolves $\eta^+$; by $y^+ = \delta^+_0$, $\Delta x^+/\eta^+ \simeq \Delta y^+_\mathrm{Tanh}/\eta^+ \lesssim 2.0$ and $\Delta z^+/\eta^+ \lesssim 1.0$. In other words, increasing $\delta^+_0$ with a Cartesian grid, further over-resolves local $\eta^+$ with $y^+$. Assessment of figure~\ref{fig:eta_Dxp_Dzp} highlights that the unstructured-grid solvers could be exploited to generate optimal grids proportional to $\eta^+$.



Setting the local grid size proportional to $\eta^+$ was the basis of \cite{yang2021grid}'s grid estimation analysis; they revised \cite{choi2012grid}'s estimation of the required number of grid points for DNS of TBL. In practice, this approach is hardly implemented in unstructured-grid solvers, as is evident through figures~\ref{fig:grids_intro_p1} and \ref{fig:grids_intro_p2}. \cite{pirozzoli2021natural} devised a $y$-grid mapping proportional to $\eta^+$. 
\begin{align}
 \Delta y^+_\mathrm{PO}(j) = \frac{1}{\left[1+(j/j_b)^2\right]^2}\left\{ \left[ 1 - \left( j/j_b \right)^2 \right] \Delta y^+_w + \frac{2}{3}\left( 0.6 C_y  \right)^{4/3} \frac{j^{7/3}}{j^2_b} \left[ 5 + 2\left( j/j_b \right)^2 \right] \right\}, \tag{1.2}  \label{eq:PO21}
 \end{align}
where $j$ varies from $1$ to $N_{y_\mathrm{PO}}$, as obtained from integrating (\ref{eq:PO21}), equation 5 in \cite{pirozzoli2021natural}
\begin{align}
 \frac{\left( 0.6 C_y \right)^{4/3}}{j^2_b} N^{10/3}_{y_\mathrm{PO}} - \frac{\delta^+_0}{j^2_b} N^2_{y_\mathrm{PO}} + \Delta y^+_w N_{y_\mathrm{PO}} = \delta^+_0. \tag{1.3} \label{eq:N_yPO21}
\end{align}
The mapping (\ref{eq:PO21}) blends a uniform $\Delta y^+_w$ near the wall with $\Delta y^+ = C_y \eta^+_\mathrm{log}$ beyond the buffer region ($y^+ \gtrsim 50$), where $\eta^+_\mathrm{log} = (\kappa y^+)^{1/4}$ with $\kappa = 0.4$ was considered as a semi-empirical fit for $\eta^+$ in the log region~\citep{jimenez2018coherent,lee2019spectral}; the parameter $j_b$ controls the blending height. \cite{pirozzoli2021natural} tested (\ref{eq:PO21}) for DNSs of smooth-wall turbulent pipe flows at $180 \lesssim \delta^+_0 \lesssim 1140$. With $C_y = 1.5$, $\Delta y^+_w = 0.05$ and $j_b = 16$, the wall-normal grid points were reduced to half of a hyperbolic-tangent $y$-grid mapping, and the time-step size was increased by two times, hence four times saving in the computational cost. The mapping (\ref{eq:PO21}) was extended to compressible flows~\citep{ceci2023natural}, and was applied to high Reynolds number incompressible turbulent pipe flows~\citep{pirozzoli2021one,pirozzoli2022dns,pirozzoli2023prandtl,pirozzoli2024streamwise}, turbulent open-channel flow~\citep{pirozzoli2023searching}, and compressible TBLs~\citep{cogo2022direct,cogo2023assessment}. These studies were conducted using finite-difference codes with structured grids, hence only the $y$-grid could be stretched proportional to $\eta^+$. To afford DNS of TBL over a bump, \cite{prakash2024streamline} implemented an unstructured grid in PHASTA, with wedge and tetrahedral elements proportionate to $\eta^+$; local $\eta^+$ was obtained from the TBL data of \cite{spalart1988direct}. The grid consisted of three layers; layer one, with fixed $\Delta x^+ = 15, \Delta z^+ = 6$, but growing $\Delta y^+$ first with a growth factor of $1.025$, and then following $\Delta y^+ = 2\eta^+$ until it reaches $\Delta z^+$; layer two, with fixed $\Delta x^+$ but growing $\Delta y^+ = \Delta z^+ = 2\eta^+$ until they reach $\Delta x^+$; layer three, with growing $\Delta x^+  = \Delta y^+ = \Delta z^+= 2\eta^+$. The resulting number of grid points was a third of the one with a structured grid.


In the present study, we formulate a grid-generation framework based on $\eta^+$, termed $\eta$-grid. We aim to make the grid implementable in widely used unstructured-grid solvers, and applicable for DNSs of turbulent flows over smooth and uneven surfaces at any Reynolds number. Towards these aims, we revise the semi-empirical fit for $\eta^+$ by processing the available DNS data of turbulent channel flow and TBL (\S\ \ref{sec:eta_fit}). We formulate $\eta$-grid for application to turbulent channel flow and TBL over a smooth wall (\S\ \ref{sec:grid_framework}), and over riblets (\S\ \ref{sec:grid_riblets}). Our formulation for riblets is based on careful study of the near-wall physics; it also satisfies the resolution constraints for DNS of rough-wall turbulent flows. We prescribe an implementation approach for $\eta$-grid in FVM (hence extendable to FEM) and SEM solvers (\S\ \ref{sec:grid_generation}); our implementation is based on hexahedral elements, to make it applicable to the commonly used SEM solvers (Nek5000, NekRS). After careful study of the grid parameters (\S\ \ref{sec:yp_study}), we test it on a campaign of cases using a recently developed SEM solver (SOD2D) and the FVM solver OpenFOAM. Our test cases are smooth-wall turbulent channel flow up to $\delta^+_0 = 1000$ (\S\ \ref{sec:grid_saving_smooth}), smooth-wall ZPG TBL up to $\delta^+_0 = 740$ (\S\ \ref{sec:smooth_tbl}), and turbulent channel flow and ZPG TBLs over various riblet geometries at $\delta^+_0 = 400$ (\S\ \ref{sec:riblet_channel}, \ref{sec:riblet_tbl}). We compare the results with the reference DNS and experimental data.
The results with the $\eta$-grid are as accurate as those with a finer grid or a Cartesian grid ($\lesssim 1\%$ difference). Any difference with the reference data is related to the different numerical schemes or tripping techniques for TBLs. For all test cases, we predict the number of grid points from $\eta$-grid ($N_\eta$), and compare with those from the Cartesian grid, either with $\Delta y^+_\mathrm{Tanh}$ (\ref{eq:tanh}) mapping ($N_\mathrm{Tanh}$) or $\Delta y^+_\mathrm{PO}$ (\ref{eq:PO21}) mapping ($N_\mathrm{PO}$), with the results presented in \S\ \ref{sec:grid_saving_smooth}, \ref{sec:grid_saving_tbl}, \ref{sec:riblet_grid_save}. For turbulent channel flow or TBL, by $\delta^+_0 = 6000$, $N_\eta/N_\mathrm{Tanh}$ drops to $0.1$ over a smooth wall, and it drops to $0.04$ over drag-reducing triangular riblets with tip angle $\ang{60}$ and $s^+ = 15$.

\section{Methodology}
\subsection{Fits for the Kolmogorov scale}\label{sec:eta_fit}
The semi-empirical fit $\eta^+_\mathrm{log} = (\kappa y^+)^{0.25}$ is obtained by assuming a balance between turbulent kinetic energy production $\mathcal{P}^+_\mathcal{K} \simeq (\kappa y^+)^{-1}$ and its dissipation rate $\varepsilon_\mathcal{K} = {\eta^+}^{-4}$ in the logarithmic region \citep{pope2000turbulent}. \cite{lee2019spectral} slightly revised the scaling of $\eta^+_\mathrm{log}$ based on their DNSs of turbulent channel flow up to $\delta^+_0 = 5200$, that is $\eta^+_\mathrm{log} = (\kappa y^+)^{0.266}$ with $\kappa = 0.384$. The agreement between the revised $\eta^+_\mathrm{log}$ and the actual $\eta^+$ from DNS persists up to $y^+ \simeq \delta^+_0/2$, as raised by \cite{anderson2021uniform}, and confirmed in figure~\ref{fig:eta_Dxp_Dzp}(\textit{a}). For $y^+ \gtrsim \delta^+_0/2$, \cite{anderson2021uniform} propose $\eta^+ \propto {y^+}^{0.5}$ for turbulent channel flow. For ZPG TBL, we processed the DNS data of \cite{schlatter2010assessment} and \cite{sillero2013one} up to $\delta^+_0 \simeq 2000$ (figure~\ref{fig:eta_Dxp_Dzp}\textit{c}). For $y^+ \le \delta^+_0/2$, $\eta^+_\mathrm{log} = (\kappa y^+)^{0.25}$ fits well with the data; for $y^+ > \delta^+_0/2$, $\eta^+ \propto {y^+}^{0.8}$ yields close agreement with the DNS profiles. Therefore, we propose the following semi-empirical fits for turbulent channel flow and ZPG TBL.
\begin{align}
 \eta^+_\mathrm{fit} = \begin{cases}
(\kappa y^+)^\beta & 20 \lesssim y^+ \le \dfrac{\delta^+_0}{2}\\
C_\eta {y^+}^\gamma &  \dfrac{\delta^+_0}{2} < y^+  \le \delta^+_0
\end{cases}, \quad C_\eta = \kappa^\beta \left(\frac{\delta^+_0}{2} \right)^{(\beta-\gamma)}, \tag{2.1\textit{a,b}} \label{eq:eta_fit}
\end{align}
where $\kappa = 0.384$, and for turbulent channel flow $\beta = 0.266, \gamma=0.5$, and for ZPG TBL $\beta = 0.25, \gamma =0.8$; $C_\eta$ ensures that $\eta^+_\mathrm{fit}$ is continuous at $y^+ = \delta^+_0/2$. Figures~\ref{fig:eta_Dxp_Dzp}(\textit{a,c}) show good agreements of $\eta^+_\mathrm{fit}$ with the DNSs of turbulent channel flow and ZPG TBL.


\begin{figure}
  \centering
 \includegraphics[width=\textwidth,trim={{-0.22\textwidth} {0.0\textwidth} {0.02\textwidth} {0.0\textwidth}},clip]{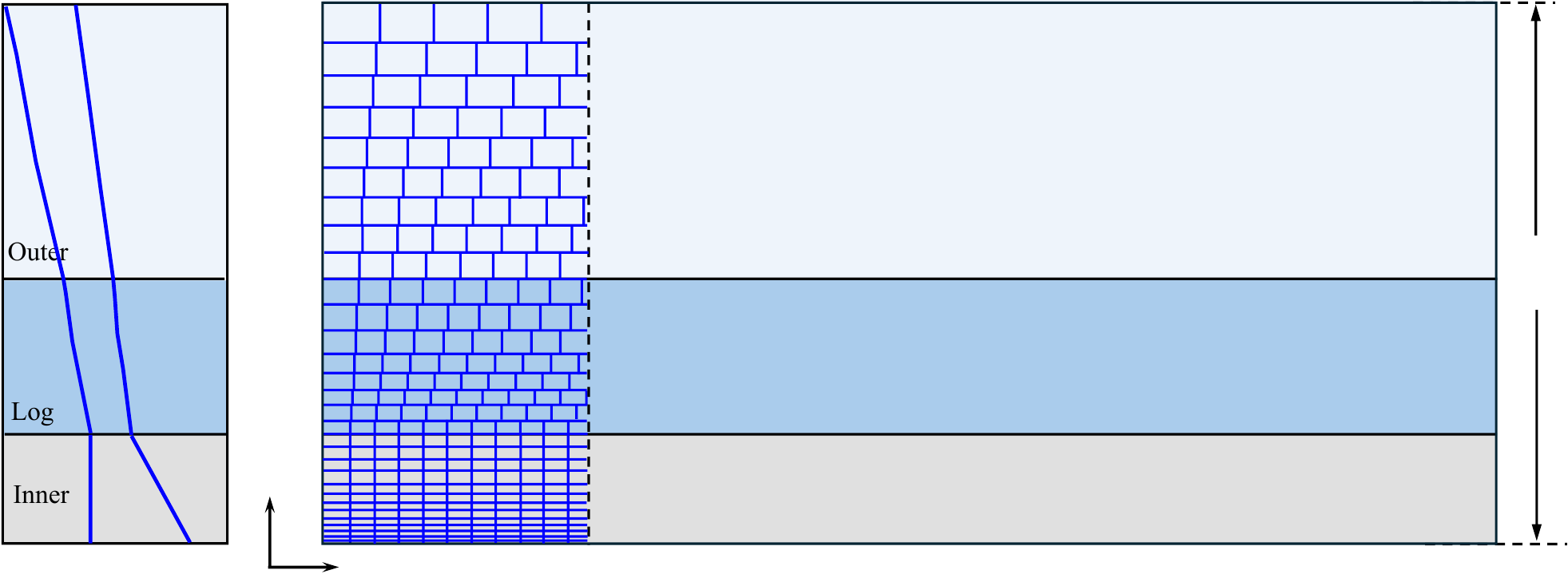}
 \put(-375,135){(\textit{a})}
 \put(-290,135){(\textit{b})}
   \put(-280,-5){$\Scale[0.78]{z}$}
 \put(-296,15){$\Scale[0.78]{y}$}
 \put(-358,5){$\Scale[0.75]{0}$}
 \put(-362,30){$\Scale[0.75]{y^+_\mathrm{in}}$}
 \put(-362,65){$\Scale[0.75]{\dfrac{\delta^+_0}{2}}$}
 \put(-360,128){$\Scale[0.75]{\delta^+_0}$}
 \put(-350,48){$\Scale[0.75]{\Delta z^+_\eta}$}
 \put(-322,48){$\Scale[0.75]{\Delta y^+_\eta}$}
 \put(-160,135){$\Scale[0.78]{N_{{yz}_\eta}}$}
 \put(-7,67){$\Scale[0.75]{\delta^+_0}$}
  \put(-195,18){$\Scale[0.75]{\dfrac{L_z/\delta_0}{C_z (\kappa y^+_\mathrm{in})^\beta}\left( \dfrac{1}{r} \right)\ln \left( \dfrac{C_y (\kappa y^+_\mathrm{in})}{\Delta y^+_w} \right)\delta^+_0}$}
  \put(-195,48){$\Scale[0.75]{\dfrac{L_z/\delta_0}{C_y C_z \kappa^{2\beta}(1-2\beta)}\left[ \left(\dfrac{\delta^+_0}{2}\right)^{(1-2\beta)} - {y^+_\mathrm{in}}^{(1-2\beta)} \right]\delta^+_0}$}
  \put(-211,88){$\Scale[0.75]{\mbox{TBL: }\dfrac{L_z/\delta_0}{C_y C_z \kappa^{2\beta}(1-2\gamma)}\left[ \dfrac{1}{2^{(2\gamma-2\beta)}} - \dfrac{1}{2^{(1-2\beta)}} \right]{\delta^+_0}^{(2-2\beta)}}$}
  \put(-211,115){$\Scale[0.75]{\mbox{Channel: }\dfrac{(L_z/\delta_0)\ln(2)}{C_y C_z \kappa^{2\beta}2^{(1-2\beta)}}{\delta^+_0}^{(2-2\beta)}}$}   
  \caption{Schematic illustration of $\eta$-grid. (\textit{a}) Profiles of $\Delta y^+_\eta$ (\ref{eq:Dyp}) and $\Delta z^+_\eta$ (\ref{eq:Dzp}), and (\textit{b}) the $\eta$-grid arrangement on a $yz$-plane; the number of grid points $N_{{yz}_\eta}$ on the $yz$-plane, as obtained from (\ref{eq:Nyz}), is reported for the inner, log and outer layers.}
  \label{fig:idealised_grid}
\end{figure}
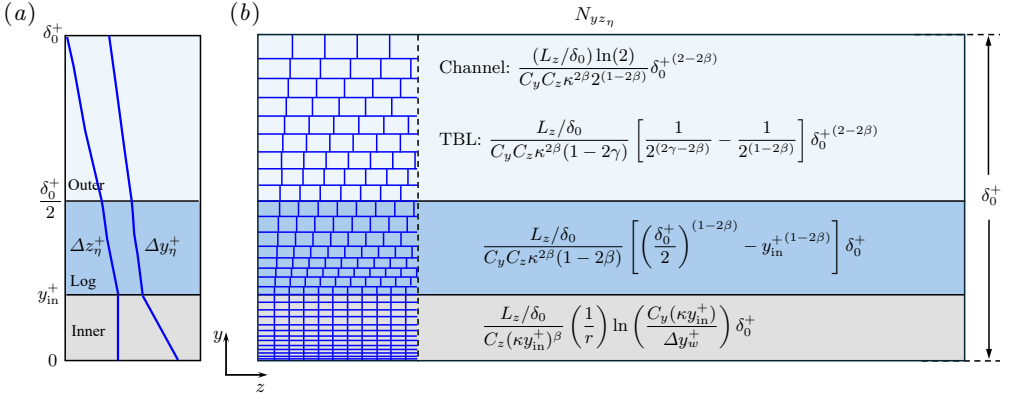

\subsection{Proposed grid}\label{sec:grid_framework}
We propose an unstructured $yz$-grid (figure~\ref{fig:idealised_grid}), termed $\eta$-grid, with the grid sizes denoted as $\Delta y^+_\eta$ and $\Delta z^+_\eta$. The grid consists of an inner layer with a fine grid and thickness $y^+_\mathrm{in}$, followed by increasing $\Delta y^+_\eta$ and $\Delta z^+_\eta$ proportional to $\eta^+_\mathrm{fit}$ (\ref{eq:eta_fit}).
\begin{align}
 \Delta y^+_\eta &= \begin{cases}
 \Delta y^+_w + ry^+ & 0 < y^+ \le y^+_\mathrm{in} \quad \mbox{Inner} \\
C_y (\kappa y^+)^\beta &  y^+_\mathrm{in} < y^+ \le \dfrac{\delta^+_0}{2} \quad \mbox{Log}  \\
C_y C_\eta {y^+}^\gamma &  \dfrac{\delta^+_0}{2} < y^+ \le \delta^+_0 \quad \mbox{Outer}
\end{cases}\tag{2.2\textit{a}} \label{eq:Dyp} \\
 \Delta z^+_\eta &= \begin{cases}
C_z (\kappa y^+_\mathrm{in})^\beta & 0 < y^+ \le y^+_\mathrm{in} \quad \mbox{Inner}  \\
C_z (\kappa y^+)^\beta &  y^+_\mathrm{in} < y^+ \le \dfrac{\delta^+_0}{2} \quad \mbox{Log} \\
C_z C_\eta {y^+}^\gamma &  \dfrac{\delta^+_0}{2} < y^+ \le \delta^+_0 \quad \mbox{Outer}
\end{cases} \tag{2.2\textit{b}} \label{eq:Dzp}
\end{align}
where $r = [C_y (\kappa y^+_\mathrm{in})^\beta - \Delta y^+_w]/y^+_\mathrm{in}$, and $y^+_\mathrm{in}, \Delta y^+_w, C_y$ and $C_z$ are the grid parameters. The names inner, log and outer for the layers of $\eta$-grid are merely chosen for distinguishing them; nevertheless, the names are consistent with the naming convention in the wall-turbulence literature. In figure~\ref{fig:idealised_grid}(\textit{a}), we plot $\Delta y^+_\eta$ and $\Delta z^+_\eta$, and in figure~\ref{fig:idealised_grid}(\textit{b}) we depict the resulting grid elements distribution. In the inner layer, a Cartesian grid is generated with uniform $\Delta z^+_\eta$, and $\Delta y^+_\eta$ follows a geometric progression with common ratio $r$. Beyond $y^+_\mathrm{in}$, $\Delta y^+_\eta$ and $\Delta z^+_\eta$ increase proportional to $\eta^+_\mathrm{fit}$. The resulting grid appears as a brick wall, where the bricks' width and height increase with the wall distance.

\begin{figure}
  \centering
 \includegraphics[width=\textwidth,trim={{0.0\textwidth} {0.02\textwidth} {0.0\textwidth} {0.0\textwidth}},clip]{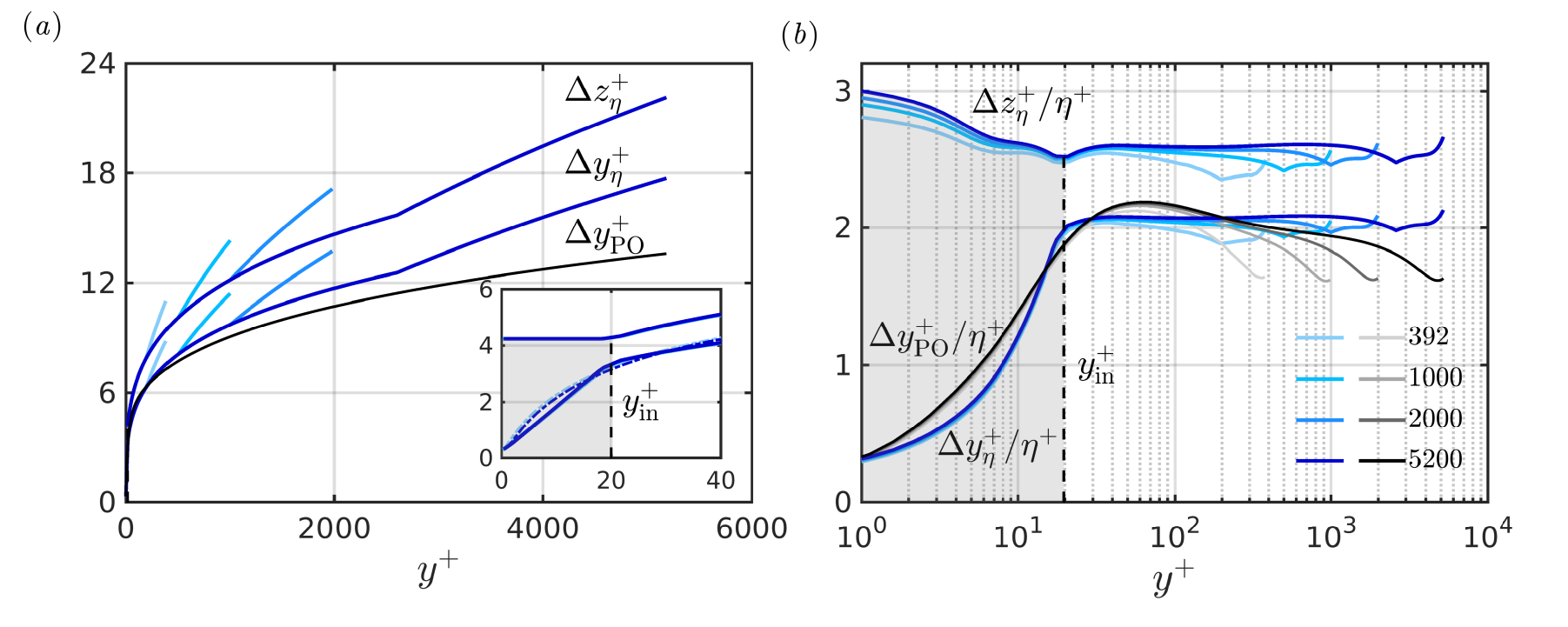}
 \includegraphics[width=\textwidth,trim={{0.0\textwidth} {0.0\textwidth} {0.0\textwidth} {0.02\textwidth}},clip]{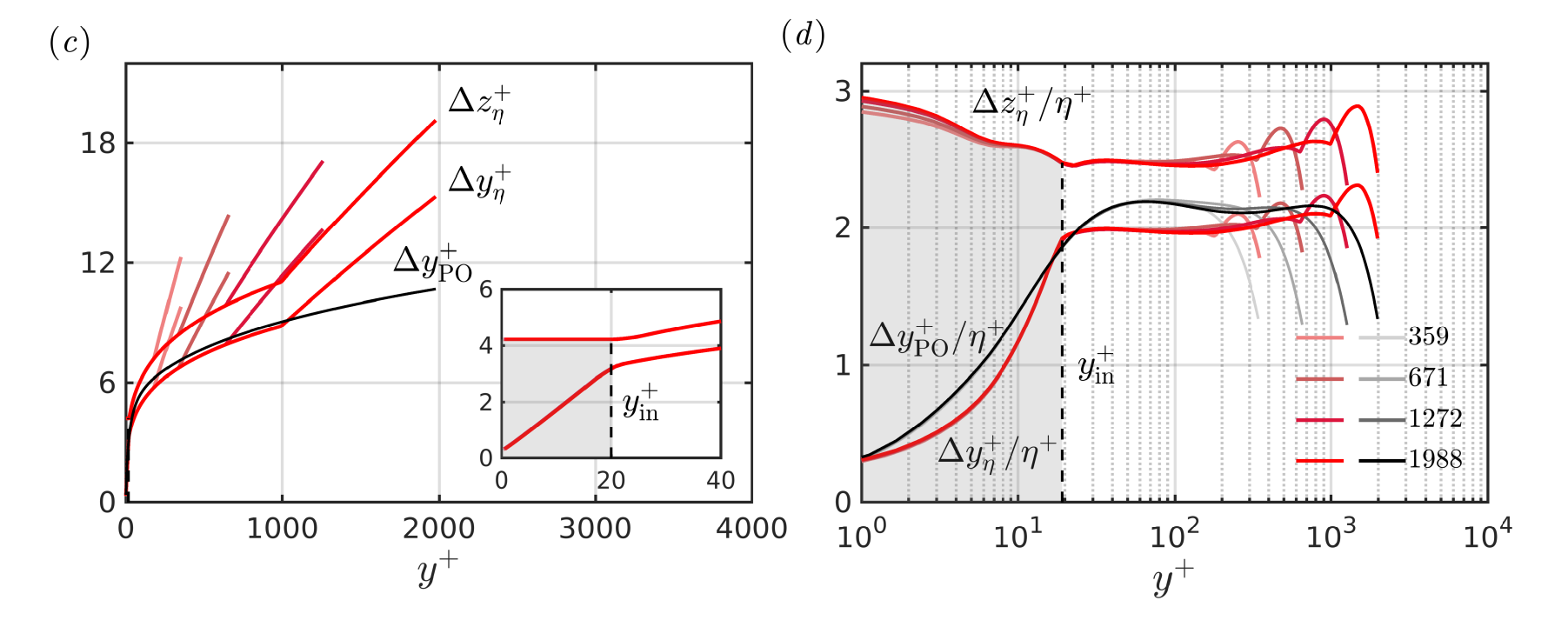} 
  \caption{Profiles of $\Delta y^+_\eta$ (\ref{eq:Dyp}) and $\Delta z^+_\eta$ (\ref{eq:Dzp}) for (\textit{a,b}) turbulent channel flow (blue profiles), and (\textit{c,d}) ZPG TBL (red profiles), with $y^+_\mathrm{in} = 20, \Delta^+_{y_w} = 0.3, C_y = 2.0$ and $C_z = 2.5$; the gray/black profiles plot $\Delta y^+_\mathrm{PO}$ (\ref{eq:PO21}) by \cite{pirozzoli2021natural}, with $\Delta y^+_w = 0.3, C_y = 2.0$ and $j_b = 16$. The grid profiles are plotted for the same $\delta^+_0$ values as the reference DNS cases from figure~\ref{fig:eta_Dxp_Dzp}. (\textit{b,d}) plot the ratios of $\Delta y^+, \Delta z^+$ over the profiles of $\eta^+$ from the reference DNS data (figures~\ref{fig:eta_Dxp_Dzp}\textit{a,c}).}
  \label{fig:Dyp_Dzp}
\end{figure}

In figure~\ref{fig:Dyp_Dzp}, we plot (2.2\textit{a,b}) for turbulent channel flow (figure~\ref{fig:Dyp_Dzp}\textit{a}) and ZPG TBL (figure~\ref{fig:Dyp_Dzp}\textit{c}) for the same $\delta^+_0$ values as those from the reference DNSs (figure~\ref{fig:eta_Dxp_Dzp}). In figures~\ref{fig:Dyp_Dzp}(\textit{b,d}) we plot the ratios $\Delta y^+_\eta/\eta^+, \Delta z^+_\eta/\eta^+$, where $\eta^+$ is from the reference DNSs. Our grid parameters are $y^+_\mathrm{in} = 20, \Delta y^+_w = 0.3, C_y = 2.0$ and $C_z = 2.5$. We also overlay $\Delta y^+_\mathrm{PO}$ (\ref{eq:PO21}) with $\Delta y^+_w = 0.3, C_y = 2.0$ and $j_b = 16$, which are comparable parameters to those for $\Delta y^+_\eta$. Both $\Delta y^+_\eta$ and $\Delta y^+_\mathrm{PO}$ grow from $0.2\eta^+$ at $y^+ = 1$ to a plateau of about $2.0 \eta^+$ in the log region. Differences between the two grids are noticeable for $y^+ > \delta^+_0/2$, especially for ZPG TBL (figures~\ref{fig:Dyp_Dzp}\textit{c,d}), owing to the different semi-empirical fit that we propose for $\eta^+$ (\ref{eq:eta_fit}). For $y^+> \delta^+_0/2$, $\Delta y^+_\eta/\eta^+ \simeq 2.0 - 2.3 \simeq C_y$, whereas $\Delta y^+_\mathrm{PO}/\eta^+$ drops to $1.3$. 

The number of grid points on a $yz$-plane with $\eta$-grid (2.2\textit{a,b}) can be derived from
\begin{align}
 N_{{yz}_\eta} = \bigintssss_0^{L^+_z} \bigintssss_0^{\delta^+_0} \frac{dy^+ dz^+}{\Delta y^+_\eta \Delta z^+_\eta} = L^+_z \bigintssss_0^{\delta^+_0} \frac{dy^+}{\Delta y^+_\eta \Delta z^+_\eta}, \tag{2.3} \label{eq:Nyz}
\end{align}
where $(dy^+ dz^+)/(\Delta y^+_\eta \Delta z^+_\eta)$ is the number of grid points in a square with area $dy^+ dz^+$. This is a common approach for obtaining the number of grid points for DNS~\citep{choi2012grid,yang2021grid}. In figure~\ref{fig:idealised_grid}(\textit{b}), we express $N_{{yz}_\eta}$ in each layer in terms of $\delta^+_0$. We will show that our predicted $N_{{yz}_\eta}$ (\ref{eq:Nyz}) is in excellent agreement with the actual number of grid points from our grid generation approach (\S\ \ref{sec:grid_generation}), for turbulent channel flow (\S\ \ref{sec:grid_saving_smooth}) and TBL (\S\ \ref{sec:grid_saving_tbl}).

 \begin{figure}
  \centering
 \includegraphics[width=\textwidth,trim={{0.0\textwidth} {0.02\textwidth} {0.0\textwidth} {0.0\textwidth}},clip]{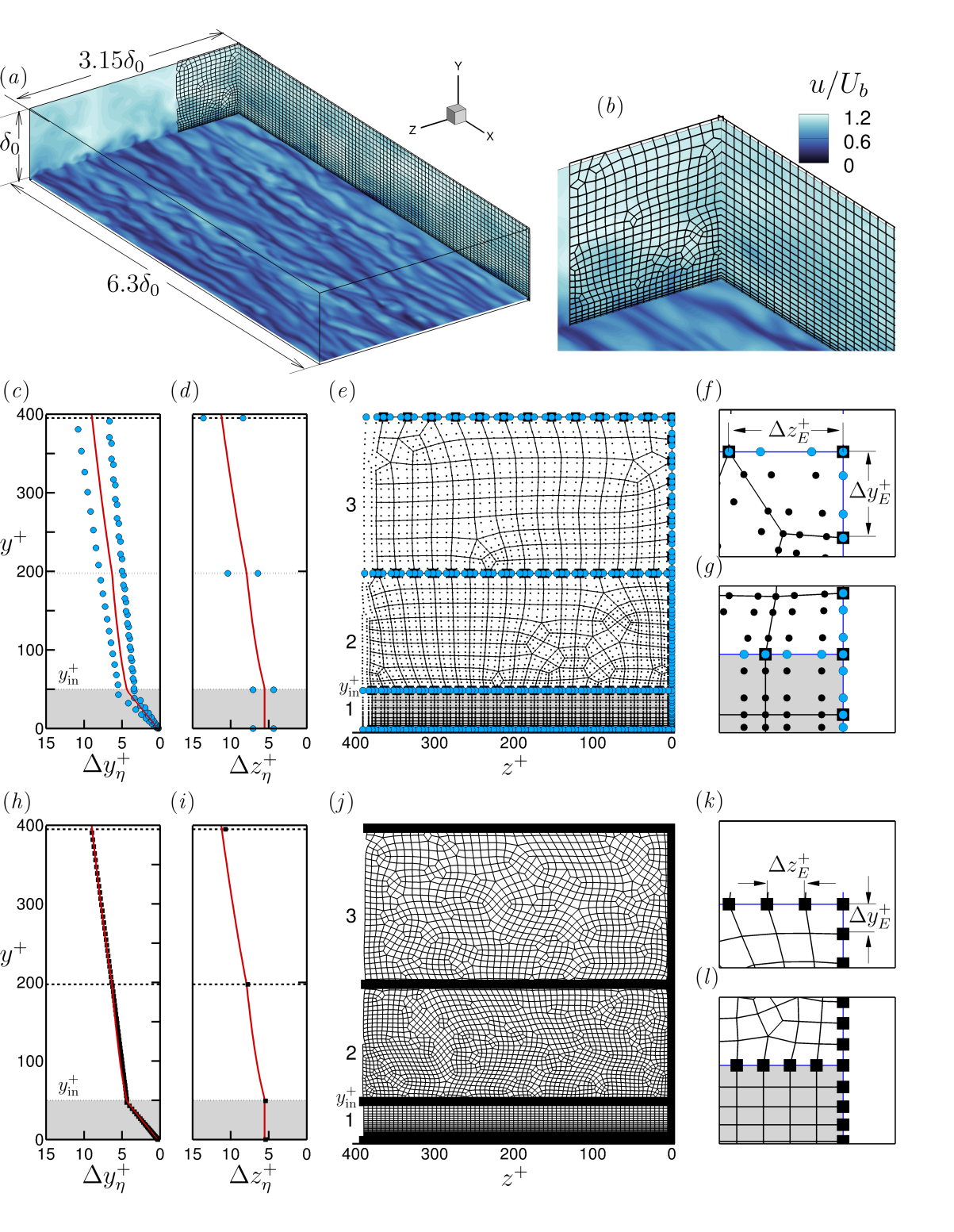} 
  \caption{Grid generation of (2.2\textit{a,b}) with $\Delta y^+_w = 0.3, y^+_\mathrm{in} = 50, C_y = 2.0$ and $C_z = 2.5$ for turbulent open-channel flow at $\delta^+_0 = 395$. (\textit{a,b}) Computational domain and the spectral elements on $xy$ and $yz$ planes. (\textit{c-g}) Distribution of the spectral elements, their vertices (black squares) and the 3\textsuperscript{rd}-order polynomial points (blue and black dots) for the SEM grid. In (\textit{c}), the solid line plots $\Delta y^+_\eta$ (\ref{eq:Dyp}), and the blue dots are $\Delta y^+$ between the polynomial points along the vertical edge at $z^+ = 0$. In (\textit{d}), the solid line plots $\Delta z^+_\eta$ (\ref{eq:Dzp}), and the blue dots are $\Delta z^+$ between the polynomial points along the horizontal edges at $y^+ = 0, y^+_\mathrm{in}, \delta^+_0/2$ and $\delta^+_0$. (\textit{f,g}) are the close-up views of (\textit{e}) near $z^+ = 0, y^+ = \delta^+_0/2$ (\textit{f}) and $z^+ = 0, y^+ = y^+_\mathrm{in}$ (\textit{g}); in (\textit{f}), $\Delta y^+_E, \Delta z^+_E$ are respectively the vertical and horizontal spacing between the element vertices. (\textit{h-l}) are similar to (\textit{c-g}), but illustrate the distribution of the elements and their vertices (black squares) for the FVM (and FEM) grid.}
  \label{fig:grid_viz}
\end{figure}

\subsection{Grid generation}\label{sec:grid_generation}
We generate $\eta$-grid (2.2\textit{a,b}) using the open-source grid-generation package Gmsh~\citep{geuzaine2009gmsh}. Figure~\ref{fig:grid_viz} demonstrates our grid-generation approach for turbulent open-channel flow at $\delta^+_0 = 395$. We generate grids for SEM (figures~\ref{fig:grid_viz}\textit{c-g}), and FVM (figures~\ref{fig:grid_viz}\textit{h-l}), which is also applicable for FEM. Figures~\ref{fig:grid_viz}(\textit{c,d,h,i}) plot $\Delta y^+_\eta$ (\ref{eq:Dyp}) and $\Delta z^+_\eta$ (\ref{eq:Dzp}) with $\Delta y^+_w = 0.3, y^+_\mathrm{in} = 50, C_y = 2.0$ and $C_z = 2.5$ (solid lines). To control the grid size following these profiles, we divide the domain into several blocks in the $y$-direction (here 3 blocks). The aim is to generate a grid with a close topology as the ideal grid in figure~\ref{fig:idealised_grid}. We control the distribution of the elements' vertices along the vertical and horizontal edges of each block (black squares in figures~\ref{fig:grid_viz}\textit{e,f,g,j,k,l}). We divide the open channel into block 1 $[0,y^+_\mathrm{in}]$, 2 $[y^+_\mathrm{in}, \delta^+_0/2]$ and 3 $[\delta^+_0/2, \delta^+_0]$, covering the inner layer, log layer and outer layer, respectively. We decide the number of blocks such that in each block, $\Delta y^+_\eta$ shows an almost linear variation with $y^+$. Then, we distribute the elements vertices along the vertical edges following a geometric progression $\Delta y^+_E(j) = \Delta y^+_{E_0} a^{(j-1)}$, where $\Delta y^+_E$ is the viscous spacing between the elements vertices (figures~\ref{fig:grid_viz}\textit{f,k}), $\Delta y^+_{E_0}$ is the initial spacing at the start of the block, and $a$ is the common ratio. For the FVM grid, $\Delta y^+_E = \Delta y^+_\eta$ (\ref{eq:Dyp}), as shown in figure~\ref{fig:grid_viz}(\textit{h}). For the SEM grid, $\Delta y^+_E = (N_p - 1)\Delta y^+_\eta$, where $N_p$ is the number of Gauss–Lobatto–Legendre (GLL) integration points (in 1D) for polynomial order $p$. Here we pick $N_p = 4$, hence $\Delta y^+_E = 3\Delta y^+_\eta$; that means the average spacing between the polynomial nodes of each element is $\Delta y^+_\eta$ (\ref{eq:Dyp}), as is the case in figure~\ref{fig:grid_viz}(\textit{c}). For $\Delta z^+_\eta$, we uniformly distribute the vertices along the horizontal edges of each block. Based on the $y^+$ of each horizontal edge, we set the viscous spacing between the elements' vertices to $\Delta z^+_E = \Delta z^+_\eta$ (\ref{eq:Dzp}) for the FVM grid (figure~\ref{fig:grid_viz}\textit{i}), and $\Delta z^+_E = (N_p - 1)\Delta z^+_\eta = 3\Delta z^+_\eta$ for the SEM grid, hence the average spacing between its polynomial nodes is $\Delta z^+_\eta$ (figure~\ref{fig:grid_viz}\textit{d}). After distributing the elements' vertices, we generate quadrilateral mesh elements on a $yz$-plane using the Blossom-Quad algorithm~\citep{remacle2012blossom}, followed by extruding the mesh in the $x$-direction, with a fixed $\Delta x^+$, to generate hexahedral elements. Throughout the manuscript, $\Delta x^+$ denotes the actual value of $\Delta x^+_E$ for the OpenFOAM runs (FVM grid), and it denotes the actual value of $\Delta x^+_E/3$ for the SOD2D runs (SEM grid); the same notation applies to $\Delta y^+$ and $\Delta z^+$. Except figure~\ref{fig:grid_viz}, all the grid visualisations in this manuscript show the spectral elements with $N_p = 4$, without showing the polynomial points (e.g.\ figure~\ref{fig:yp_study}\textit{c}). 


\subsection{Governing equations and computational solvers}
The governing equations are the continuity and momentum equations for an incompressible fluid with density $\rho$ and kinematic viscosity $\nu$
\begin{align}
 \boldsymbol{\nabla}\boldsymbol{\cdot}\mathbf{u} = 0, \quad
 \frac{\partial \mathbf{u}}{\partial t} + \boldsymbol{\nabla}\boldsymbol{\cdot}\mathbf{(uu)} = -\frac{1}{\rho}\boldsymbol{\nabla}p + \nu \nabla^2 \mathbf{u}. \tag{2.4\textit{a,b}} \label{eq:cont_mom}
\end{align}
In our notation, $\mathbf{u} = (u,v,w)$ is the velocity vector and $p$ is pressure. We solve (\ref{eq:cont_mom}) using an SEM solver SOD2D~\citep{gasparino2024sod2d} and the FVM solver OpenFOAM. In SOD2D, spatial discretisation is based on spectral formulation of the Continuous Galerkin Finite Element method~\citep{zienkiewicz2005finite}, with GLL spacing between the polynomial nodes. An anti-aliasing operator splitting is applied to the convection terms~\citep{kennedy2008reduced}. Time advancement is via 4\textsuperscript{th}-order Runge-Kutta scheme, and equations (\ref{eq:cont_mom}) are marched using fractional-step algorithm~\citep{chorin1967numerical,perot1993analysis}. For OpenFOAM, spatial discretisation is via a second-order central differencing scheme, time advancement is via second-order Crank-Nicolson scheme, and (\ref{eq:cont_mom}) are marched via the pressure-implicit splitting operators (PISO) algorithm~\citep{issa1986solution}.

\section{Smooth wall turbulent channel flow}\label{sec:smooth_channel}


\begin{table}
\centering
 \begin{tabular}{c|ccc|ccc|ccc}
              \multicolumn{10}{c}{open-channel cases at $\delta^+_0 = 395$ (\S\ \ref{sec:yp_study})} \vspace{0.2cm}  \\
          &  \multicolumn{3}{c|}{Set 1} & \multicolumn{3}{c|}{Set 2} & \multicolumn{3}{c}{Set 3} \\
         & \multicolumn{3}{c|}{SOD $C_y = 2.0, Cz = 2.5$} & \multicolumn{3}{c|}{OF $C_y = 2.0, Cz = 2.5$} & \multicolumn{3}{c}{OF $C_y = 1.6, Cz = 2.0$} \\
         $y^+_\mathrm{in}$ & $\Delta y^+$ & $\Delta z^+$ & $\varepsilon_{C_f}$ &
                                    $\Delta y^+$ & $\Delta z^+$ & $\varepsilon_{C_f}$&
                                    $\Delta y^+$ & $\Delta z^+$ & $\varepsilon_{C_f}$  \\ 
         $10$ & $[0.2, 8.3]$ & $[3.6, 10.9]$ & $0.1\%$ &
                                            $[0.3, 8.9]$ & $[3.7, 11.1]$ & $2.6\%$ &
                                         $-$   &        $-$       &     $-$    \\
   $20$  & $[0.3, 8.7]$ & $[4.2, 10.9]$ & $2.3\%$ &
                                      $[0.3, 9.0]$ & $[4.2, 10.9]$ & $2.0\%$ &
                                      $[0.3, 7.0]$ & $[3.4, 8.8]$ & $2.3\%$  \\
   $50$ & $[0.3, 9.4]$ & $[5.4, 10.6]$ & $0.7\%$ &
                                 $[0.3, 9.0]$ & $[5.4, 10.9]$ & $1.8\%$ &
                                 $[0.3, 7.1]$ & $[4.3, 8.6]$ & $1.3\%$  \\
   $100$  & $[0.3, 8.6]$ & $[6.5, 10.9]$ & $0.3\%$ &
                                 $[0.3, 8.9]$ & $[6.4, 10.9]$ & $3.0\%$ &
                                 $[0.3, 7.3]$ & $[5.2, 8.8]$ & $1.9\%$  \\ \\
        & \multicolumn{3}{c|}{SOD Cartesian} & \multicolumn{3}{c|}{OF Cartesian} &
                                                 \multicolumn{3}{c}{} \\
           &  $\Delta y^+$ & $\Delta z^+$ & $\varepsilon_{C_f}$ &
                     $\Delta y^+$ & $\Delta z^+$ & $\varepsilon_{C_f}$ &
                                                 \multicolumn{3}{c}{} \\ 
          &  $[0.3, 8.4]$ & $4.9$ & $0.4\%$ &
                                    $[0.3, 8.2]$ & $4.9$ & $2.2\%$ &
                                                 \multicolumn{3}{c}{}  \\ \hline                                    
          \multicolumn{10}{c}{full-channel cases at $\delta^+_0 = 1000$ (\S\ \ref{sec:grid_saving_smooth})} \vspace{0.2cm}   \\
         & \multicolumn{3}{c|}{SOD $C_y = 2.0, Cz = 2.5$} & \multicolumn{3}{c|}{OF $C_y = 2.0, Cz = 2.5$} & \multicolumn{3}{c}{} \\
         $y^+_\mathrm{in}$ &  $\Delta y^+$ & $\Delta z^+$ & $\varepsilon_{C_f}$ &
                                    $\Delta y^+$ & $\Delta z^+$ & $\varepsilon_{C_f}$&
                                     &  &   \\ 
         $50$  & $[0.3, 11.5]$ & $[5.5, 14.0]$ & $0.3\%$ &
                                            $[0.3, 11.4]$ & $[5.5, 14.3]$ & $1.9\%$ &
                                           &              &       \\          
    \end{tabular}
\caption{Simulation cases of smooth-wall turbulent channel flow. Upper cases are turbulent open-channel flow at $\delta^+_0 = 395$ for studying the grid parameters $y^+_\mathrm{in}, C_y$ and $C_z$ (\S\ \ref{sec:yp_study}, figure~\ref{fig:yp_study}). Lower cases are turbulent full-channel flow at $\delta^+_0 = 1000$ for showing the accouracy of $\eta$-grid (\S\ \ref{sec:grid_saving_smooth}, figure~\ref{fig:channel_re1000}). The cases are grouped based on the computational solver, SOD2D (SOD) or OpenFOAM (OF), and the type of grid, either Cartesian or $\eta$-grid with a specified $C_y, C_z$. For the $\eta$-grid cases, each row presents the cases with the same $y^+_\mathrm{in}$. For each case, $\varepsilon_{C_f}$ reports the percentage difference in $C_f$ relative to the DNSs of \cite{moser1999direct} ($\delta^+_0 = 395$) and \cite{lee2015direct} ($\delta^+_0 = 1000$).   
}
\label{tab:yp_study}
\end{table}


\subsection{Grid parameters study}\label{sec:yp_study}
We consider turbulent open-channel flow at $ \delta^+_0 = 395$ for testing the grid parameters $y^+_\mathrm{in}, C_y$ and $C_z$ in (2.2\textit{a,b}). We fix $\Delta x^+ = 10$ and $\Delta y^+_w = 0.3$; up to fourth-order statistics have shown small sensitivity to $\Delta y^+_w \le 1.0$~\citep{pirozzoli2021natural}. Figure~\ref{fig:grid_viz}(\textit{a}) illustrates the computational configuration and domain dimensions. Periodic boundary conditions are applied to the streamwise and spanwise directions, and no-slip ($u=v=w=0$) and free-slip ($\partial u/\partial y = v = \partial w/ \partial y = 0$) conditions to the bottom and top boundaries, respectively. We conduct three sets of calculations, arranged as separate columns in table~\ref{tab:yp_study}; set 1 is SOD2D with $C_y = 2.0, C_z = 2.5$, and sets 2 and 3 are OpenFOAM with $C_y = 2.0, C_z = 2.5$ and $C_y = 1.6, C_z = 2.0$, respectively. Each set consists of several cases with $y^+_\mathrm{in} = 10, 20, 50$ and $100$. Our choices for $C_y $ and $C_z $ are based on the thresholds from the past DNSs of turbulent channel flow and TBL, e.g.\ $\Delta y/\eta \lesssim 2.4, \Delta z/\eta \lesssim 3.6$ in \cite{pirozzoli2016passive}, or $\Delta y/\eta \lesssim 2.2$ in \cite{ceci2023natural}. With our choices for $C_z$ and $y^+_\mathrm{in}$, $3.4 \le \Delta z^+_\eta \le 6.5$ below $y^+_\mathrm{in}$, which is the typical range for DNS (figures~\ref{fig:grids_intro_p1} and \ref{fig:grids_intro_p2}). For each case, we report $\varepsilon_{C_f}$, the difference in skin-friction coefficient $C_f$ relative to the DNS of \cite{moser1999direct}. Also, with each solver we conduct a reference DNS case with a Cartesian grid, with $\Delta x^+ = 10, \Delta z^+ = 4.9$ and $\Delta y^+_\mathrm{Tanh}$ (\ref{eq:tanh}) with $\Delta y^+_w = 0.3, \Delta y^+_{\delta_0} = 8.0$. Comparison with the Cartesian grids allows us to evaluate the net effect of the $\eta$-grid on the solution accuracy, and circumvent the uncertainties due to the discretisation schemes.

\begin{figure}
  \centering
  \includegraphics[width=\textwidth,trim={{0.02\textwidth} {0.0\textwidth} {0.02\textwidth} {0.0\textwidth}},clip]{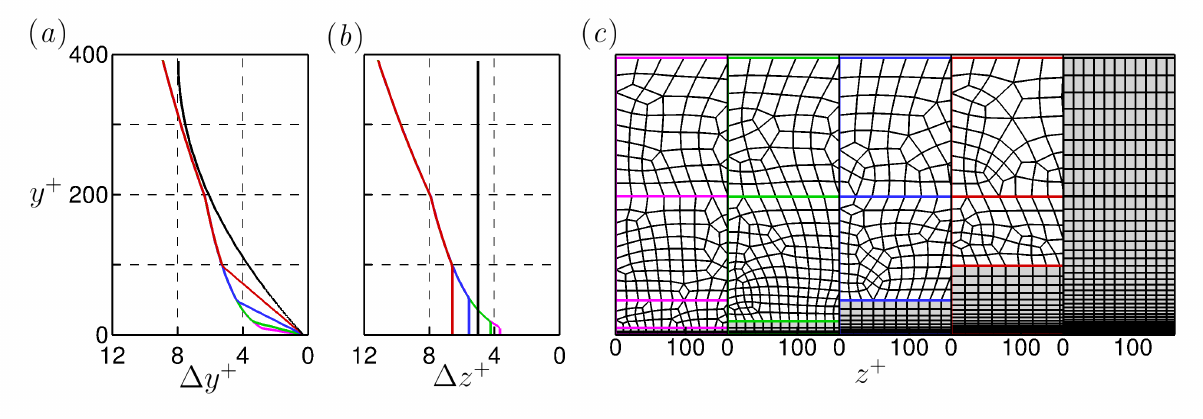}   
  \includegraphics[width=\textwidth,trim={{0.0\textwidth} {0.0\textwidth} {0.0\textwidth} {0.0\textwidth}},clip]{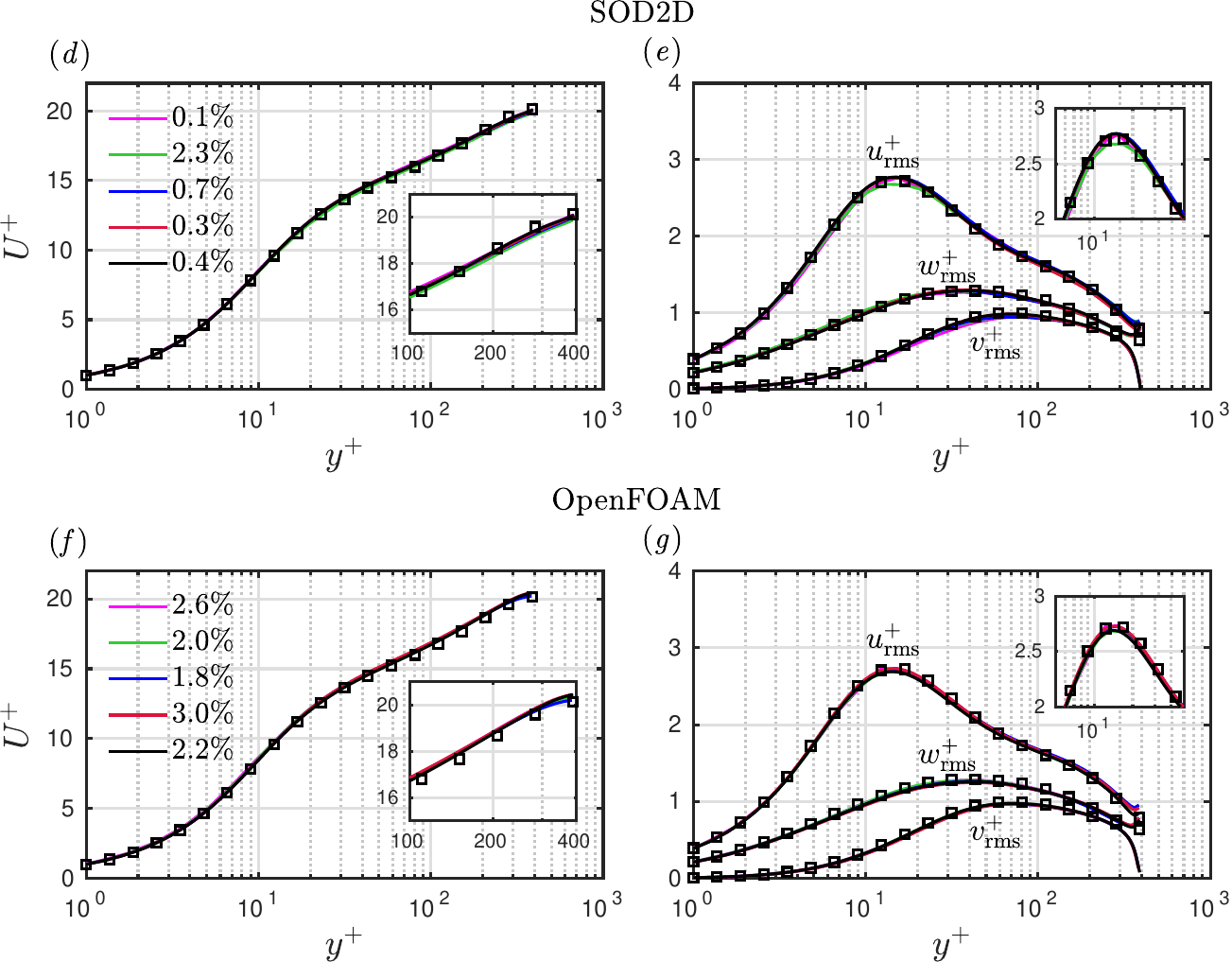} 
  \caption{Results from the simulation cases of turbulent open-channel flow at $\delta^+_0 = 395$ (table~\ref{tab:yp_study}), with the Cartesian grid (black), and $\eta$-grid (2.2\textit{a,b}) with $\Delta y^+_w = 0.3, C_y = 2.0, C_z = 2.5$ and $y^+_\mathrm{in} = 10$ (magenta), $20$ (green), $50$ (blue), and $100$ (red). Profiles of (\textit{a}) $\Delta y^+$ and (\textit{b}) $\Delta z^+$, and (\textit{c}) the resulting grid, presenting the spectral elements for SOD2D; the inner (Cartesian) layer is shaded in grey. Profiles of (\textit{d,f}) $U^+$, and (\textit{e,g}) $u^+_\mathrm{rms}, v^+_\mathrm{rms}, w^+_\mathrm{rms}$, from SOD2D runs (\textit{d,e}) and OpenFOAM runs (\textit{f,g}). The legends report the percentage difference in $C_f$ relative to the DNS of \cite{moser1999direct}.}
  \label{fig:yp_study}
\end{figure}

\begin{figure}
  \centering
  \includegraphics[width=\textwidth,trim={{0.0\textwidth} {0.0\textwidth} {0.0\textwidth} {0.0\textwidth}},clip]{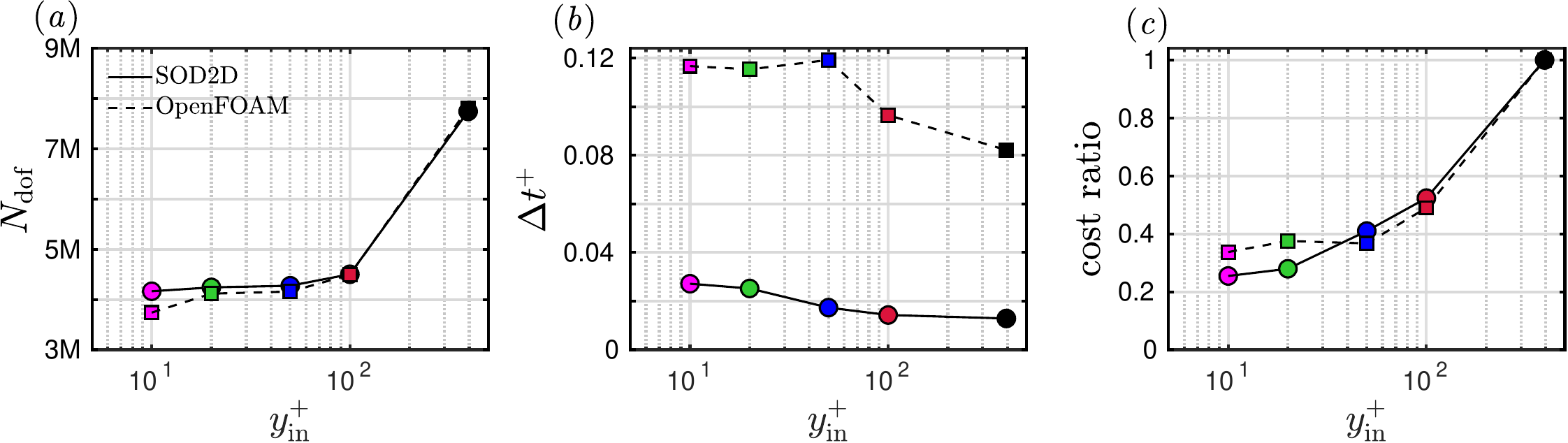}  
  \caption{Cost analysis for the cases from figure~\ref{fig:yp_study}, with $\eta$-grid and $C_y = 2.0, C_z = 2.5$, and $y^+_\mathrm{in} = 10$ (magenta), $20$ (green), $50$ (blue) and $100$ (red), as well as with the Cartesian grid (black). SOD2D and OpenFOAM cases are respectively presented as circles and squares. (\textit{a}) Total number of grid points $N_\mathrm{dof}$. (\textit{b}) Viscous-scaled time-step $\Delta t^+$ with CFL $= 0.9$ for SOD2D, and $0.5$ for OpenFOAM. (\textit{c}) Ratio of the computational cost relative to the Cartesian grid $(N_\mathrm{dof}/\Delta t^+)/(N_\mathrm{dof}/\Delta t^+)_\mathrm{Cart}$.}
  \label{fig:cost_re400}
\end{figure}

Figure~\ref{fig:yp_study} presents the statistics with $C_y = 2.0, C_z = 2.5$ (sets 1 and 2 from table~\ref{tab:yp_study}). Increasing $y^+_\mathrm{in}$, refines $\Delta y^+$ but coarsens $\Delta z^+$ within the inner (Cartesian) layer (figures~\ref{fig:yp_study}\textit{a,b}). Therefore, we expect grid-resolution error to emerge when $y^+_\mathrm{in}$ is too low (coarse $\Delta y^+$ below $y^+_\mathrm{in}$) or too high (coarse $\Delta z^+$ below $y^+_\mathrm{in}$). With SOD2D (figures~\ref{fig:yp_study}\textit{d,e}), cases with $y^+_\mathrm{in} = 10, 50$ and $100$ have statistics in close agreement with the Cartesian case, as well as DNS of \cite{moser1999direct}; for these cases $\varepsilon_{C_f} < 1\%$. The exception is $y^+_\mathrm{in} = 20$, where a slight deviation appears in the inner peak of $u^+_\mathrm{rms}$ (figure~\ref{fig:yp_study}\textit{e}), and $\varepsilon_{C_f} = 2.3\%$ (figure~\ref{fig:yp_study}\textit{d}). With OpenFOAM (figures~\ref{fig:yp_study}\textit{f,g}), cases with $y^+_\mathrm{in} = 20$ and $50$ have identical statistics to the Cartesian case, and they all yield $\varepsilon_{C_f} = 2.0 \pm 0.2 \%$. However, $\varepsilon_{C_f}$ exceeds $2.5\%$ with $y^+_\mathrm{in} = 10$ (coarse $\Delta y^+$ below $y^+_\mathrm{in}$), and with $y^+_\mathrm{in} = 100$ (coarse $\Delta z^+$ below $y^+_\mathrm{in}$). Overall, with $C_y = 2.0, C_z = 2.5$ and $y^+_\mathrm{in} = 50$ (blue curves in figure~\ref{fig:yp_study}) we obtain identical statistics to the Cartesian case, with less than $0.5\%$ deviation in $\varepsilon_{C_f}$. With these grid parameters, the number of grid points is almost half of the Cartesian number of grid points (figure~\ref{fig:cost_re400}\textit{a}). These conclusions are valid for both SOD2D and OpenFOAM. 


The OpenFOAM cases with $\eta$-grid and $C_y = 2.0, C_z = 2.5$, as well as with the Cartesian grid, have $\varepsilon_{C_f} \simeq 2.0\% - 3.0\%$ (figure~\ref{fig:yp_study}\textit{f}). We conducted finer resolution cases with $C_y = 1.6, C_z = 2.0$, and $\varepsilon_{C_f}$ stayed at $\simeq 2.0\%$ for $y^+_\mathrm{in} = 20, 100$, and was reduced to $1.3\%$ for $y^+_\mathrm{in} = 50$ (set 3, table~\ref{tab:yp_study}). We anticipate that our inability to reduce $\varepsilon_{C_f}$ below $1\%$ with OpenFOAM is related to the effect of grid aspect ratio and its discretisation schemes, as extensively studied by \cite{meyers2007plane,rezaeiravesh2018effect,rezaeiravesh2021numerical} and \cite{o2024quantifying}. With a fixed $y$-grid, \cite{rezaeiravesh2021numerical} generated maps of $\varepsilon_{u_\tau}(\Delta x^+, \Delta z^+)$, an analogue of $\varepsilon_{C_f}(\Delta x^+, \Delta z^+)$, with Nek5000 and OpenFOAM. They showed that $\varepsilon_{u_\tau}(\Delta x^+, \Delta z^+)$ is unique to each set of discretisation schemes. With Nek5000, $\varepsilon_{u_\tau} \lesssim 1\%$ for any combination of $(\Delta x^+, \Delta z^+)$ within the range $\Delta x^+ \lesssim 110, \Delta z^+ \lesssim 25$ (figure~1\textit{a} in \citealt{rezaeiravesh2021numerical}). However, with OpenFOAM and its second-order discretisation schemes, the map of $\varepsilon_{u_\tau}(\Delta x^+, \Delta z^+)$ has a steep gradient, and $\varepsilon_{u_\tau} < 1\%$ is achieved for stringent combinations of $(\Delta x^+, \Delta z^+)$, see figure~1(\textit{d}) in \cite{rezaeiravesh2021numerical}. In Appendix~\ref{sec:dxp_study}, we support these findings by conducting additional OpenFOAM cases with $C_y = 2.0, C_z = 2.5$ and $y^+_\mathrm{in} = 50$, while changing $6 \le \Delta x^+ \le 18$ (figure~\ref{fig:dxp_study}). A minimum of $\varepsilon_{C_f} = 0.7\%$ is achieved with $\Delta x^+ = 14$, but $\varepsilon_{C_f}$ rises to $2.7\%$ with $\Delta x^+ = 6$.

In figure~\ref{fig:cost_re400}, we assess the computational cost of the cases from figure~\ref{fig:yp_study}, relative to the Cartesian grid. The number of grid points $N_\mathrm{dof}$ drops from $\sim 8.0$ M with the Cartesian grid to $\sim 4.0$ M with $\eta$-grid and $y^+_\mathrm{in} = 50$ (figure~\ref{fig:cost_re400}\textit{a}); thinning of $y^+_\mathrm{in}$ below $50$, hence $y_\mathrm{in} / \delta_0$ below $0.12$, has marginal influence on $N_\mathrm{dof}$. Using $\eta$-grid also increases the time-step size $\Delta t^+$ relative to the Cartesian grid (figure~\ref{fig:cost_re400}\textit{b}). With OpenFOAM and SOD2D, $\Delta t^+$ respectively increases by $\sim 1.5$ times (up to $y^+_\mathrm{in} = 50$) and $\sim 2.1$ times (up to $y^+_\mathrm{in} = 20$). Overall, the computational costs of the cases with $\eta$-grid and $y^+_\mathrm{in} \le 50$ drop below $40\%$ of the one with the Cartesian grid (figure~\ref{fig:cost_re400}\textit{c}); we obtain the cost ratios from the ratios of $(N_\mathrm{dof}/\Delta t^+)$. Considering the level of accuracy and computational cost, we choose the $\eta$-grid parameters $C_y = 2.0, C_z = 2.5$ and $y^+_\mathrm{in} = 50$ for the rest of our study.


\subsection{Grid saving with Reynolds number}\label{sec:grid_saving_smooth}
\begin{figure}
  \centering
  \includegraphics[width=\textwidth,trim={{0.0\textwidth} {0.0\textwidth} {0.0\textwidth} {0.0\textwidth}},clip]{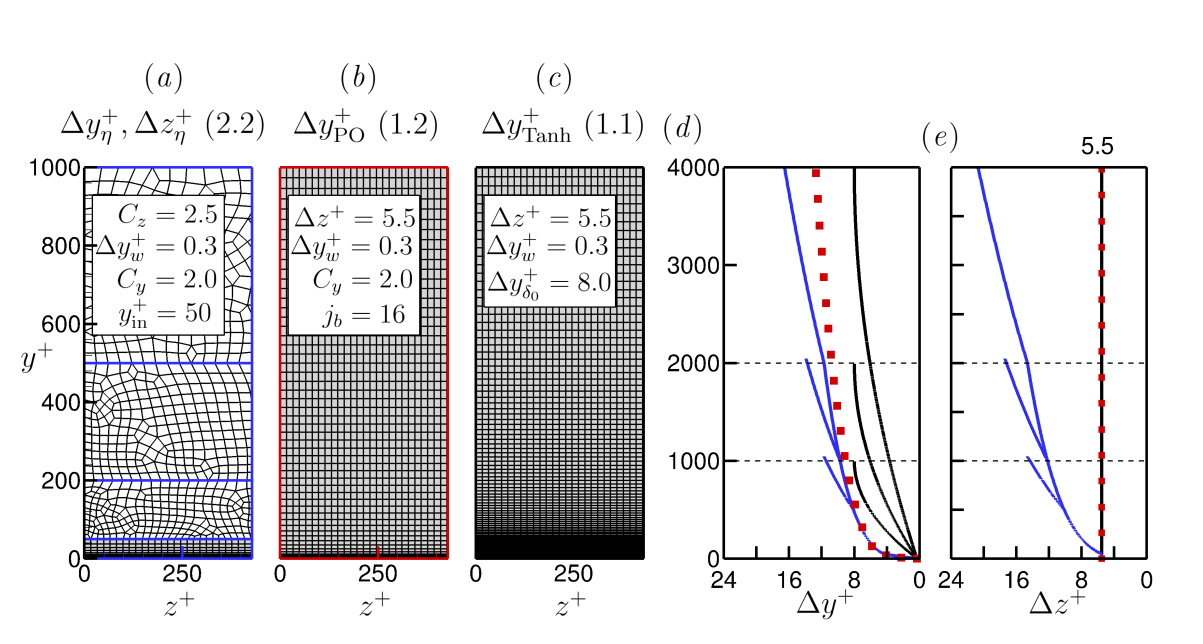}
  \includegraphics[width=\textwidth,trim={{0.9\textwidth} {0.0\textwidth} {0.0\textwidth} {0.0\textwidth}},clip]{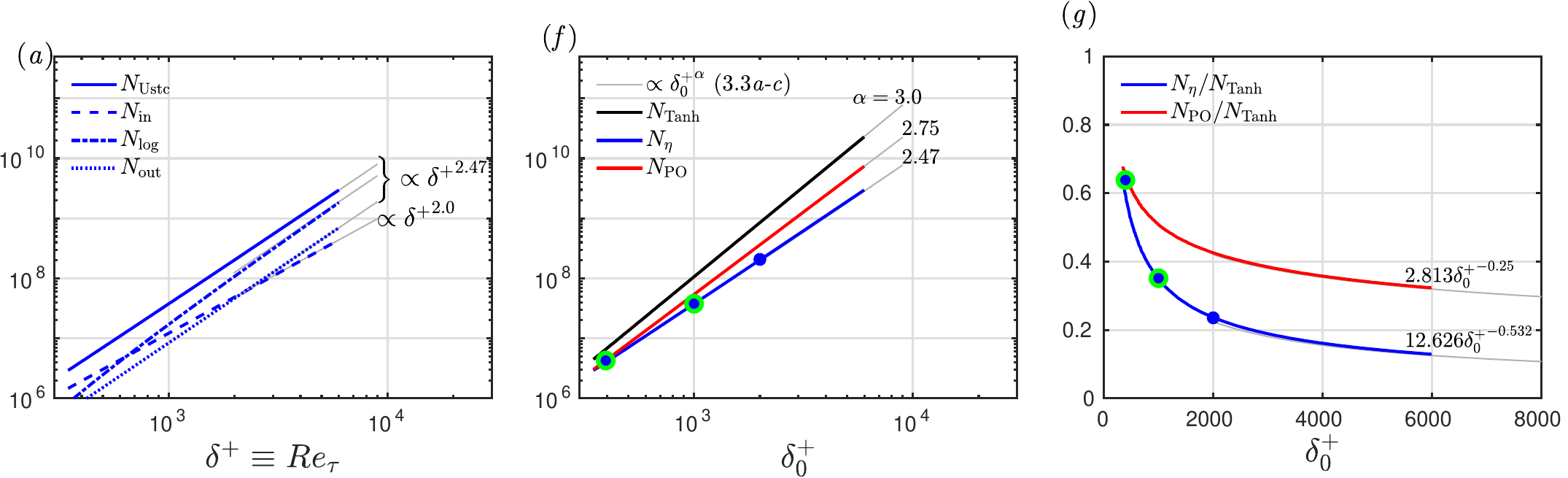}  
  \caption{Comparison of the number of grid points from three grid-generation approaches for a turbulent open-channel flow with $L_x \times L_z = 6.3 \delta_0 \times 3.15 \delta_0$. (\textit{a}) $\eta$-grid (blue curves) with $\Delta x^+ = 10$ and $\Delta y^+_\eta, \Delta z^+_\eta$ (2.2\textit{a,b}). (\textit{b,c}) Cartesian grids with $\Delta x^+ = 10, \Delta z^+ = 5.5$, and (\textit{b}) $\Delta y^+_\mathrm{PO}$(\ref{eq:PO21}, red curves), and (\textit{c}) $\Delta y^+_\mathrm{Tanh}$ (\ref{eq:tanh}, black curves). The grid parameters are provided in (\textit{a-c}). (\textit{d,e}) Profiles of $\Delta y^+$ and $\Delta z^+$ from the three grid generation approaches for $\delta^+_0 = 1000, 2000$ and $4000$, with line (symbol) colours consistent with (\textit{a-c}). (\textit{f}) Number of grid points $N_\eta,N_\mathrm{PO},N_\mathrm{Tanh}$ (3.1\textit{a-c}) versus $\delta^+_0$; the gray lines plot (3.2\textit{a-c}). (\textit{g}) Ratios $N_\eta/N_\mathrm{Tanh}$ and $N_\mathrm{PO}/N_\mathrm{Tanh}$ versus $\delta^+_0$. The blue bullets in (\textit{f,g}) are the actual $N_\eta$ from our generated grids, following figure~\ref{fig:grid_viz} (\S\ \ref{sec:grid_generation}). The bullets with green outline correspond to the tested $\eta$-grids for DNS of turbulent open-channel flow at $\delta^+_0 = 395$ (blue curves in figure~\ref{fig:yp_study}), and DNS of turbulent full-channel flow at $\delta^+_0 = 1000$ (figure~\ref{fig:channel_re1000}). }
  \label{fig:Ndof_smooth_channel}
\end{figure}

In figure~\ref{fig:Ndof_smooth_channel}, we compare the number of grid points from three grid-generation approaches for a turbulent open-channel flow with $L_x \times L_z = 6.3 \delta_0 \times 3.15 \delta_0$; $\eta$-grid (figure~\ref{fig:Ndof_smooth_channel}\textit{a}), and two Cartesian grids (figures~\ref{fig:Ndof_smooth_channel}\textit{b,c}), one with $\Delta y^+_\mathrm{PO}$ (\ref{eq:PO21}), and another one with $\Delta y^+_\mathrm{Tanh}$ (\ref{eq:tanh}). For a fair comparison, we generate an almost identical near-wall grid resolution for the three grids (figures~\ref{fig:Ndof_smooth_channel}\textit{a-c}). We set $\Delta x^+ = 10, \Delta y^+_w = 0.3$ for all three grids, and $\Delta z^+ = 5.5$ for the Cartesian grids, which is matched with $\Delta z^+_\eta$ for the inner layer of $\eta$-grid (figure~\ref{fig:Ndof_smooth_channel}\textit{e}). The total number of grid points are
\begin{align}
 N_\eta &= N_x N_{{yz}_\eta} =  \frac{(L_x/\delta_0)}{\Delta x^+}  N_{{yz}_\eta} \delta^+_0 \tag{3.1\textit{a}} \label{eq:N_eta_1} \\
 N_\mathrm{PO} &= N_x N_{y_\mathrm{PO}} N_z =  \frac{(L_x L_z/\delta^2_0)}{\Delta x^+ \Delta z^+} N_{y_\mathrm{PO}} {\delta^+_0}^2 \tag{3.1\textit{b}}  \label{eq:N_PO21_1} \\ 
 N_\mathrm{Tanh} &= N_x N_{y_\mathrm{Tanh}} N_z = \frac{(L_x L_z/\delta^2_0)}{\Delta x^+ \Delta z^+} N_{y_\mathrm{Tanh}} {\delta^+_0}^2. \tag{3.1\textit{c}} \label{eq:N_Tanh_1} 
\end{align}
The relations for $N_{{yz}_\eta}$, $N_{y_\mathrm{Tanh}}$ and $N_{y_\mathrm{PO}}$ are respectively presented in figure~\ref{fig:idealised_grid}, (1.1\textit{c}) and (\ref{eq:N_yPO21}). In figure~\ref{fig:Ndof_smooth_channel}(\textit{f}), we plot (3.1\textit{a-c}) versus $\delta^+_0$. At high Reynolds numbers ($\delta^+_0 \gtrsim \mathcal{O}(10^3)$), (3.1\textit{a-c}) approach the following relations
\begin{align}
  N_\eta &\simeq \frac{L_x L_z/\delta^2_0}{\Delta x^+ C_y C_z \kappa^{2\beta}2^{(1-2\beta)}} \left[ \ln(2) + \frac{1}{1-2\beta} \right] {\delta^+_0}^{(3-2\beta)} \simeq 1.351 {\delta^+_0}^{2.468} \tag{3.2\textit{a}} \label{eq:N_eta_2} \\
   N_\mathrm{PO} &\simeq \frac{L_x L_z/\delta^2_0}{\Delta x^+ \Delta z^+} \frac{1}{0.6 C_y}{\delta^+_0}^{2.75} \simeq 0.301 {\delta^+_0}^{2.75} \tag{3.2\textit{b}} \label{eq:N_PO21_2} \\
 N_\mathrm{Tanh} &=\frac{L_x L_z/\delta^2_0}{\Delta x^+ \Delta z^+} \frac{\alpha}{\Delta y^+_\delta\tanh(\alpha)} {\delta^+_0}^3 \simeq 0.107 {\delta^+_0}^3. \tag{3.2\textit{c}} \label{eq:N_hyp_2}
\end{align}
When $\delta^+_0 \gg y^+_\mathrm{in}$, $N_{{yz}_\eta}$ is dominated by the number of grid points in the log and outer regions (figure~\ref{fig:idealised_grid}), leading to (\ref{eq:N_eta_2}). Also, when $\delta^+_0 \gtrsim \mathcal{O}(10^3)$, (\ref{eq:N_yPO21}) yields $N_{y_\mathrm{PO}} = {\delta^+_0}^{0.75}/(0.6 C_y)$, leading to (\ref{eq:N_PO21_2}). The gray lines in figure~\ref{fig:Ndof_smooth_channel}(\textit{f}) confirm the accuracy of (3.2\textit{a-c}). The number of grid points from the Cartesian grids $N_\mathrm{Tanh}$ and $N_\mathrm{PO}$ respectively grow proportional to $ {\delta^+_0}^{3.0}$ and ${\delta^+_0}^{2.75}$, as discussed by \cite{pirozzoli2021natural}. Our proposed grid drops the growth rate to $N_\eta \propto {\delta^+_0}^{2.47}$. In figure~\ref{fig:Ndof_smooth_channel}(\textit{g}), we assess the grid-points saving with Reynolds number by plotting $N_\eta/N_\mathrm{Tanh}$ and $N_\mathrm{PO}/N_\mathrm{Tanh}$; the ratio $N_\eta/N_\mathrm{Tanh} \simeq 12.6 {\delta^+_0}^{-0.53}$ drops to $0.32$ by $\delta^+_0 = 1000$, whereas $N_\mathrm{PO}/N_\mathrm{Tanh} \simeq 2.8 {\delta^+_0}^{-0.25}$ drops to the same ratio by $\delta^+_0 \simeq 6000$. By $\delta^+_0 = 6000$, $N_\eta/N_\mathrm{Tanh} \simeq 0.12$.


 \begin{figure}
  \centering
  \includegraphics[width=.45\textwidth,trim={{0.0\textwidth} {0.0\textwidth} {0.0\textwidth} {0.0\textwidth}},clip]{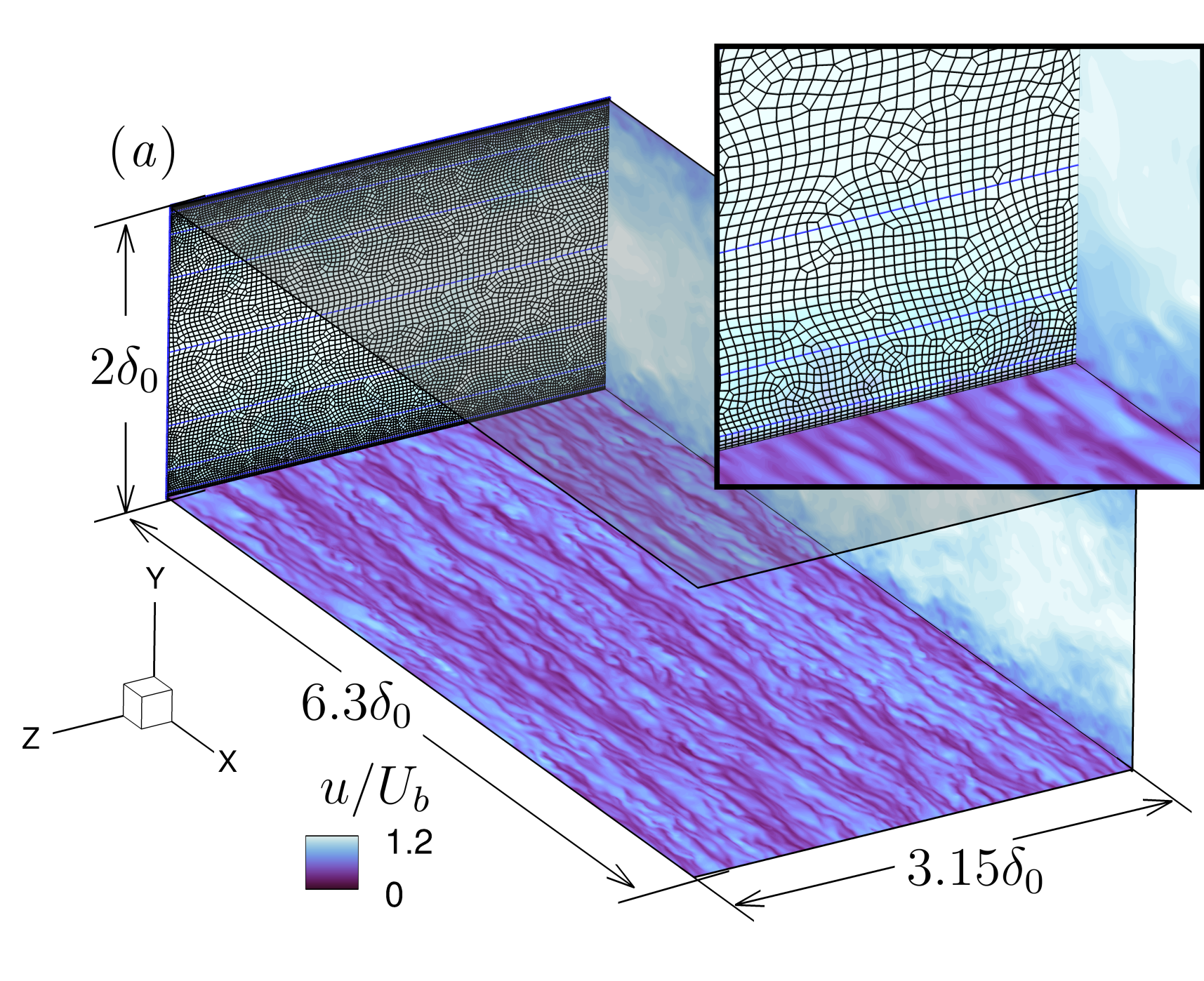} \hspace{0.5cm}
  \includegraphics[width=.48\textwidth,trim={{1.0\textwidth} {0.7\textwidth} {0.2\textwidth} {0.0\textwidth}},clip]{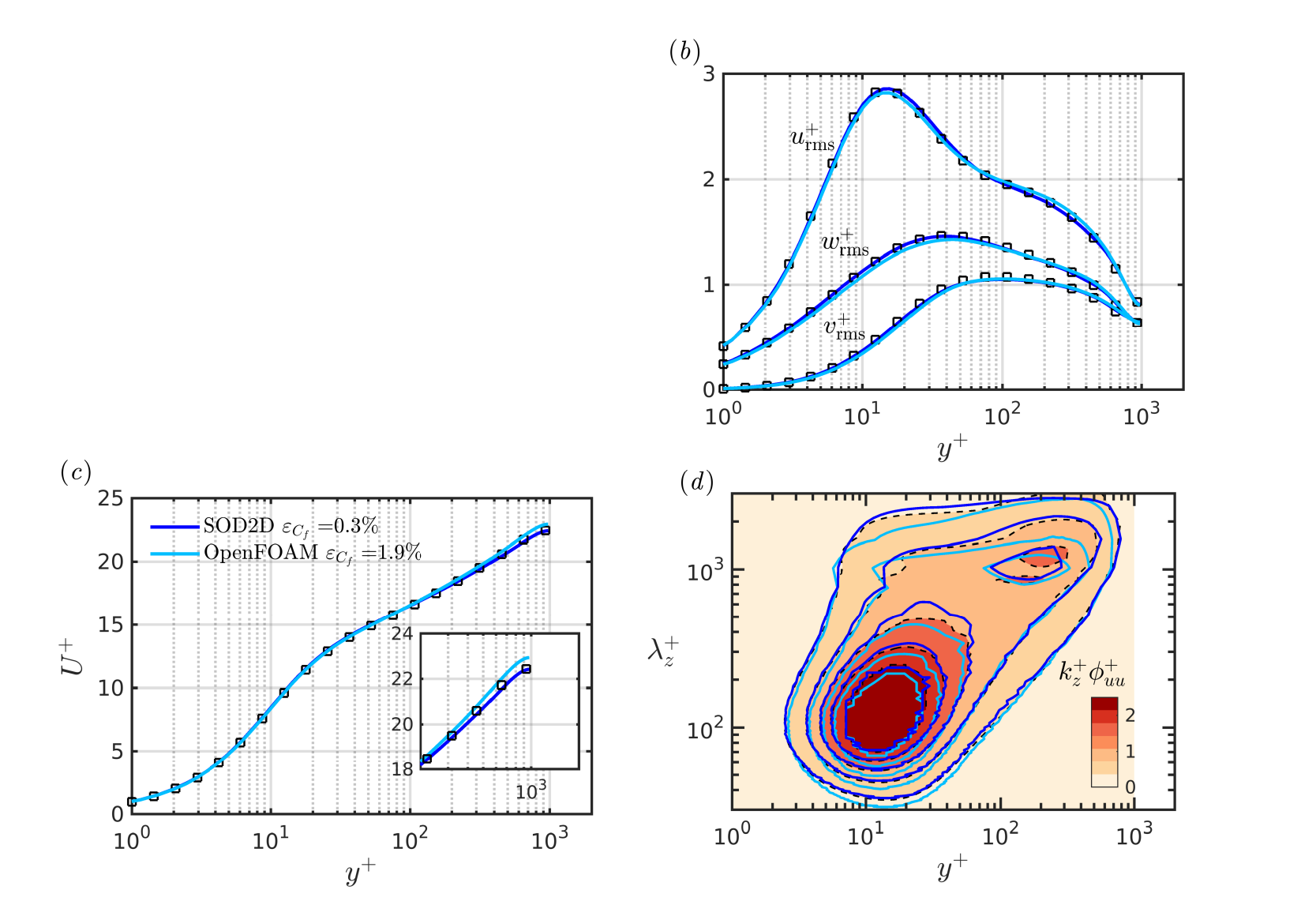}\\
  \includegraphics[width=\textwidth,trim={{0.09\textwidth} {0.0\textwidth} {0.2\textwidth} {0.74\textwidth}},clip]{figures/channel_re1000} 
  \caption{Application of $\eta$-grid to turbulent full-channel flow at $\delta^+_0 = 1000$; the grid details are reported in figure~\ref{fig:Ndof_smooth_channel}(\textit{a}) and table~\ref{tab:yp_study} (lower cases). (\textit{a}) Computational domain and visualisation of the $u$ field and spectral elements on a $yz$-plane. Profiles of (\textit{b}) r.m.s of velocity fluctuations, and (\textit{c}) $U^+$. (\textit{d}) Pre-multiplied spectrograms of the streamwise velocity fluctuation $k^+_z \phi^+_{uu}$ versus the spanwise wavelength $\lambda^+_z$ and $y^+$; yellow-red contour field is the reference \cite{lee2015direct}, and the blue contour lines are SOD2D and OpenFOAM cases with line colours consistent with (\textit{b,c}). }
  \label{fig:channel_re1000}
\end{figure}

The bullets in figures~\ref{fig:Ndof_smooth_channel}(\textit{f,g}) are from the explicit generation of $\eta$-grid at $\delta^+_0 = 395, 1000$ and $2000$, following the procedure in \S\ \ref{sec:grid_generation} (figure~\ref{fig:grid_viz}); they are in excellent agreement with the analytical expression for $N_\eta$ (\ref{eq:N_eta_1}), following ${\delta^+_0}^{2.47}$ scaling. We already assessed the accuracy of $\eta$-grid at $\delta^+_0 = 395$ (\S\ \ref{sec:yp_study}, figure~\ref{fig:yp_study}). We further assess the accuracy of $\eta$-grid at $\delta^+_0 = 1000$ for a turbulent full-channel flow ($L_x \times L_y \times L_z = 6.30\delta_0 \times 2.0\delta_0 \times 3.15 \delta_0$), figure~\ref{fig:channel_re1000}(\textit{a}). We conduct DNSs with both SOD2D and OpenFOAM (lower cases in table~\ref{tab:yp_study}). The $yz$-grid details for the lower half of the domain are presented in figure~\ref{fig:Ndof_smooth_channel}(\textit{a}); we mirror the grid on the upper half. The number of grid points $2N_\eta \simeq 76$ M is roughly only a third of that with a Cartesian grid and $\Delta y^+_\mathrm{Tanh}$ ($2N_\mathrm{Tanh} \simeq 215$ M; blue curve in figure~\ref{fig:Ndof_smooth_channel}\textit{g}). In figures~\ref{fig:channel_re1000}(\textit{b-d}), we compare the statistics and spectrograms against the DNS of \cite{lee2015direct}. Results from SOD2D are in excellent agreement with the reference data, in terms of the statistics and spectrograms, with $\varepsilon_{C_f} = 0.3\%$ (figure~\ref{fig:channel_re1000}\textit{c}). The r.m.s.\ profiles from OpenFOAM as well as the pre-multiplied spectrogram of streamwise velocity fluctuations $k^+_z \phi^+_{uu}$ agree well with the reference counterparts (figures~\ref{fig:channel_re1000}\textit{b,d}). The $U^+$ profile from OpenFOAM slightly exceeds the reference profile by a maximum of $\epsilon_{U^+} \simeq 2\%$ at $y^+ \simeq \delta^+_0$, and $\varepsilon_{C_f} = 1.9\%$ (figure~\ref{fig:channel_re1000}\textit{c}). Such level of difference was also observed from our open-channel flow cases with OpenFOAM at $\delta^+_0 = 395$, with both $\eta$-grid and the Cartesian grid (figures~\ref{fig:yp_study}\textit{f,g}).



 


 \begin{figure}
  \centering
  \includegraphics[width=\textwidth,trim={{0.1\textwidth} {0.1\textwidth} {0.0\textwidth} {0.0\textwidth}},clip]{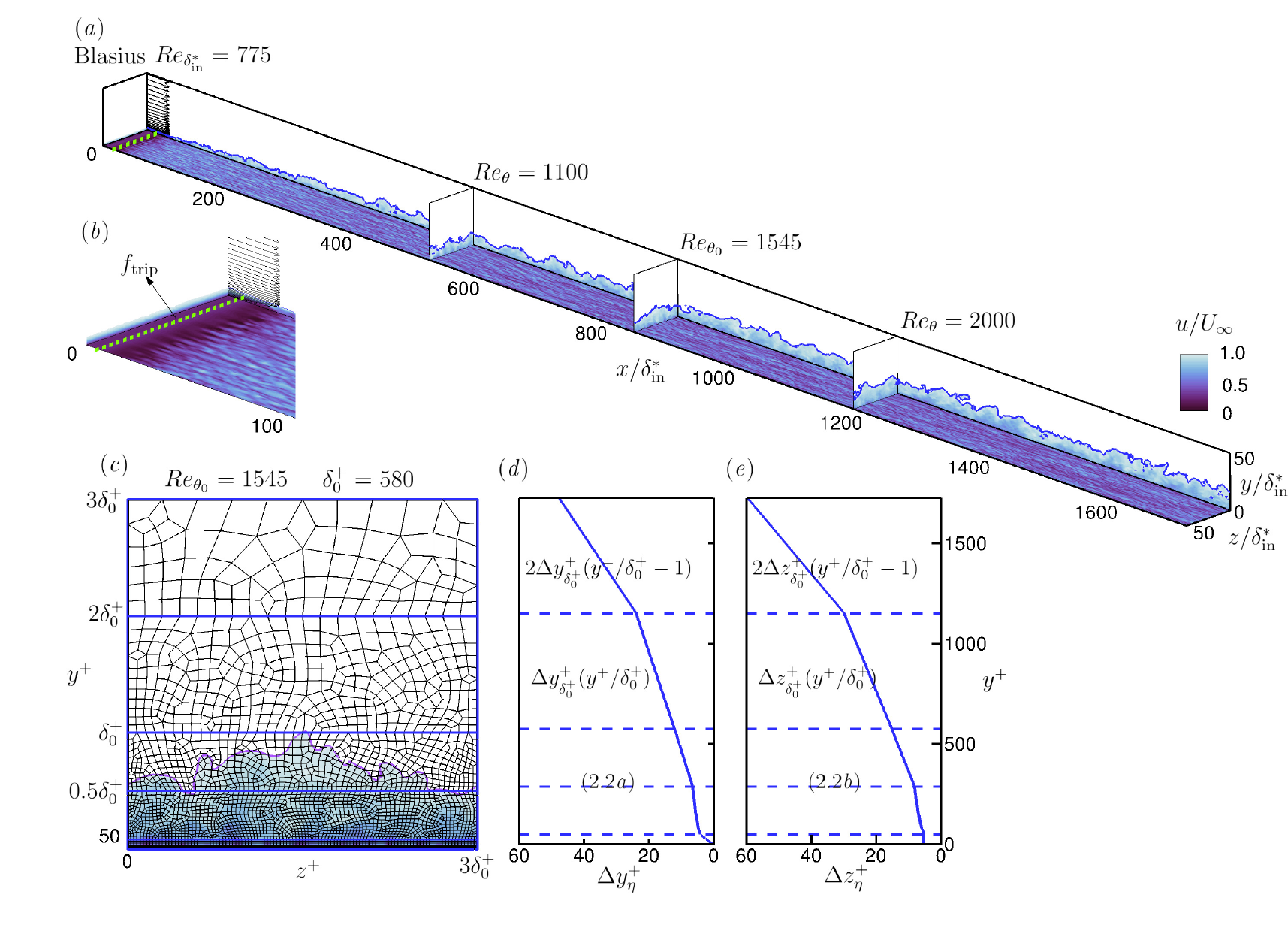}
  \caption{Setup and $\eta$-grid for the smooth-wall TBL based on $\delta^+_0 = 580$ (table~\ref{tab:bl_table}, second row). (\textit{a}) Computational domain and visualisation of the $u$ field, with (\textit{b}) showing the close-up view near the parametric tripping $f_\mathrm{trip}$ (\ref{eq:force}). (\textit{c}) Spectral elements on a $yz$-plane. (\textit{d,e}) Profiles of $\Delta y^+_\eta$ and $\Delta z^+_\eta$. }
  \label{fig:flowviz_bl}
\end{figure}

\section{Smooth wall turbulent boundary layer}\label{sec:smooth_tbl}
\subsection{Setup and results}\label{sec:bl_setup}
Figure~\ref{fig:flowviz_bl} presents the application of $\eta$-grid for DNS of smooth-wall TBL. We conduct the calculations using SOD2D. A laminar Blasius velocity profile with displacement thinkness Reynolds number $Re_{\delta^*_\mathrm{in}} \equiv U_{\infty} \delta^*_\mathrm{in}/\nu = 775$ is applied at the inlet (figure~\ref{fig:flowviz_bl}\textit{a}); all the setup dimensions are in units of $\delta^*_\mathrm{in}$, the inlet displacement thickness. We trip the boundary layer via a parametric forcing, added to (2.4\textit{b})
\begin{align}
 f_x = -\frac{1}{2}\rho C_D u |u|/l_x, \quad  f_y = 0, \quad f_z = 0. \tag{4.1} \label{eq:force}
\end{align}
Here $C_D$ is a constant factor. The forcing is applied for $x_\mathrm{trip} \le x \le x_\mathrm{trip}+l_x$, $0 \le y \le l_y$ and the entire spanwise width. Equation (\ref{eq:force}) mimics a drag force by a trip wire with streamwise and wall-normal thicknesses $l_x,l_y$, respectively. From a series of preliminary calculations, we arrived at the suitable tripping parameters $C_D = 3.0,l_x = l_y = 1.45\delta^*_\mathrm{in}$ and $x_\mathrm{trip} = 5.8\delta^*_\mathrm{in}$. We apply the no-slip condition at the bottom boundary, the zero-vorticity condition at the top boundary, and periodic boundary conditions in the spanwise direction. At the outlet, we apply a buffer region with length $40 \delta^*_\mathrm{in}$, where the flow is forced to re-transition to its laminar profile at the inlet, following a similar approach to the previous TBL simulations~\citep{schlatter2009turbulent,schlatter2012turbulent}.

\begin{table}
\centering
 \begin{tabular}{cccc|ccccc}
         \multicolumn{4}{c|}{For setup} & \multicolumn{5}{c}{From simulation} \\
         Case & $(L_x,L_y,L_z)/\delta^*_\mathrm{in}$ & $\delta^+_0$ & $\delta_0/\delta^*_\mathrm{in}$ & $\delta^+$ & $\delta/\delta^*_\mathrm{in}$ & $Re_{\theta}$ & $x/\delta^*_\mathrm{in}$ & $\Delta \hat{x}^+ \quad (\Delta \hat{y}^+_w, \Delta \hat{y}^+_{\delta_0}) \quad (\Delta \hat{z}^+_\mathrm{in}, \Delta \hat{z}^+_{\delta_0})$ \\ 
        Short & $(840,36,36)$ & $400$ & $11.8$ & $436$ & $11.95$ & $1100$ & $515$ & $13.14 \quad (0.28, 11.63) \quad (5.80, 14.57)$  \\
        &&&&&&&&\\
        Long & $(1740,54,54)$ & $580$ & $17.8$ & $580$ & $16.45$ & $1545$ & $829$ & $13.18 \quad (0.28, 12.59) \quad (5.77, 16.71)$  \\
          &   &   &                            & $737$ & $21.26$ & $2000$ & $1177$ & $12.95 \quad (0.28, 12.37) \quad (5.67, 16.41)$  \\        
    \end{tabular}
\caption{Simulation details for smooth wall TBL with SOD2D. The left side presents the domain dimensions $L_x > x_0, L_y = L_z \simeq 3\delta_0$, based on the target $\delta^+_0$, and the predictive correlations (\ref{eq:bl_corr}). The right side presents the calculated TBL characteristics, as well as the local viscous-scaled grid sizes from the simulations.}
\label{tab:bl_table}
\end{table}

 \begin{figure}
  \centering
  \includegraphics[width=.98\textwidth,trim={{-0.06\textwidth} {0.\textwidth} {0.0\textwidth} {0.0\textwidth}},clip]{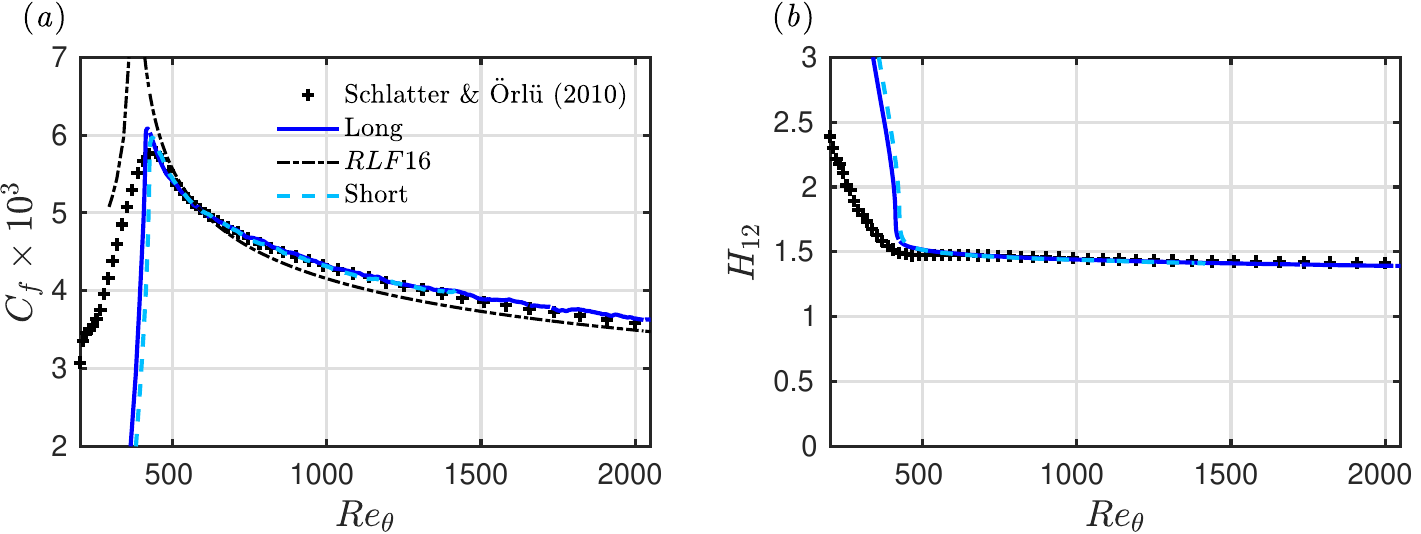}
  \includegraphics[width=\textwidth,trim={{0.\textwidth} {0.\textwidth} {0.0\textwidth} {0.0\textwidth}},clip]{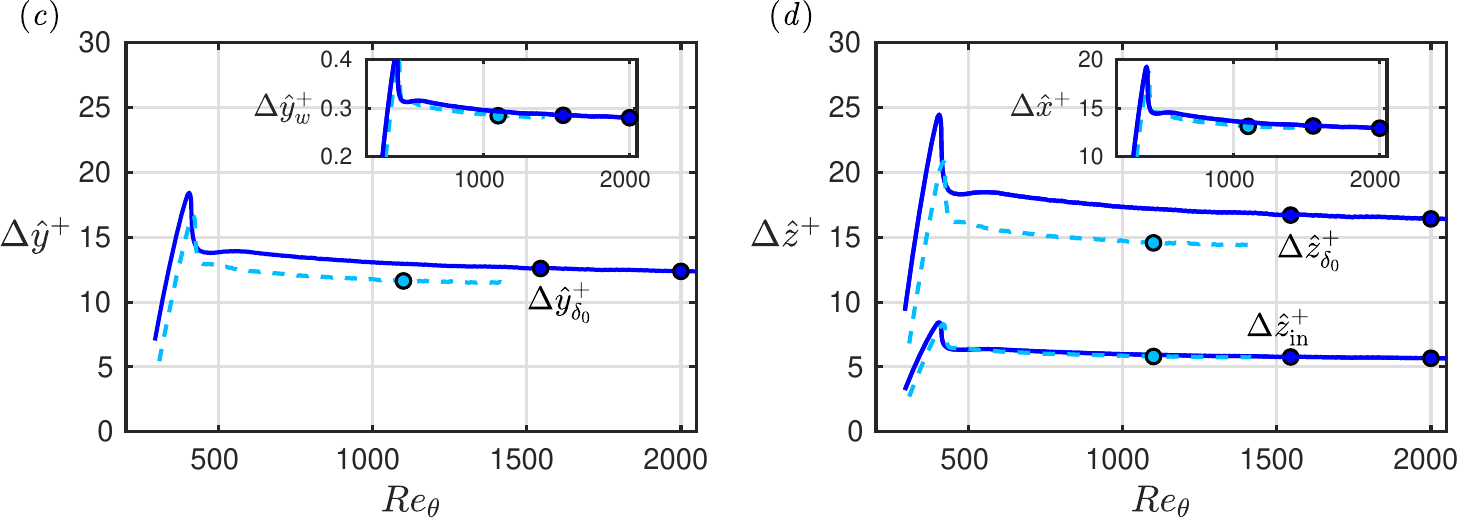}  
  \caption{Variations of different parameters with $Re_\theta$. (\textit{a}) $C_f$ and (\textit{b}) shape factor $H_{12} \equiv \delta^*/\theta$. The dashed-dotted line in (\textit{a}) is the empirical relation $C_f = 0.0134 (Re_\theta - 373.83)^{-2/11}$ by \cite{rezaeiravesh2016grid}, RLF16. In (\textit{c,d}) we plot the viscous-scaled grid sizes based on the local $u_\tau$. (\textit{c}) $\Delta \hat{y}^+_{\delta_0}$ at $y = \delta_0$, and $\Delta \hat{y}^+_w$ at the wall (inset). (\textit{d}) $\Delta \hat{z}^+_\mathrm{in}$ in the inner layer, and $\Delta \hat{z}^+_{\delta_0}$ at $y = \delta_0$, as well as $\Delta \hat{x}^+$ (inset). The bullets in (\textit{c,d}) are the selected locations ($Re_\theta = 1100, 1545, 2000$), where we report the viscous-scaled grid sizes in table~\ref{tab:bl_table}.}
  \label{fig:Cf_H12}
\end{figure}


We conduct two cases, with details provided in table~\ref{tab:bl_table}. We set up the domain dimensions based on resolving the TBL up to a target $\delta^+_0$. We estimate $\delta_0/\delta^*_\mathrm{in}$ and its downstream location $x_0/\delta^*_\mathrm{in}$ from the power-law correlations by \cite{rezaeiravesh2016grid}.  
\begin{align}
 Re_{x_0} \equiv \frac{U_\infty x_0}{\nu} = 218.6864 {\delta^+_0}^{1.2726}, \quad Re_{\delta_0} \equiv \frac{U_\infty \delta_0}{\nu} = 12.7605 {\delta^+_0}^{1.098} \tag{4.2\textit{a,b}} \label{eq:bl_corr}
\end{align}
The correlations (\ref{eq:bl_corr}) agree well with the available DNS and experimental data for a broad range of momentum thickness Reynolds numbers ($500 \lesssim Re_\theta \lesssim 10^5$), figure~1 in \cite{rezaeiravesh2016grid}. Figure~\ref{fig:Cf_H12}(\textit{a}) further supports the agreement between \cite{rezaeiravesh2016grid}'s correlation for $C_f$ and the DNS data for $Re_\theta \gtrsim 500$. From (\ref{eq:bl_corr}), we can estimate $\delta_0/\delta^*_\mathrm{in} = Re_{\delta_0} / Re_{\delta^*_\mathrm{in}}$ and $x_0/\delta^*_\mathrm{in} = Re_{x_0} / Re_{\delta^*_\mathrm{in}}$. We set the domain height and width $L_y = L_z \simeq 3 \delta_0$, as recommended by the previous TBL studies~\citep{schlatter2009turbulent,schlatter2010simulations,schlatter2012turbulent}, and we ensure that the domain length is larger than $x_0$ ($L_x > x_0$). In table~\ref{tab:bl_table} (left side), we adjust the dimensions of the short and long cases to resolve the TBL up to $\delta^+_0 \simeq 400$ and $580$, respectively, equivalent to $Re_\theta \simeq 1000$ and $1500$. On the right side of table~\ref{tab:bl_table}, we report the calculated TBL characteristics from the simulations at the target locations (where $\delta^+ \simeq \delta^+_0$); at theses locations, the values of $\delta/\delta^*_\mathrm{in}$ from the simulations are in good agreement with the predicted values for $\delta_0/\delta^*_\mathrm{in}$. We calculate $\delta$ based on $0.99 U_\infty$. 






Based on the target $\delta^+_0$, we set up the $\eta$-grid (figures~\ref{fig:flowviz_bl}\textit{c,d,e}); that means, our $\eta$-grid sizes are viscous-scaled based on $u_{\tau_0}$ at the target location (e.g.\ $\Delta x^+ \equiv u_{\tau_0} \Delta x/\nu$). The grid size up to $y^+ = \delta^+_0$ follows (2.2\textit{a,b}); we set $\Delta y^+_w = 0.3, C_y = 2.0, C_z = 2.5$ and $y^+_\mathrm{in} = 50$, as concluded from channel flow DNSs (\S\ \ref{sec:smooth_channel}). Beyond $y^+ = \delta^+_0$, we expand $\Delta y^+_{\delta_0}, \Delta z^+_{\delta_0}$ linearly with $y^+$ to $2\Delta y^+_{\delta_0}, 2\Delta z^+_{\delta_0}$ at $y^+ = 2\delta^+_0$, and further to $4\Delta y^+_{\delta_0}, 4\Delta z^+_{\delta_0}$ at $y^+ = L^+_y$ (figures~\ref{fig:flowviz_bl}\textit{d,e}). Beyond $y^+ = \delta^+_0$ is the inviscid region, and does not demand a fine grid. We generate a uniform grid in the streamwise direction with $\Delta x^+ = 13$, by setting the number of grid points $N_x = (L_x/\delta_0)\delta^+_0/\Delta x^+ $. The viscous-scaled grid sizes based on the local $u_\tau$ vary along the $x$-direction; we denote them with $\hat{(.)}$ (e.g.\ $\Delta \hat{x}^+ \equiv  u_\tau \Delta x/\nu$), to distinguish them from the viscous-scaled grid sizes based on the target $u_{\tau_0}$.
In figures~\ref{fig:Cf_H12}(\textit{c,d}), we plot the local viscous-scaled grid sizes, and report their values at selected locations in table~\ref{tab:bl_table} (right side). The obtained grid sizes are what we expect with our chosen grid parameters. Figures~\ref{fig:Cf_H12}(\textit{c,d}) also highlight the disparities in $\Delta \hat{y}^+$ and $\Delta \hat{z}^+$ between the inner layer ($\Delta \hat{y}^+_w, \Delta \hat{z}^+_\mathrm{in}$) and the outer region ($\Delta \hat{y}^+_{\delta_0}, \Delta \hat{z}^+_{\delta_0}$). According to (2.2\textit{b}), for TBL with $y^+_\mathrm{in} = 50$, $\Delta z^+_{\delta_0}/\Delta z^+_\mathrm{in} \simeq 0.6 {\delta^+_0}^{0.25}$. That means at $\delta^+_0 \simeq 580$ ($Re_\theta = 1545$), $\Delta z^+_{\delta_0}/\Delta z^+_\mathrm{in} \simeq 2.9$ (figure~\ref{fig:Cf_H12}\textit{d}), and for $Re_\theta \gtrsim 20,000$ ($\delta^+_0 \gtrsim 6000$), $\Delta z^+_{\delta_0}/\Delta z^+_\mathrm{in} \gtrsim 5.0$. Such disparities significantly save the number of grid points, as discussed in \S\ \ref{sec:grid_saving_tbl}. 

\begin{figure}
  \centering
  \includegraphics[width=\textwidth,trim={{0.\textwidth} {0.\textwidth} {0.0\textwidth} {0.0\textwidth}},clip]{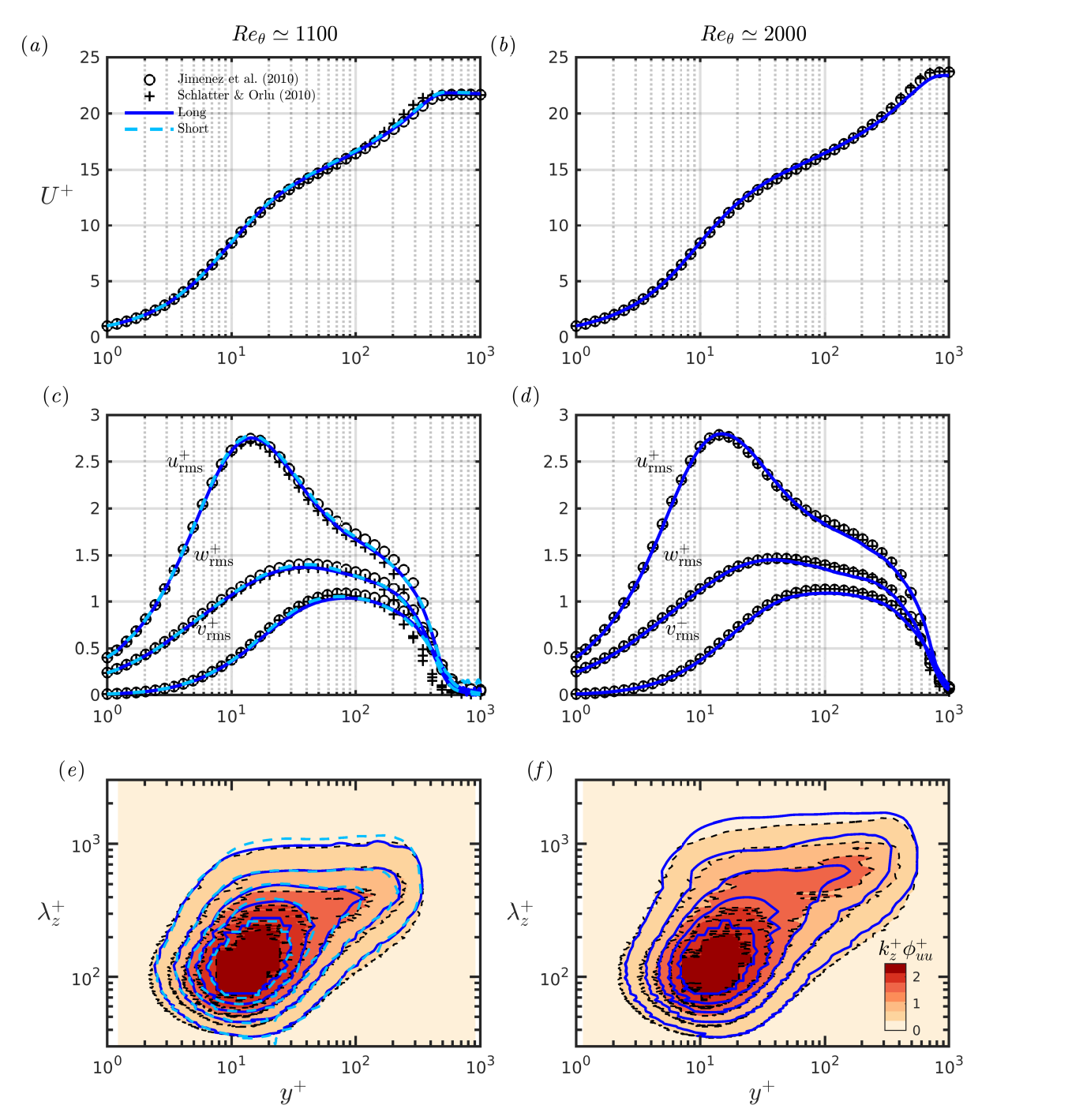}  
  \caption{Comparison of statistics and spectrograms between our cases (table~\ref{tab:bl_table}) and the reference DNSs~\citep{schlatter2010assessment,jimenez2010turbulent} at $Re_\theta \simeq 1100$ (\textit{a,c,e}) and $Re_\theta \simeq 2000$ (\textit{b,d,f}). Profiles of (\textit{a,b}) $U^+$ and (\textit{c,d}) r.m.s. of velocity fluctuations. (\textit{e,f}) Pre-multiplied spectrograms $k^+_z \phi^+_{uu}$, where filled contour fields are from \cite{schlatter2010assessment}, and the blue line contours are our cases, with line colours consistent with (\textit{a-d}).}
  \label{fig:bl_stats}
\end{figure}

In figures~\ref{fig:Cf_H12} and \ref{fig:bl_stats}, we compare our simulation results with the reference DNSs by \cite{schlatter2010assessment} and \cite{jimenez2010turbulent}. The variations of $C_f$ and shape-factor $H_{12} \equiv \delta^*/\theta$ from our present cases are in excellent agreement with the DNS of \cite{schlatter2010assessment} for $Re_\theta \gtrsim 700$ (figures~\ref{fig:Cf_H12}\textit{a,b}). The differences up to $Re_\theta \simeq 700$ are due to the different inflow conditions and tripping techniques, as extensively discussed by \cite{schlatter2012turbulent}. At $Re_\theta \simeq 1100$ (figures~\ref{fig:bl_stats}\textit{a,c,e}), profiles of $U^+$ and r.m.s.\ of velocity fluctuations, as well as the pre-multiplied spectrograms $k^+_z \phi^+_{uu}$ are in good agreement between our cases and the reference DNSs. Some slight differences are seen in the r.m.s. profiles for $y^+ \gtrsim 100$, even between the two reference DNSs. Such differences are related to the different upstream conditions. \cite{schlatter2012turbulent} discover that the upstream effects would vanish for $Re_\theta > 2000$. For the long case, we set the target $\delta^+_0 = 580$ ($Re_\theta \simeq 1545$), yet our domain and grid sizes can resolve the TBL up to $Re_\theta = 2000$ (table~\ref{tab:bl_table}, bottom row on the right). Consistent with the findings by \cite{schlatter2012turbulent}, at $Re_\theta \simeq 2000$ the upstream history effects vanish, and we observe very good agreements in the statistics and spectrograms between our long case and the reference DNSs (figures~\ref{fig:bl_stats}\textit{b,d,f}).

\begin{figure}
  \centering
  \includegraphics[width=\textwidth,trim={{0.0\textwidth} {0.0\textwidth} {0.0\textwidth} {0.0\textwidth}},clip]{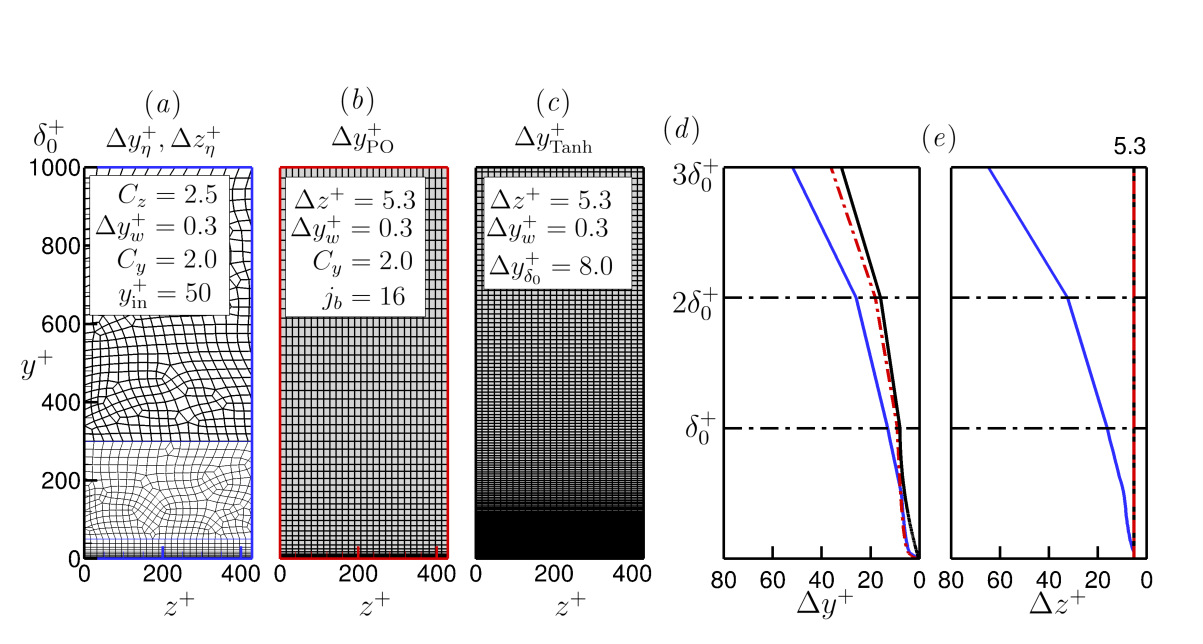}
  \includegraphics[width=\textwidth,trim={{0.0\textwidth} {0.0\textwidth} {0.0\textwidth} {0.0\textwidth}},clip]{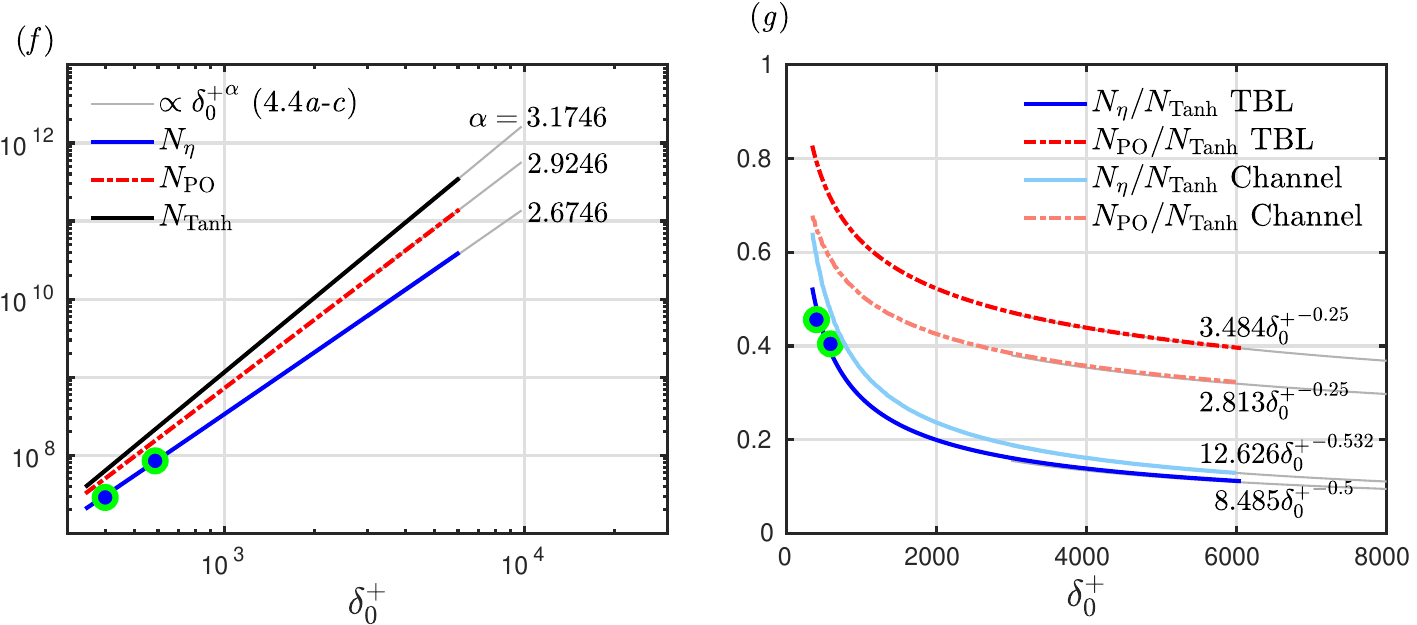} 
  \caption{Analysis of the number of grid points similar to figure~\ref{fig:Ndof_smooth_channel}, but for smooth wall TBL with domain dimensions as in (\ref{eq:domain_bl}). (\textit{a-c}) Visualisation of the grid elements with the grid parameters for $\delta^+_0 = 1000$. (\textit{d,e}) profiles of $\Delta y^+$ and $\Delta z^+$. (\textit{f}) Number of grid points $N_\eta, N_\mathrm{PO}, N_\mathrm{Tanh}$ versus $\delta^+_0$ as obtained from (4.4\textit{a-c}); the gray lines are the asymptotic relations (4.5\textit{a-c}). The bullets are the actual values of $N_\eta$ from the simulated cases at $\delta^+_0 = 400$ and $580$ (table~\ref{tab:bl_table}); we count $N_\eta$ up to the target distance where $\delta^+ = \delta^+_0$. (\textit{g}) The ratios $N_\eta/N_\mathrm{Tanh}$ and $N_\mathrm{PO}/N_\mathrm{Tanh}$ for TBL and channel flow (copied from figure~\ref{fig:Ndof_smooth_channel}\textit{g}); the gray lines are from the asymptotic relations (4.5\textit{a-c}) and (3.2\textit{a-c}).}
  \label{fig:Ndof_smooth_bl}
\end{figure}

\subsection{Grid saving with Reynolds number}\label{sec:grid_saving_tbl}
We conduct a similar analysis as the one for turbulent channel flow (\S\ \ref{sec:grid_saving_smooth}). We compare $\eta$-grid (figure~\ref{fig:Ndof_smooth_bl}\textit{a}) with the Cartesian grids and $y$-grid mappings $\Delta y^+_\mathrm{PO}$ (figure~\ref{fig:Ndof_smooth_bl}\textit{b}) and $\Delta y^+_\mathrm{Tanh}$ (figure~\ref{fig:Ndof_smooth_bl}\textit{c}), with comparable grid parameters; we set $\Delta x^+ = 12, \Delta y^+_w = 0.3$ for all grids, and $\Delta z^+ = 5.3$ for the Cartesian grids, which is matched with $\Delta z^+_\eta$ in the inner region of $\eta$-grid (figure~\ref{fig:Ndof_smooth_bl}\textit{e}). Configuration is a spatially developing TBL with a maximum target $\delta^+_0$, a setup as in figure~\ref{fig:flowviz_bl}(\textit{a}). We consider domain dimensions
\begin{align}
 \frac{L_x}{\delta_0} = 17.1378 {(\delta^+_0)}^{0.1746}, \frac{L_y}{\delta_0} = \frac{L_z}{\delta_0} = 3. \tag{4.3\textit{a,b}} \label{eq:domain_bl}
\end{align}
The relation (4.3\textit{a}) is obtained by dividing (4.2\textit{a}) over (4.2\textit{b}). The grid parameters from the wall up to $y^+ = \delta^+_0$ (figures~\ref{fig:Ndof_smooth_bl}\textit{a-c}) are similar to our analysis for turbulent channel flow (figures~\ref{fig:Ndof_smooth_channel}\textit{a-c}). In the inviscid region ($y^+ > \delta^+_0$), for all grids, $\Delta y^+$ expands following figure~\ref{fig:flowviz_bl}(\textit{d}), i.e.\ linear expansion to $2\Delta y^+_{\delta_0}$ by $y^+ = 2\delta^+_0$, and then to $4\Delta y^+_{\delta_0}$ by $y^+ = 3\delta^+_0$ (figure~\ref{fig:Ndof_smooth_bl}\textit{d}). Similarly, $\Delta z^+_\eta$ expands following figure~\ref{fig:flowviz_bl}(\textit{e}), but for the Cartesian grids, $\Delta z^+ = 5.3$ remains fixed across the domain (figure~\ref{fig:Ndof_smooth_bl}\textit{e}). The equations for the number of grid points are similar to the ones for turbulent channel flow (3.1\textit{a-c}), except we add the number of grid points for $y^+ > \delta^+_0$
\begin{align}
 N_\eta &= N_x \left( N_{{yz}_\eta} + \frac{5}{8} \frac{\delta^+_0 L^+_z}{\Delta y^+_{\delta_0} \Delta z^+_{\delta_0}} \right), \tag{4.4\textit{a}} \label{eq:N_eta_bl_1} \\
 N_\mathrm{PO} &= N_x N_z \left( N_{y_\mathrm{PO}} + \frac{3\ln(2)}{2} \frac{\delta^+_0}{\Delta y^+_{\delta_0}} \right), \tag{4.4\textit{b}} \label{eq:N_PO21_bl_1} \\
 N_\mathrm{Tanh} &= N_x N_z \left( N_{y_\mathrm{Tanh}} + \frac{3\ln(2)}{2} \frac{\delta^+_0}{\Delta y^+_{\delta_0}} \right). \tag{4.4\textit{c}} \label{eq:N_hyp_bl_1}
\end{align}
Compared to (3.1\textit{a-c}), the extra terms on the right-hand-side of (4.4\textit{a-c}) are from the integration of $\Delta y^+, \Delta z^+$ (figures~\ref{fig:Ndof_smooth_bl}\textit{d,e}) from $y^+ = \delta^+_0$ to $L^+_y = 3\delta^+_0$ following (\ref{eq:Nyz}).


To plot (4.4\textit{a-c}) versus $\delta^+_0$ (figure~\ref{fig:Ndof_smooth_bl}\textit{f}), we substitute for $N_x = L^+_x/\Delta x^+ = 17.1377(\delta^+_0)^{1.1746}/\Delta x^+$ and $N_z = L^+_z/\Delta z^+ = 3\delta^+_0/\Delta z^+$ from (\ref{eq:domain_bl}), $N_{yz_\eta}$ from figure~\ref{fig:idealised_grid}, $N_{y_\mathrm{PO}}$ from (\ref{eq:N_yPO21}), and $N_{y_\mathrm{Tanh}}$ from (1.1\textit{c}). The grid sizes at $y^+ = \delta^+_0$ are $\Delta y^+_{\delta_0} = C_y C_\eta {\delta^+_0}^\gamma, \Delta z^+_{\delta_0} = C_z C_\eta {\delta^+_0}^\gamma$ in (\ref{eq:N_eta_bl_1}), $\Delta y^+_{\delta_0} = 0.8 C_y {\delta^+_0}^{0.25}$ in (\ref{eq:N_PO21_bl_1}), and $\Delta y^+_{\delta_0} = 8.0$ in (\ref{eq:N_hyp_bl_1}). In figure~\ref{fig:Ndof_smooth_bl}(\textit{f}), the actual values of $N_\eta$ from our TBL cases (bullets), from table~\ref{tab:bl_table}, agree well with (\ref{eq:N_eta_bl_1}). At high $\delta^+_0$, the number of grid points from the log region and beyond have the dominant contribution, and (4.4\textit{a-c}) approach the following asymptotic relations. 
\begin{align}
  N_\eta &\simeq \frac{L_x L_z/\delta^2_0}{\Delta x^+ C_y C_z \kappa^{2\beta}2^{(1-2\beta)}} \left[ \underbrace{\frac{1}{1-2\beta} + \frac{1-2^{(1-2\gamma)}}{2\gamma-1}}_{\mathrm{TBL,0.86}} + \underbrace{\frac{5}{2^{(2\gamma+2)}}}_{\mathrm{inv,0.14}} \right] {\delta^+_0}^{(3-2\beta)} \nonumber \\
         & \simeq 3.496 {\delta^+_0}^{2.6746} \tag{4.5\textit{a}} \label{eq:N_eta_bl_2} \\
  N_\mathrm{PO} &\simeq \frac{L_x L_z/\delta^2_0}{\Delta x^+ \Delta z^+ C_y} \left[ \underbrace{\frac{1}{0.6}}_{\mathrm{TBL,0.56}} + \underbrace{\frac{3\ln(2)}{1.6}}_{\mathrm{inv,0.44}} \right]{\delta^+_0}^{2.75} \simeq 1.439 {\delta^+_0}^{2.9246} \tag{4.5\textit{b}} \label{eq:N_PO21_bl_2} \\
 N_\mathrm{Tanh} &=\frac{L_x L_z/\delta^2_0}{\Delta x^+ \Delta z^+ \Delta y^+_{\delta_0}} \left[ \underbrace{\frac{\alpha}{\tanh(\alpha)}}_{\mathrm{TBL,0.69}} + \underbrace{\frac{3\ln(2)}{2}}_{\mathrm{inv,0.31}} \right] {\delta^+_0}^3 \simeq 0.413 {\delta^+_0}^{3.1746}. \tag{4.5\textit{c}} \label{eq:N_hyp_bl_2}
\end{align}
The gray lines in figure~\ref{fig:Ndof_smooth_bl}(\textit{f}) support the validity of (4.5\textit{a-c}). The number of grid points for TBL grows with a steeper power of $\delta^+_0$ compared to its counterpart for turbulent channel flow (figure~\ref{fig:Ndof_smooth_channel}); this is due to the increase in $L_x/\delta_0$ with $\delta^+_0$ (4.3\textit{a}). In (4.5\textit{a-c}), we assess the fraction of the grid points inside the TBL ($y^+ \le \delta^+_0$) versus the inviscid region ($y^+ > \delta^+_0$). With $\eta$-grid (\ref{eq:N_eta_bl_2}), only $0.14 N_\eta$ falls inside the inviscid region, owing to the coarsening of both $\Delta y^+_\eta, \Delta z^+_\eta$ (figures~\ref{fig:Ndof_smooth_bl}\textit{d,e}). However, with the Cartesian grids (4.5\textit{b,c}), significant fractions of the total grid points ($0.44 N_\mathrm{PO}, 0.31 N_\mathrm{Tanh}$) fall inside the inviscid region, due to the fixed $\Delta z^+$ (figure~\ref{fig:Ndof_smooth_bl}\textit{e}). As a result, the ratio $N_\eta/N_\mathrm{Tanh}$ is lower for TBL compared to turbulent channel flow; conversely, $N_\mathrm{PO}/N_\mathrm{Tanh}$ is higher for TBL compared to channel flow (figure~\ref{fig:Ndof_smooth_bl}\textit{g}). By $\delta^+_0 \simeq 6000$, $N_\eta/N_\mathrm{Tanh}$ drops to $0.11$, whereas $N_\mathrm{PO}/N_\mathrm{Tanh}$ drops to $0.40$. 

\begin{figure}
  \centering
  \includegraphics[width=\textwidth,trim={{-0.22\textwidth} {0.0\textwidth} {0.02\textwidth} {0.0\textwidth}},clip]{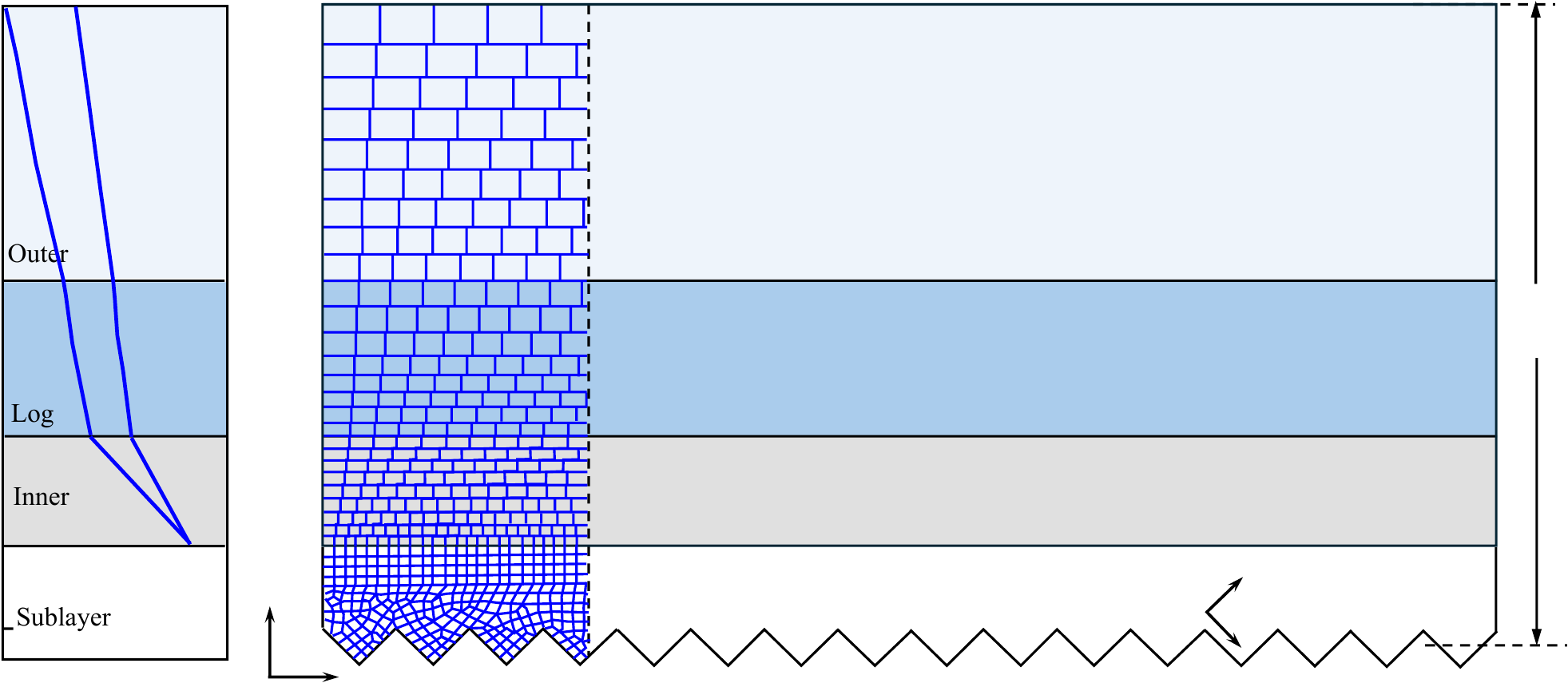}
 \put(-370,165){$\Scale[0.78]{(\textit{a})}$} 
 \put(-285,165){$\Scale[0.78]{(\textit{b})}$} 
 \put(-280,-5){$\Scale[0.78]{z}$}
 \put(-296,15){$\Scale[0.78]{y}$}
 \put(-71,10){$\Scale[0.78]{\xi}$}
 \put(-70,21){$\Scale[0.78]{n}$}
 \put(-358,11){$\Scale[0.75]{0}$}
 \put(-367,2){$\Scale[0.75]{-k^+}$}
 \put(-360,30){$\Scale[0.75]{\delta^+_r}$}
 \put(-362,56){$\Scale[0.75]{y^+_\mathrm{in}}$}
 \put(-379,90){$\Scale[0.75]{\dfrac{\delta^+_0}{2}-\dfrac{k^+}{2}}$}
 \put(-379,152){$\Scale[0.75]{\delta^+_0-\dfrac{k^+}{2}}$}
 \put(-350,75){$\Scale[0.75]{\Delta z^+_\eta}$}
 \put(-325,75){$\Scale[0.75]{\Delta y^+_\eta}$}
 \put(-316,24){$\Scale[0.75]{\Delta \ell^+}$}
 \put(-160,162){$\Scale[0.78]{N_{{yz}_\eta}}$}
 \put(-7,79){$\Scale[0.75]{\delta^+_0}$}
  \put(-191,19){$\Scale[0.75]{\dfrac{(L_z/\delta_0)(\delta^+_r + k^+/2)}{{\Delta \ell^+}^2}\delta^+_0}$}
  \put(-191,43){$\Scale[0.75]{\dfrac{(L_z/\delta_0)}{\Delta \ell^+} \dfrac{(y^+_\mathrm{in}-\delta^+_r)\ln(C_z/C_y)}{(C_z-C_y)(\kappa y^+_\mathrm{in})^\beta}\delta^+_0}$}
  \put(-191,75){$\Scale[0.75]{\dfrac{(L_z/\delta_0)}{C_y C_z \kappa^{2\beta}(1-2\beta)} \left[ \left(\dfrac{\delta^+_0}{2}\right)^{(1-2\beta)} - {y^+_\mathrm{in}}^{(1-2\beta)} \right] \delta^+_0}$}
  \put(-209,110){$\Scale[0.75]{\mbox{TBL: }\dfrac{L_z/\delta_0}{C_y C_z \kappa^{2\beta}(1-2\gamma)}\left[ \dfrac{1}{2^{(2\gamma-2\beta)}} - \dfrac{1}{2^{(1-2\beta)}} \right]{\delta^+_0}^{(2-2\beta)}}$}
  \put(-209,140){$\Scale[0.75]{\mbox{Channel: }\dfrac{(L_z/\delta_0)\ln(2)}{C_y C_z \kappa^{2\beta}2^{(1-2\beta)}}{\delta^+_0}^{(2-2\beta)}
  }$} 
  \caption{Same as figure~\ref{fig:idealised_grid}, but for turbulent flows over riblets. (\textit{a}) Profiles of $\Delta y^+_\eta$ (\ref{eq:Dyp_riblet}) and $\Delta z^+_\eta$ (\ref{eq:Dzp_riblet}) from the inner layer. (\textit{b}) Idealised representation of $\eta$-grid over riblets (5.1\textit{a-c}), and number of grid points on a $yz$-plane in different layers of $\eta$-grid. We obtain $N_{{yz}_\eta}$ in the log and outer regions by assuming $k^+ \ll \delta^+_0$. We draw the azimuthal and normal coordinates $(\xi,n)$ over the riblet surface.}
  \label{fig:idealised_grid_riblets}
\end{figure}
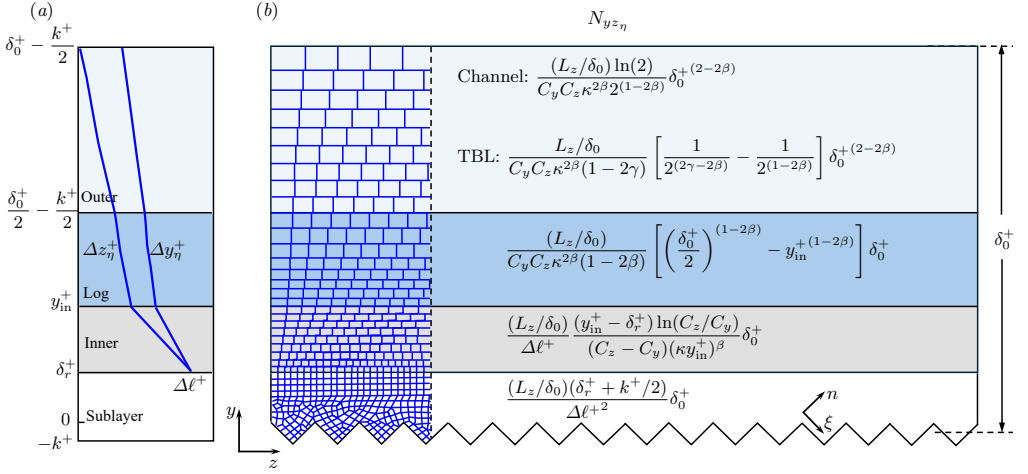

\section{Turbulent flows over riblets}
\subsection{Grid formulation}\label{sec:grid_riblets}
We extend the formulation of $\eta$-grid (2.2\textit{a,b}) for DNS of turbulent flows over riblets with the following formulation:
\begin{align}
& \mbox{square cells with size} \; \Delta \ell^+  \quad  -k^+ \le y^+ \le \delta^+_{r} \quad \mbox{Sublayer} \tag{5.1\textit{a}} \label{eq:Dlp_riblet} \\
 & \Delta y^+_\eta = \begin{cases}
 \Delta \ell^+ + r_y(y^+-\delta^+_{r}) & \delta^+_{r} < y^+ \le y^+_\mathrm{in} \quad \mbox{Inner} \\
C_y (\kappa y^+)^\beta &  y^+_\mathrm{in} < y^+ \le \dfrac{\delta^+_0}{2} \quad \quad \quad \mbox{Log} \\
C_y C_\eta {y^+}^\gamma &  \dfrac{\delta^+_0}{2} < y^+ \le \delta^+_0 - \dfrac{k^+}{2} \quad \mbox{Outer}
\end{cases}\tag{5.1\textit{b}} \label{eq:Dyp_riblet} \\
& \Delta z^+_\eta = \begin{cases}
\Delta \ell^+ + r_z(y^+-\delta^+_{r}) & \delta^+_{r} < y^+ \le y^+_\mathrm{in} \quad \mbox{Inner}  \\
C_z (\kappa y^+)^\beta &  y^+_\mathrm{in} < y^+ \le \dfrac{\delta^+_0}{2} \quad \quad \quad \mbox{Log} \\
C_z C_\eta {y^+}^\gamma &  \dfrac{\delta^+_0}{2} < y^+ \le \delta^+_0 - \dfrac{k^+}{2} \quad \mbox{Outer}
\end{cases} \tag{5.1\textit{c}} \label{eq:Dzp_riblet}
\end{align}
The above formulation and its parameters are illustrated in figure~\ref{fig:idealised_grid_riblets}. The origin $y^+ = 0$ is at the riblet crest, and $k^+$ and $s^+$ are respectively the viscous-scaled riblets height and spacing. Up to the riblet sublayer $y^+ = \delta^+_{r} \simeq 0.62s^+$~\citep{modesti2021dispersive}, we fill the space with square cells with size $s^+/30 \lesssim \Delta \ell^+ \lesssim s^+/20$ to well resolve the riblet groove area (\ref{eq:Dlp_riblet}). For $\delta^+_r < y^+ \le y^+_\mathrm{in}$, we place an inner layer, where $\Delta y^+_\eta$ and $\Delta z^+_\eta$ grow up to their values at the beginning of the log region, $C_y (\kappa y^+_\mathrm{in})^\beta, C_z (\kappa y^+_\mathrm{in})^\beta$. The growth factors are $r_y = [C_y (\kappa y^+_\mathrm{in})^\beta - \Delta \ell^+ ]/(y^+_\mathrm{in} - \delta^+_{r})$ and $r_z = [C_z (\kappa y^+_\mathrm{in})^\beta - \Delta \ell^+ ]/(y^+_\mathrm{in} - \delta^+_{r})$. The grid formulations in the log and outer regions are identical to the ones for the smooth wall (2.2\textit{a,b}). We follow the definition of \cite{endrikat2021influence} for the nominal $\delta^+_0$ (right-hand side of figure~\ref{fig:idealised_grid_riblets}\textit{b}), which is measured from the riblet mean height up to the top boundary (for open-channel flow), or up to $0.99U_\infty$ (for TBL). Based on this definition, the open-channel top boundary or the edge of the TBL is located at $y^+ = \delta^+_0-k^+/2$. For turbulent flows over riblets, friction velocity is obtained from
\begin{align}
 \mbox{Channel:} \quad u^2_{\tau_0} &= \frac{\tau_{w_0}}{\rho} = \dfrac{\bigintss_0^T \bigintss_0^{L_\xi} \bigintss_0^{L_x} (\nu \partial u/\partial n) dx d\xi dt}{L_x L_z T}, \tag{5.2\textit{a}} \label{eq:utau_ch} \\
 \mbox{TBL:} \quad u^2_\tau &= \frac{\tau_w}{\rho} = \dfrac{\bigintss_0^T \bigintss_0^{L_\xi} (\nu \partial u/\partial n) d\xi dt}{L_z T}. \tag{5.2\textit{b}} \label{eq:utau_bl}
\end{align}
Equation (\ref{eq:utau_ch}) presents the global $\tau_{w_0}$ for channel flow, as the total wall drag divided by the planar area, and (\ref{eq:utau_bl}) presents the local $\tau_w$ for TBL, as the local wall drag, at each $x$-location, divided by the domain width; $T$ is the time-averaging period, $\xi$ is the azimuthal coordinate over the riblets (figure~\ref{fig:idealised_grid_riblets}\textit{b}), and $\partial u/\partial n$ is the streamwise velocity gradient normal to the riblet surface.

\begin{figure}
  \centering
  \includegraphics[width=\textwidth,trim={{0.0\textwidth} {0.0\textwidth} {0.0\textwidth} {0.0\textwidth}},clip]{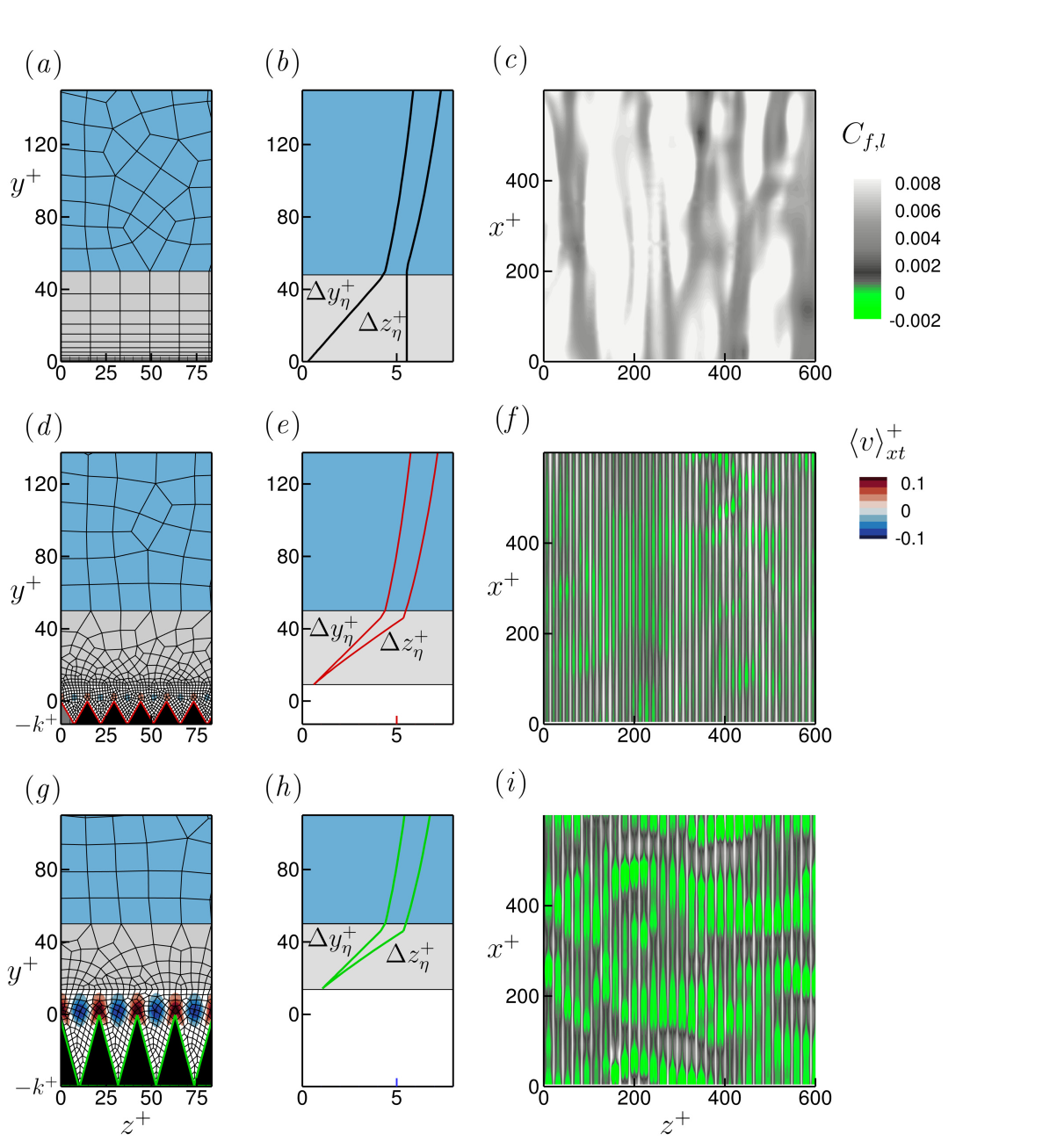} 
  \caption{Illustration of the unique flow mechanisms over riblets compared to over a smooth wall. Turbulent open-channel flow at $\delta^+_0 \simeq 400$ over (\textit{a-c}) smooth wall (from table~\ref{tab:yp_study} with $y^+_\mathrm{in} = 50$), and over riblet cases T615 (\textit{d--f}) and T321 (\textit{g--i}) from table~\ref{tab:riblets_channel}. The grid in (\textit{a}) is $\eta$-grid for smooth wall (2.2\textit{a,b}), and the grids in (\textit{d,g}) are $\eta$-grids for riblets (5.1\textit{a--c}). In (\textit{d,g}), the blue-to-red contour fields plot the $xt$-averaged vertical velocity $\left< v \right>^+_{xt}$, highlighting the vertical motions of the secondary flows. (\textit{b,e,h}) Profiles of $\Delta y^+_\eta, \Delta z^+_\eta$; (\textit{b}) plots (2.2\textit{a,b}) with $C_y = 2.0, C_z = 2.5, y^+_\mathrm{in} = 50$ and $\Delta y^+_w = 0.3$; (\textit{e,h}) plot (5.1\textit{a-c}) with $C_y = 2.0, C_z = 2.5, y^+_\mathrm{in} = 50$ and $\Delta \ell^+ \simeq 0.6 \simeq s^+/25$ (\textit{e}), and $\Delta \ell^+ \simeq 1.1 \simeq s^+/20$ (\textit{h}). (\textit{c,f,i}) Fields of local instantaneous skin-friction coefficient $C_{f,l}$; regions of $C_{f,l} \le 0$ are coloured in green, to highlight KH rollers.}
  \label{fig:riblets_Cf_v_uvp_grid}
\end{figure}

We formulated (5.1\textit{a-c}) by taking into account some prominent flow mechanisms over riblets. We explain these mechanisms via figure~\ref{fig:riblets_Cf_v_uvp_grid}, presenting turbulent open-channel flow simulations over two riblet geometries at $\delta^+_0 = 400$ (from table~\ref{tab:riblets_channel}). These cases are essentially the DNSs of \cite{endrikat2021influence}, reproduced with $\eta$-grid (5.1\textit{a-c}). The riblets are isosceles triangles, with tip angle $\alpha = \ang{60}$ and $s^+ \simeq 15$ (T615, figures~\ref{fig:riblets_Cf_v_uvp_grid}\textit{d-f}), and with $\alpha = \ang{30}$ and $s^+ \simeq 21$ (T321, figures~\ref{fig:riblets_Cf_v_uvp_grid}\textit{g-i}); the former riblet geometry reduces drag, while the latter one increases drag. The important flow mechanisms are itemised below.

\begin{itemize}
 \item $\:$ \textit{Virtual origin} is an equivalent smooth wall as perceived by the turbulent eddies above the riblet crest. Various approaches have been proposed to locate the virtual origin. Regardless of the approach, the virtual origin is below the riblet crest~\citep{bechert1989viscous}; this is also supported by the comprehensive DNSs of \cite{endrikat2021influence} over riblets with triangular, trapezoidal and rectangular grooves, tip angles $\alpha = \ang{30}, \ang{60}, \ang{90}$, spacing $10 \lesssim s^+ \lesssim 62$ and height $7 \lesssim k^+ \lesssim 41$; they locate the virtual origin through the Reynolds shear-stress profiles, following~\cite{luchini1996reducing}. From their DNS campaign, we found the virtual origin falls close to the riblet crest (between $0.04k$ to $0.26k$ below the crest). Therefore, it is reasonable to place $y^+ = 0$ at the riblet crest (figures~\ref{fig:idealised_grid_riblets} and \ref{fig:riblets_Cf_v_uvp_grid}\textit{d,g}).
 \item $\:$ \textit{Riblet sublayer $\delta_r$} is the region of direct influence by riblets~\citep{modesti2021dispersive}; it is the distance from the virtual origin up to which the time-averaged flow field is heterogeneous in wall-parallel directions. The heterogeneity is due to the riblet-generated secondary flows; in figures~\ref{fig:riblets_Cf_v_uvp_grid}(\textit{d,g}), the vertical motions of secondary flows are identified through the streamwise and time averaged wall-normal velocity $\left< v \right>^+_{xt}$, following figure~9 in \cite{modesti2021dispersive}. \cite{modesti2021dispersive} quantified $\delta^+_r$ for \cite{endrikat2021influence}'s riblet cases; the data points fitted well with $\delta^+_r = 0.62 s^+$. In figures~\ref{fig:riblets_Cf_v_uvp_grid}(\textit{d,g}), the lobes of positive (red) and negative (blue) $\left< v \right>^+_{xt}$ fall below $y^+ \simeq 0.6 s^+$. Therefore, the space up to $y^+ = \delta^+_r \simeq 0.6 s^+$ (called sublayer in figure~\ref{fig:idealised_grid_riblets}\textit{a}) is filled with square cells with size $\Delta \ell^+ \lesssim s^+/20$, to well resolve the secondary flows (\ref{eq:Dlp_riblet}).
 \item $\:$ \textit{Kelvin-Helmholtz (KH) rollers} are spanwise-aligned coherent structures that emerge near the riblet crest, but disturb the flow down to the riblet groove~\citep{endrikat2021influence}. In figures~\ref{fig:riblets_Cf_v_uvp_grid}(\textit{f,i}), the footprint of KH rollers are evident in the patches of negative local instantaneous skin-friction coefficient $C_{f,l} = 2(\nu \partial u/\partial n)/U^2_b$. These patches indicate more coherent KH rollers over the drag-increasing T321 (figure~\ref{fig:riblets_Cf_v_uvp_grid}\textit{i}) compared to the drag-reducing T615 (figure~\ref{fig:riblets_Cf_v_uvp_grid}\textit{f}). Our formulation (5.1\textit{a-c}) naturally resolves the KH rollers. These rollers are negligibly weak or non-existent over drag-reducing riblets~\citep{garcia2011hydrodynamic}, and not all drag-increasing riblets trigger KH rollers~\citep{rowin2025experimental,camobreco2025only}. Triangular riblets with $\alpha = \ang{30}$, and blade riblets are the prominent geometries that promote KH rollers~\citep{endrikat2021influence,rouhi2022riblet}. In these cases, the KH rollers have a spanwise length $\lambda^+_z \simeq 1000- 1500$~\citep{garcia2012scaling}. Also, they do not protrude beyond $0.5s^+$ above the riblet crest, as evident from the co-spectra of Reynolds shear stress by \cite{endrikat2021influence} (their figure~9). Therefore, filling $y^+ \lesssim 0.6s^+ $ with square elements of size $\Delta \ell^+ \lesssim s^+/20$, well resolves the spanwise and vertical lengths of KH rollers. The streamwise length of KH rollers $65 \lesssim \lambda^+_x \lesssim 290$~\citep{garcia2011hydrodynamic} is a constraint for the streamwise grid size.    
\end{itemize}\vspace{0.1cm}
We believe that $\eta$-grid (5.1\textit{a-c}) is also applicable to turbulent flows over roughness, provided that an estimate of the roughness sublayer $\delta^+_r$ is available. Similar to riblets, $\delta^+_r$ scales with the roughness geometrical characteristics, e.g.\ the roughness height~\citep{raupach1991rough,yuan2018topographical}, or its spanwise wavelength~\citep{chan2018secondary}.

\begin{table}
\centering
 \begin{tabular}{ccccccccccc}
 & Case & Code & $C_y, C_z, y^+_\mathrm{in}$ & $\Delta x^+$ & $\Delta \ell^+$ & $n_s$ & $\Delta y^+$ & $\Delta z^+$ & $N_\mathrm{dof}$ & $\Delta U^+$  \\ 
 \multirow{3}{*}{\includegraphics[width=.07\textwidth,trim={{0.0\textwidth} {1.3\textwidth} {0.0\textwidth} {0.0\textwidth}},clip]{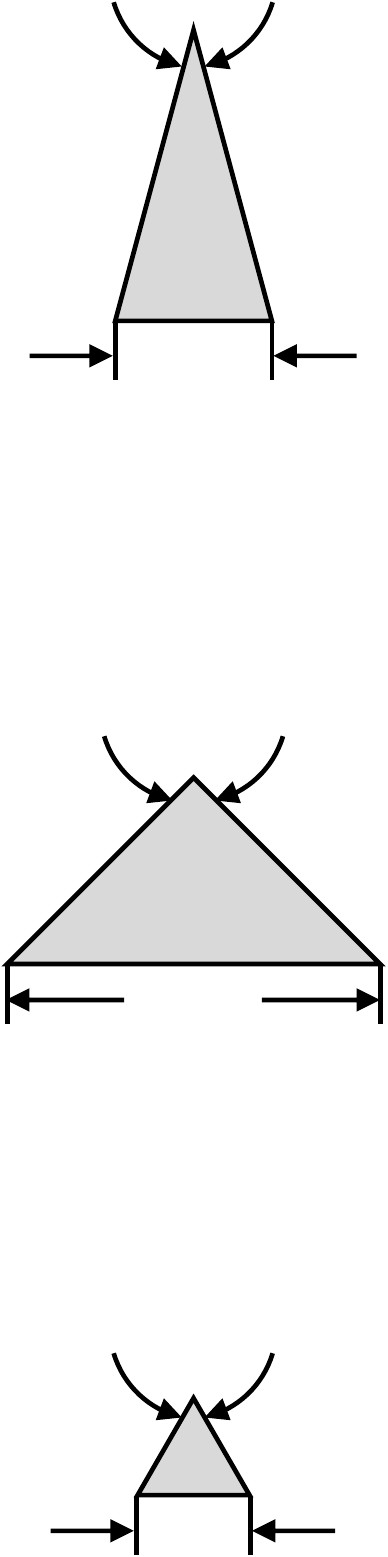}
\put(-18.0,35){\scriptsize $ \ang{30}$}
\put(-20.0,3){\scriptsize $ 21.1$}}
&  &  &  & & & &  & & \\
 & T321\_S & SOD & $2.0, 2.5, 50$ & $10.0$ & $1.04$ & $20$ & $1.04 - 8.57$ & $1.04 - 10.54$ & $17.9$ M & $0.83$ \\ 
 & T321F\_S& SOD & $-$             & $10.0$ & $0.50$ & $42$ & $0.50 - 7.92$ & $0.50 - 3.10$ & $80.3$ M & $0.77$ \\
 & T321\_C& Cliff & $-$            & $6.0$ & $-$ & $29$ & $0.023 - 6.9$ & $0.12 - 3.2$ & $-$ & $0.83$ \\ 
  &  &  &  & & & &  &  & & \\
\multirow{2}{*}{\includegraphics[width=.07\textwidth,trim={{0.0\textwidth} {0.0\textwidth} {0.0\textwidth} {1.6\textwidth}},clip]{figures/T321_T615_T950}
\put(-18.0,15){\scriptsize $ \ang{60}$}
\put(-20.0,-6){\scriptsize $ 14.7$}} 
 & T615\_S & SOD & $2.0, 2.5, 50$ & $10.0$ & $0.54$ & $27$ & $0.54 - 8.36$ & $0.54 - 11.07$ & $29.2$ M & $-0.79$ \\ 
 & T615F\_S & SOD & $-$           & $7.0$ & $0.38$ & $39$ & $0.38-5.16$ & $0.38-8.09$ & $87.8$ M & $-0.89$\\ 
 & T615\_C& Cliff & $-$           & $6.0$ & $-$    & $29$ & $0.041 - 7.0$ & $0.083 - 2.2$ & $-$ & $-0.82$ \\
 &  &  &  & & & &  &  & \\  
& T950\_S & SOD & $2.0, 2.5, 60$ & $10$ & $1.47$ & $34$ & $1.47 - 8.91$ & $1.47 - 11.68$ & $11.6$ M & $0.85$ \\
\multirow{3}{*}{\includegraphics[width=.07\textwidth,trim={{0.0\textwidth} {0.3\textwidth} {0.0\textwidth} {0.9\textwidth}},clip]{figures/T321_T615_T950}
\put(-18.0,40){\scriptsize $ \ang{90}$}
\put(-17.5,15){\scriptsize $  50$}}
& T950C\_S & SOD & $2.0, 2.5, 60$ & $10$ & $2.36$ & $21$ & $2.36 - 8.92$ & $2.36 - 11.68$ & $6.2$ M & $0.77$ \\ 
& T950F\_S & SOD & $-$          & $10$ & $0.98$& $51$ & $0.98 - 7.86$ & $0.98 - 4.48$ & $28.0$ M & $0.81$ \\
& T950\_O & OF & $2.0, 2.5, 60$ & $10$ & $1.54$ & $32$ & $1.54 - 8.99$ & $1.54 - 10.83$ & $10.6$ M & $0.81$ \\
& T950\_C& Cliff & $-$          & $6.0$ & $-$ & $33$ & $0.029 - 7.00$ & $0.30 - 7.1$ & $-$ & $0.78$ \\
    \end{tabular}
\caption{Simulation cases for turbulent open-channel flow over riblets with $\delta^+_0 = 400$. The leftmost column shows the riblets' geometries, including their tip angles and viscous-scaled spacing. The domain setup and the grids for the cases with SOD2D are depicted in figure~\ref{fig:riblets_setup}. For comparison, we add the reference cases by \cite{endrikat2021influence} (cases ending with `\_C'). We report the total number of grid points $N_\mathrm{dof}$, as well as the number of grid points per riblet spacing $n_s$. The velocity difference $\Delta U^+ = U^+_\mathrm{smooth} - U^+_\mathrm{riblet}$ is calculated consistent with \cite{endrikat2021influence}. $U^+_\mathrm{riblet}$ is plotted versus $y^+ - y^+_\mathrm{vo}$, and is subtracted from $U^+_\mathrm{smooth}$; $\Delta U^+$ is their difference at $y^+ - y^+_\mathrm{vo} = 100$.}
\label{tab:riblets_channel}
\end{table}

\begin{figure}
  \centering
  \includegraphics[width=\textwidth,trim={{0.0\textwidth} {0.0\textwidth} {0.0\textwidth} {0.0\textwidth}},clip]{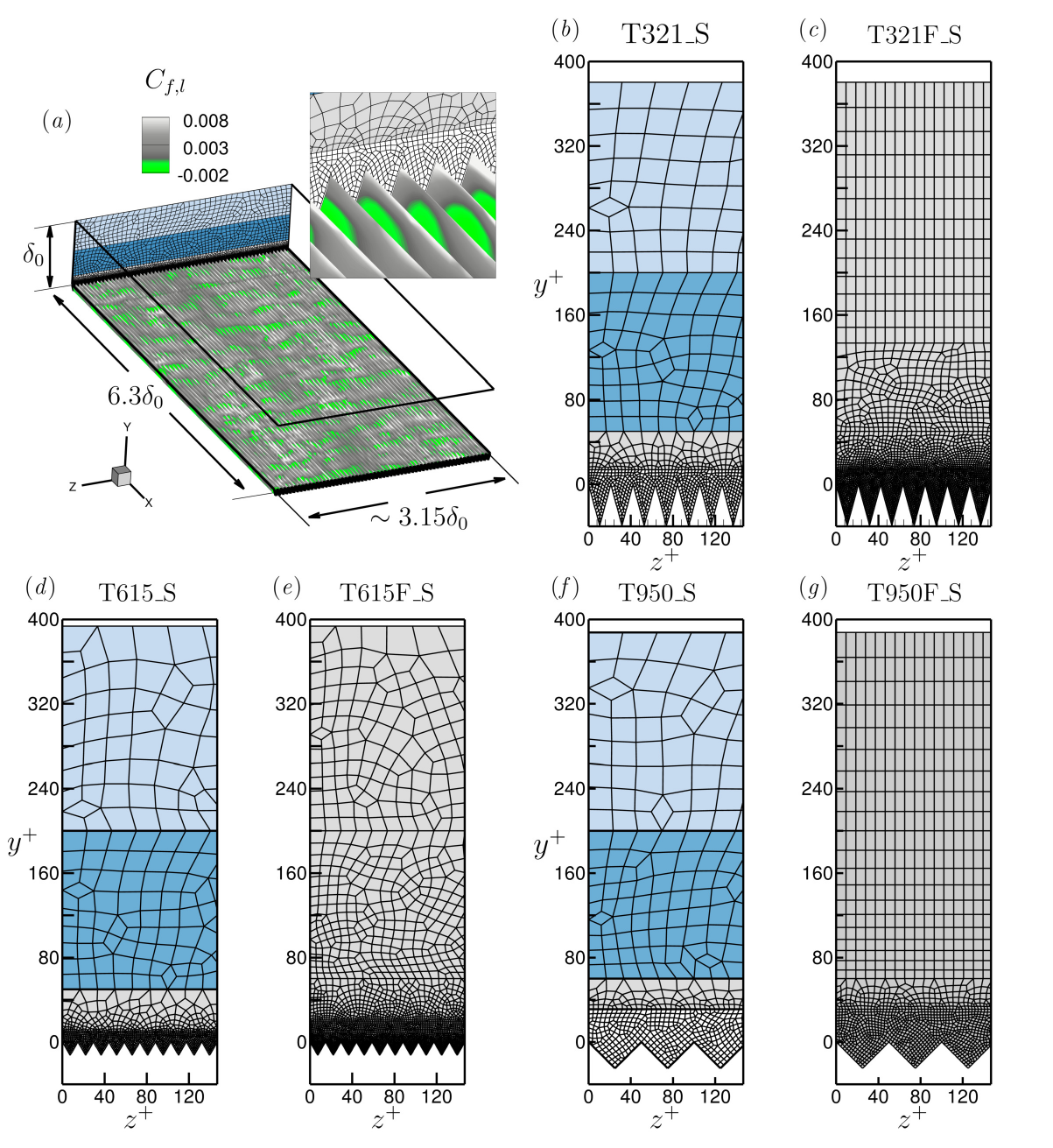} 
  \caption{Setup and grids for turbulent open-channel flow over riblets at $\delta^+_0 = 400$ (table~\ref{tab:riblets_channel}). (\textit{a}) Domain dimensions for a representative case T321\_S; local instantaneous $C_{f,l}$ is visualised over the riblets, with $C_{f,l} \le 0$ in green. (\textit{b--g}) Spectral elements for the cases with SOD2D; (\textit{b,d,f}) are the cases with $\eta$-grid (5.1\textit{a--c}), and (\textit{c,e,g}) are the finer grid cases.}
  \label{fig:riblets_setup}
\end{figure}

\subsection{Turbulent open-channel flow}\label{sec:riblet_channel}
We test the accuracy of $\eta$-grid (5.1\textit{a-c}) by replicating three cases of \cite{endrikat2021influence} (table~\ref{tab:riblets_channel}). They are turbulent open-channel flows over two drag-increasing riblet geometries (T321, T950) and a drag-reducing one (T615); geometrical details of these riblets are provided in table~\ref{tab:riblets_channel} (left-hand side). The computational domain is presented in figure~\ref{fig:riblets_setup}(\textit{a}). We conduct DNSs with $\eta$-grid and SOD2D as the solver (cases ending with `\_S'). We set the grid parameters $C_y = 2.0, C_z = 2.5$ and $y^+_\mathrm{in} = 50 - 60$, found to be suitable based on DNSs of smooth wall turbulent channel flow and TBL (\S\ \ref{sec:smooth_channel}, \ref{sec:smooth_tbl}). We denote these cases and their grids as our production runs and grids (figures~\ref{fig:riblets_setup}\textit{b,d,f}). For these cases, $\Delta \ell^+$ is adjusted such that the number of grid points per riblet spacing $n_s = s^+/\Delta \ell^+$ is between $n_s = 20$ (T321\_S) and $34$ (T950\_S). We also test $\eta$-grid with OpenFOAM for T950 (T950\_O), with the same grid parameters as the ones for T950\_S. To study grid convergence, we repeat T950\_S with $1.6$ times coarser $\Delta \ell^+$, hence lower $n_s$, named T950C\_S. Also, for each riblet geometry, we conduct a finer grid case (cases ending with `F\_S'), with their grids presented in figures~\ref{fig:riblets_setup}(\textit{c,e,g}). Compared to the production grids, the finer grids have $2.5$ to $4.5$ times more number of grid points ($N_\mathrm{dof}$). For the fine grids, $\Delta \ell^+$ is reduced by $1.5$ times (T615F\_S, T950F\_S) and $2$ times (T321F\_S), and $\Delta y^+$ follows the hyperbolic tangent mapping (\ref{eq:tanh}). For T321F\_S and T950F\_S (figures~\ref{fig:riblets_setup}\textit{c,g}), the maximum $\Delta z^+$ is $3.1$ and $4.48$, respectively, about three times finer than the maximum $\Delta z^+_\eta$ for their production counterparts with $\eta$-grid, T321\_S, T950\_S (figures~\ref{fig:riblets_setup}\textit{b,f}). For T615F\_S (figure~\ref{fig:riblets_setup}\textit{e}), the grid sizes are about $1.5$ times finer in all three directions compared to its production counterpart T615\_S (figure~\ref{fig:riblets_setup}\textit{d}). For comparison, we add the reference cases by \cite{endrikat2021influence} to table~\ref{tab:riblets_channel} (cases ending with `\_C'); they used the second-order FVM solver Cliff by Cascade Technologies Inc.\ \citep{ham2004energy,ham2006accurate}. They set a finer streamwise grid size ($\Delta x^+ = 6$) compared to our production cases ($\Delta x^+ = 10$). Their $yz$-grid is the conformal mapping of a smooth-wall grid with a non-uniform $\Delta \ell^+$ over riblets (figure~3 in \citealt{endrikat2021influence}). Their $n_s$ values are close to the ones from our production grids, e.g.\ $n_s = 29$ in T615\_C versus $n_s = 27$ in T615\_S. Their maximum $\Delta z^+$ is $3$ to $5$ times finer than our production grids, comparing T321\_C with T321\_S, and T615\_C with T615\_S.



\begin{figure}
  \centering
  \includegraphics[width=\textwidth,trim={{0.0\textwidth} {0.0\textwidth} {0.0\textwidth} {0.0\textwidth}},clip]{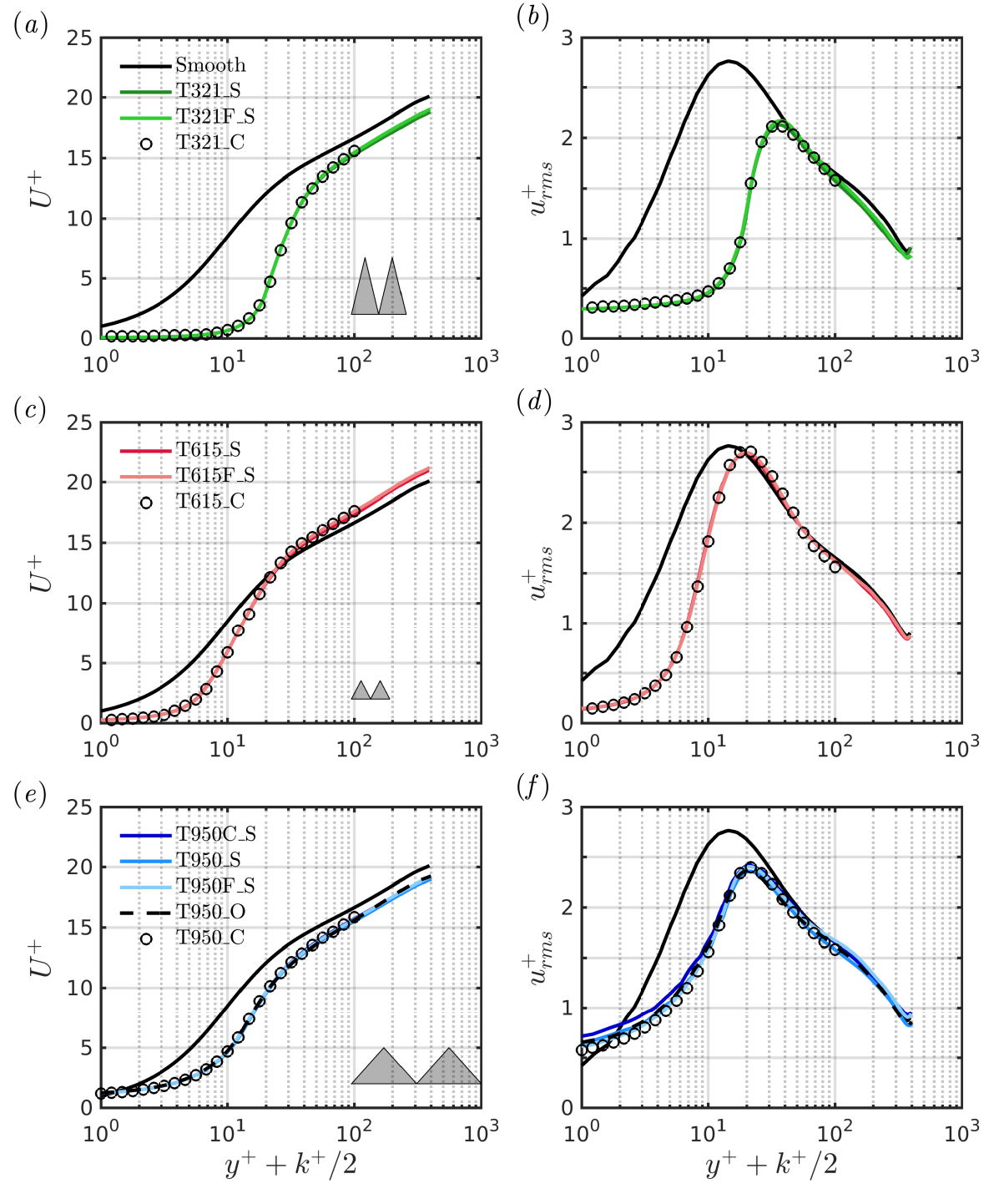} 
  \caption{Statistics of turbulent open-channel flow DNSs over riblets at $\delta^+_0 = 400$ (table~\ref{tab:riblets_channel}). Profiles of (\textit{a,c,e}) $U^+$ and (\textit{b,d,f}) $u^+_{rms}$ for (\textit{a,b}) T321, (\textit{c,d}) T615, and (\textit{e,f}) T950. Following \cite{endrikat2021influence}, the origin of plots is placed at the riblets mean height ($y^+ = -k^+/2$). The smooth profiles correspond to our smooth wall turbulent open-channel flow with $C_y = 2.0, C_z = 2.5$ and $y^+_\mathrm{in} = 50$ (Set 1, table~\ref{tab:yp_study}). Reference profiles by \cite{endrikat2021influence} are shown up to $y^+ + k^+/2 = 100$; their profiles beyond this range are not resolved by the minimal channel domain.}
  \label{fig:riblets_channel_stats}
\end{figure}

Results from different grids and solvers (table~\ref{tab:riblets_channel}) are compared in terms of the profiles of $U^+$ (figures~\ref{fig:riblets_channel_stats}\textit{a,c,e}) and $u^+_{rms}$ (figures~\ref{fig:riblets_channel_stats}\textit{b,d,f}). \cite{endrikat2021influence} conducted their DNSs in a minimal channel unit, resolving the flow up to $y^+ \simeq 100$; therefore, we discard their profiles beyond this range (circle symbols).
For all riblet geometries, profiles from our production cases (ending with `\_S') are in excellent agreement with the finer cases (ending with `F\_S'), and they are in great agreement with the reference profiles by \cite{endrikat2021influence} (ending with `\_C'). We achieve the same level of agreement when we test $\eta$-grid with OpenFOAM (T950\_O in figures~\ref{fig:riblets_channel_stats}\textit{e,f}). The accuracy of our production grid is also evident in the velocity difference $\Delta U^+$ (table~\ref{tab:riblets_channel}). The values of $\Delta U^+$ from the production cases have maximum $0.07$ difference compared to the ones by \cite{endrikat2021influence}; this difference is within the uncertainty range of $\pm 0.1$ that \cite{endrikat2021influence} report for their $\Delta U^+$ values. The case T950C\_S $(\Delta \ell^+ = 2.36, n_s = 21)$ has a coarser grid within the riblet sublayer $(y^+ \le \delta^+_r)$ compared to the production case T950\_S $(\Delta \ell^+ = 1.47, n_s = 34)$. As a result, $u^+_{rms}$ from T950C\_S is slightly higher than the finer cases for $(y^+ + k^+/2) \lesssim 10$ (figure~\ref{fig:riblets_channel_stats}\textit{f}). Nevertheless, its $U^+$ profile (figure~\ref{fig:riblets_channel_stats}\textit{e}), as well as its $\Delta U^+$ (table~\ref{tab:riblets_channel}), are in very good agreement with the finer cases.

We conclude that similar to the smooth wall $\eta$-grid (2.2\textit{a,b}), its extended formulation for riblets (5.1\textit{a--c}), with the grid parameters $C_y = 2.0, C_z = 2.5, y^+_\mathrm{in} = 50$, yields grid-converged statistics, and in great agreement with the reference DNS data~\citep{endrikat2021influence}. These grid parameters perform accurately with both SOD2D and OpenFOAM. For the additional parameter $\Delta \ell^+$, it is not trivial to have a general prescription, as it depends on the riblet geometry. For instance, for T321\_S, with $\Delta \ell^+ = 1.04$, we allocate $n_s = 20$ grid cells per $s^+$, which yields grid convergence in the $U^+$ and $u^+_{rms}$ profiles (figures~\ref{fig:riblets_channel_stats}\textit{a,b}). On the other hand, for T950C\_S, with $\Delta \ell^+ = 2.36$, we allocate $n_s = 21$ grid cells per $s^+$, yet the $u^+_{rms}$ profile is slightly under-resolved within the riblet sublayer (figure~\ref{fig:riblets_channel_stats}\textit{f}). Therefore, a grid-convergence study for $\Delta \ell^+$ is recommended for new geometries.

\subsection{Turbulent boundary layer}\label{sec:riblet_tbl}
\begin{figure}
  \centering
  \includegraphics[width=\textwidth,trim={{0.0\textwidth} {0.05\textwidth} {0.28\textwidth} {0.0\textwidth}},clip]{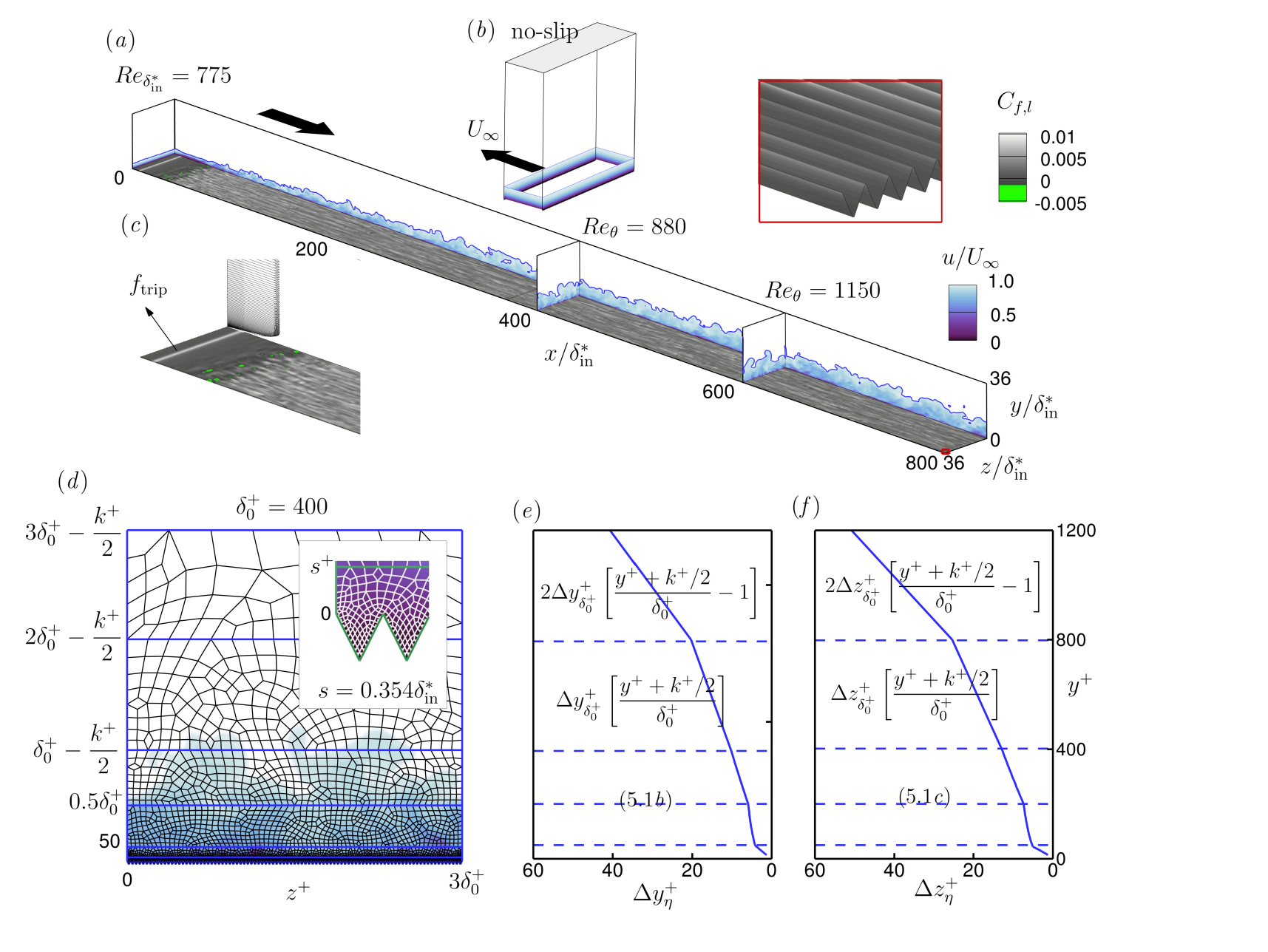}
  \caption{Setup and grid for ZPG TBL over triangular riblets with $s = k$, $\delta^+_0 = 400$ and $s^+ = 12$ (table~\ref{tab:riblets_bl_table}). (\textit{a}) Computational domain, and visualisations of $u$ and $C_{f,l}$. (\textit{b}) Temporal boundary layer setup for the inlet laminar boundary layer over riblets at $Re_{\delta^*_\mathrm{in}} = 775$. (\textit{c}) Close-up view of the inlet with the parametric forcing trip. (\textit{d}) Visualisation of the spectral elements, and (\textit{e,f}) profiles of $\Delta y^+_\eta$ and $\Delta z^+_\eta$ for $\delta^+_0 = 400$, that follow (5.1\textit{a--c}) up to $y^+ = \delta^+_0 - k^+/2$, and then linearly increase with $y^+$, similar to the grid sizes for the smooth-wall TBL (figures~\ref{fig:flowviz_bl}\textit{c--e}).}
  \label{fig:flowviz_riblets_bl}
\end{figure}

\begin{figure}
  \centering
  \includegraphics[width=\textwidth,trim={{0.0\textwidth} {0.0\textwidth} {0.0\textwidth} {0.0\textwidth}},clip]{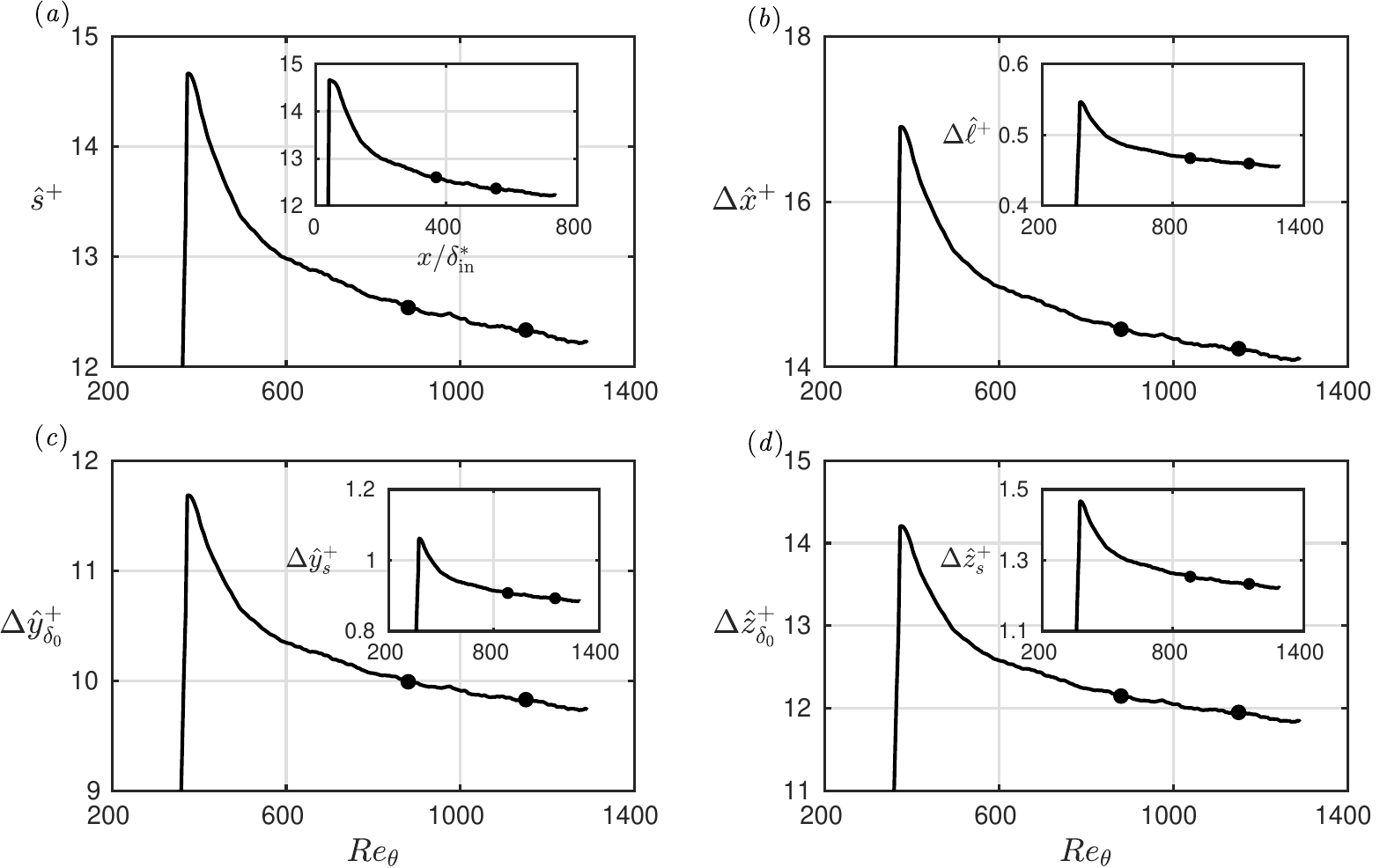}
  \caption{Variations of the viscous-scaled riblet spacing and grid sizes based on the local $u_\tau$. (\textit{a}) $\hat{s}^+$ versus $Re_\theta$ and versus $x/\delta^*_\mathrm{in}$ (inset); (\textit{b}) $\Delta \hat{x}^+$ and $\Delta \hat{\ell}^+$ (inset) versus $Re_\theta$; (\textit{c}) $\Delta \hat{y}^+_{\delta_0}$ and $\Delta \hat{y}^+_s$ (inset), and (\textit{d}) $\Delta \hat{z}^+_{\delta_0}$ and $\Delta \hat{z}^+_s$ (inset) versus $Re_\theta$, which are the grid sizes at $y = \delta_0$ and $s$, respectively. The bullets mark $Re_\theta = 880$ and $1150$, with the corresponding values of $\hat{s}^+$ and grid sizes reported in table~\ref{tab:riblets_bl_table} (right side).}
  \label{fig:riblets_bl_dxp_dyp_dzp}
\end{figure}

Our final test case is a ZPG TBL over riblets (figure~\ref{fig:flowviz_riblets_bl}). We set up our configuration to make our results comparable to the experimental studies by \cite{baron1993some} and \cite{choi1997turbulence}. Both experiments consider triangular riblets with unit aspect ratio ($s^+ = k^+$), and take measurements at similar values of $s^+$ and $Re_\theta$; $s^+ \simeq 12,Re_\theta \simeq 1150$ in \cite{baron1993some}, and $s^+ \simeq 13, Re_\theta \simeq 880$ in \cite{choi1997turbulence}. To quantify drag reduction, each study conducts measurements of a ZPG TBL over a smooth wall. Our ZPG TBL over a smooth wall with the short domain resolves the TBL up to $Re_\theta \simeq 1200$ (figure~\ref{fig:Cf_H12}). Therefore, our setup over riblets (figure~\ref{fig:flowviz_riblets_bl}\textit{a}) is almost identical to our short-domain setup, in terms of the domain dimensions (table~\ref{tab:riblets_bl_table}, left side), laminar inlet ($Re_{\delta^*_\mathrm{in}} = 775$), tripping parameters (\ref{eq:force}), and boundary conditions. Riblets up to $s^+ \simeq 35$ create a velocity shift $| \Delta U^+ | \lesssim 0.8$, equivalent to a drag change less than $\pm 5\%$, or change in $u_\tau$ less than $\pm 3\%$ for $\delta^+ \lesssim 1500$, e.g.\ figure 8 by \cite{endrikat2022reorganisation}. Therefore, under these conditions, the correlations for TBL characteristics over a smooth wall (\ref{eq:bl_corr}) are nearly valid over riblets. We set the riblet spacing in units of $\delta^*_\mathrm{in}$ based on the target $\delta^+_0 = 400$ and $s^+ \equiv s u_{\tau_0}/\nu = 12$, which yield $s/\delta^*_\mathrm{in} = (s^+/\delta^+_0)(Re_{\delta_0}/Re_{\delta^*_\mathrm{in}}) = 0.354$, where $Re_{\delta_0}$ is obtained from (4.2\textit{b}). In figure~\ref{fig:riblets_bl_dxp_dyp_dzp}(\textit{a}), we plot $\hat{s}^+ \equiv s u_\tau/\nu$ based on the local $u_\tau$ from the simulation (5.2\textit{b}), and in table~\ref{tab:riblets_bl_table} (right side) we report $\hat{s}^+$ at locations of interest. The resulting $\hat{s}^+ \simeq 12.5$ is comparable with those by \cite{baron1993some} and \cite{choi1997turbulence} at matched $Re_\theta = 880, 1150$. 
We obtain the inlet condition over riblets from a precursor temporal boundary layer simulation (figure~\ref{fig:flowviz_riblets_bl}\textit{b}). The setup consists of a periodic box in the streamwise and spanwise directions, with a moving bottom wall at the free-stream velocity $U_\infty$, and no-slip condition at the top boundary~\citep{kozul2016direct}. The simulation is initialised from zero and continues until $Re_{\delta^*_\mathrm{in}} = 775$. Over riblets, $\delta$ is measured from the riblet mean height ($y=-k/2$) up to $0.99U_\infty$, and 
\begin{align}
 \delta^* = \frac{1}{L_z} \int \int_{A_\delta} \left( 1 - \frac{\left< u \right>_t}{U_\infty} \right) dA_\delta, \quad \theta = \frac{1}{L_z} \int \int_{A_\delta} \frac{\left< u \right>_t}{U_\infty} \left( 1 - \frac{\left< u \right>_t}{U_\infty} \right) dA_\delta \tag{5.3\textit{a,b}} \label{eq:riblets_dstar_theta}
\end{align}
where $\left< u \right>_t$ is the time-averaged velocity, and $A_\delta$ is the cross-sectional area from the riblet surface up to the TBL edge ($y  = \delta - k/2$).

\begin{table}
\centering
 \begin{tabular}{cccc|ccccc}
         \multicolumn{4}{c|}{For setup} & \multicolumn{5}{c}{From simulation} \\
         $(L_x,L_y,L_z)/\delta^*_\mathrm{in}$ & $\delta^+_0$ & $s^+$ & $s/\delta^*_\mathrm{in}$ & $\delta^+$ & $Re_{\theta}$ & $\hat{s}^+$ & $x/\delta^*_\mathrm{in}$ & $\Delta \hat{\ell}^+ \quad \Delta \hat{x}^+ \quad (\Delta \hat{y}^+_s, \Delta \hat{y}^+_{\delta_0}) \quad (\Delta \hat{z}^+_s, \Delta \hat{z}^+_{\delta_0})$ \\
        $(840,36,36.108)$ & $400$ & $12.0$ & $0.354$ & $362$ & $880$ & $12.54$ & $399$ & $0.47 \quad 14.46 \quad (0.91, 9.99) \quad (1.25, 12.15)$ \\
          &   &   &                                  & $450$ & $1150$ & $12.33$ & $601$ & $0.46 \quad 14.22 \quad (0.89, 9.83) \quad (1.23, 11.95)$        
    \end{tabular}
\caption{Simulation details for ZPG TBL over triangular riblets with unit aspect ratio ($s^+ = k^+$), figure~\ref{fig:flowviz_riblets_bl}. The left side presents the domain and riblet dimensions based on the target $\delta^+_0$ and $s^+$. The right side presents the TBL characteristics at locations of interest for comparison with the reference experiments~\citep{baron1993some,choi1997turbulence}.}
\label{tab:riblets_bl_table}
\end{table}

Figures~\ref{fig:flowviz_riblets_bl}(\textit{d,e,f}) demonstrate the grid details for $\delta^+_0 = 400$. The $\eta$-grid sizes $\Delta y^+_\eta$ and $\Delta z^+_\eta$ follow (5.1\textit{a--c}) up to the TBL edge ($y^+ = \delta^+_0 - k^+/2$); we set $\Delta \ell^+ = 0.5, y^+_\mathrm{in} = 50, C_y = 2.0$ and $C_z = 2.5$, as found to be suitable parameters from our extensive DNS cases. Beyond the TBL edge, $\Delta y^+_\eta$ and $\Delta z^+_\eta$ increase linearly with $y^+$, similar to the grid for the smooth-wall TBL (figures~\ref{fig:flowviz_bl}\textit{c--e}). We generate a uniform streamwise grid with $\Delta x^+ = 15$. In figures~\ref{fig:riblets_bl_dxp_dyp_dzp}(\textit{b,c,d}), we plot the variations of the viscous-scaled grid sizes based on the local $u_\tau$, and in table~\ref{tab:riblets_bl_table} (right side) we report their values at locations of interest. The resulting grid sizes are what we expect from our $\eta$-grid parameters.

 \begin{figure}
  \centering
  \includegraphics[width=\textwidth,trim={{0.0\textwidth} {0.0\textwidth} {0.0\textwidth} {0.0\textwidth}},clip]{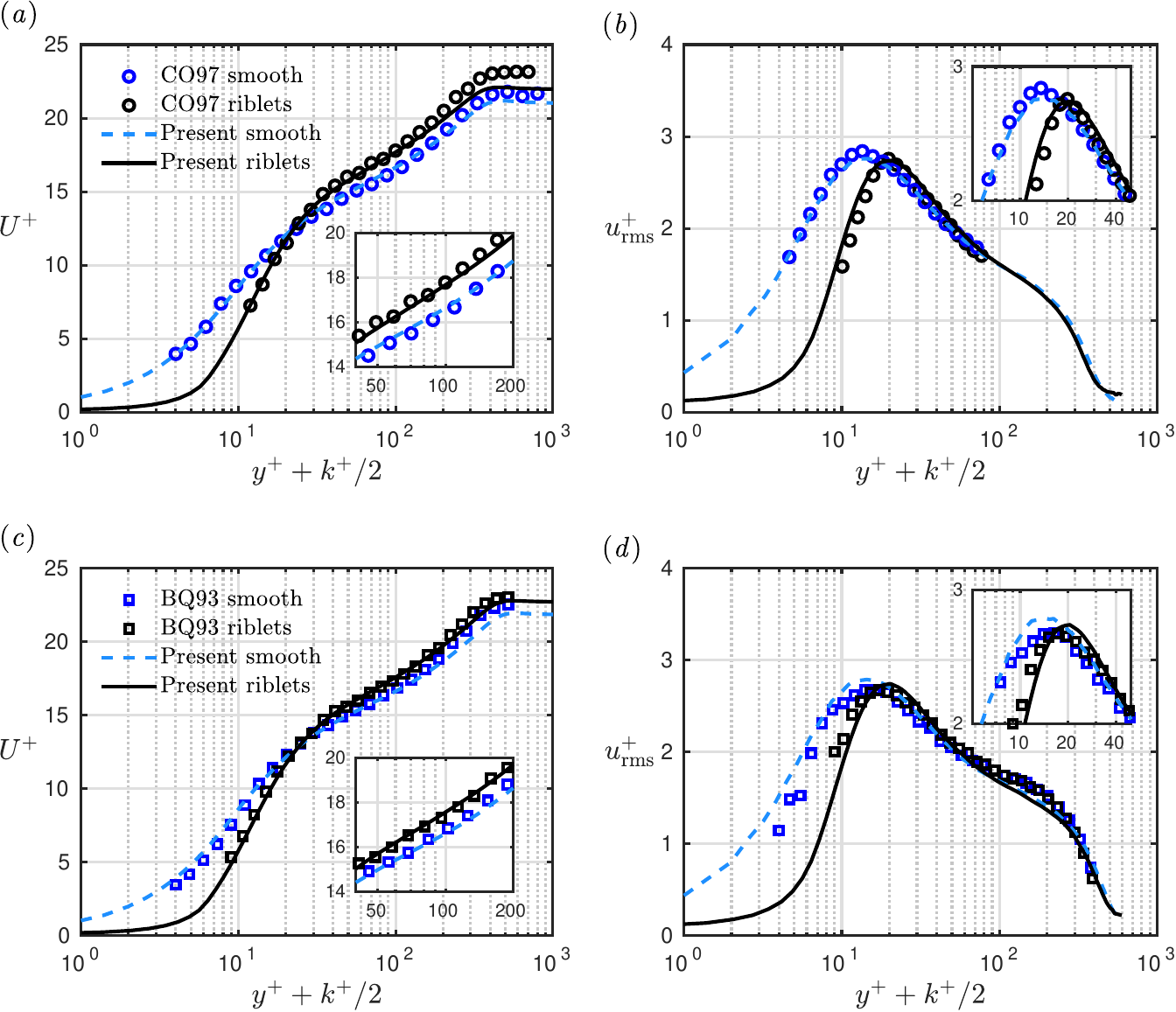}
  \caption{Profiles of (\textit{a,c}) $U^+$ and (\textit{b,d}) $u^+_\mathrm{rms}$ for comparison with the data of (\textit{a,b}) \cite{choi1997turbulence} (CO97) at $Re_{\theta_\mathrm{smooth}} = 880$, and (\textit{c,d}) \cite{baron1993some} (BQ93) at $Re_{\theta_\mathrm{smooth}} = 1150$. We place the profiles origin at the riblets mean height ($y = -k/2$).}
  \label{fig:riblets_bl_stats}
\end{figure}

We compare profiles of $U^+$ and $u^+_\mathrm{rms}$ between our DNS and the reference experiments in figure~\ref{fig:riblets_bl_stats}. In the experiments, measurements over the smooth surface and riblets are conducted at the same distance from the inlet, and $Re_\theta$ at that distance is reported for the smooth case ($Re_{\theta_\mathrm{smooth}}$). Consistent with the experiments, we plot our DNS profiles at a distance $x$, where $Re_{\theta_\mathrm{smooth}}$ is matched with the experiments. Overall, we achieve good agreement between DNS and the experiments by \cite{choi1997turbulence} (figures~\ref{fig:riblets_bl_stats}\textit{a,b}) and \cite{baron1993some} (figures~\ref{fig:riblets_bl_stats}\textit{c,d}) over both the smooth surface and riblets. Especially, up to $y^+ \simeq 100$ excellent agreement is obtained in the $U^+$ and $u^+_\mathrm{rms}$ profiles, as well as the velocity shift due to riblets (insets in figures~\ref{fig:riblets_bl_stats}\textit{a,c}). For viscous scaling, we directly calculate $u_\tau$ from the local wall drag (5.2\textit{b}), whereas the experiments use indirect techniques; \cite{baron1993some} fit the mean velocity slope in the log region, and \cite{choi1997turbulence} fit the mean velocity defect profile. The excellent agreements between the DNS and experimental profiles up to $y^+ \simeq 100$, supports the accuracy of the indirect techniques for obtaining $u_\tau$. Beyond $y^+ \simeq 100$, discrepancies appear in the wake profiles; even the profiles from the two reference experiments differ in the wake region (compare symbols in figure~\ref{fig:riblets_bl_stats}\textit{a} with those in figure~\ref{fig:riblets_bl_stats}\textit{c}). Such discrepancy is attributed to history effects of the upstream condition that persist up to $Re_\theta \simeq 2000$, as discussed in \S\ \ref{sec:bl_setup}. 

 \begin{figure}
  \centering
  \includegraphics[width=\textwidth,trim={{0.0\textwidth} {0.0\textwidth} {0.0\textwidth} {0.0\textwidth}},clip]{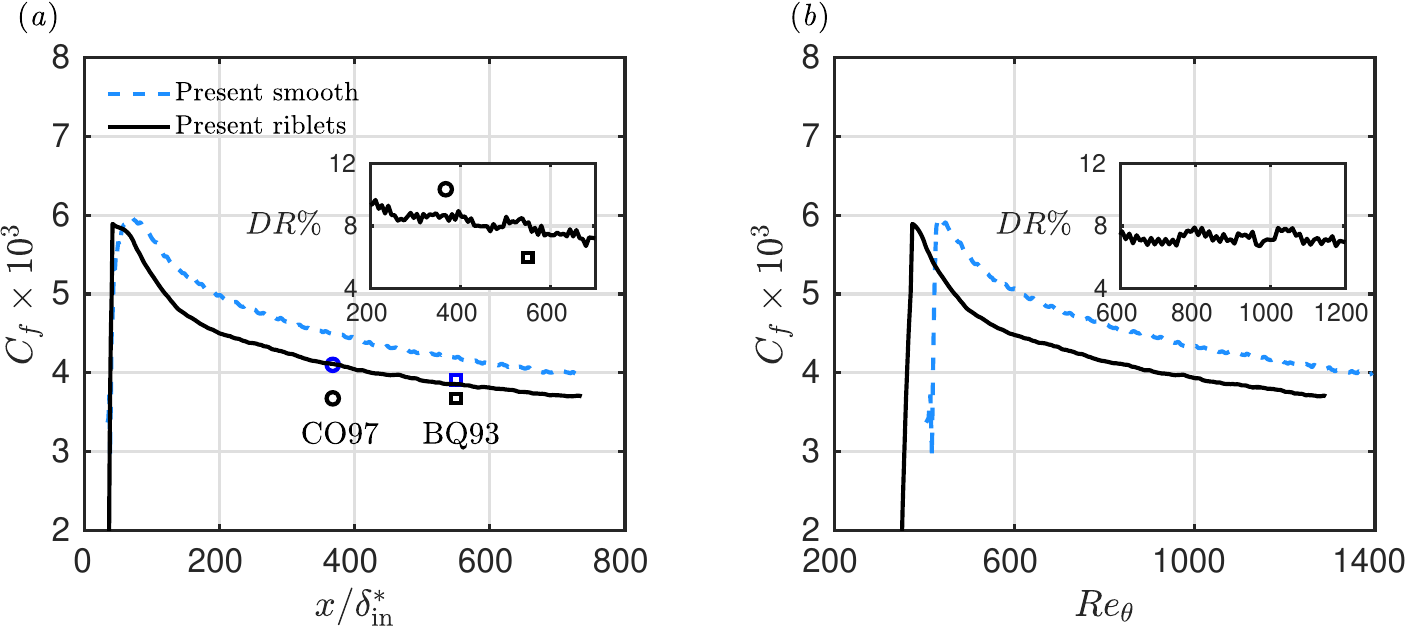}
  \caption{Variations of $C_f$ and $DR\%$ (insets) versus (\textit{a}) $x/\delta^*_\mathrm{in}$ and (\textit{b}) $Re_\theta$. In (\textit{a}), we add the data points by \cite{choi1997turbulence} (CO97) and \cite{baron1993some} (BQ93). Lines and symbols colours are consistent with figure~\ref{fig:riblets_bl_stats}.}
  \label{fig:riblets_bl_Cf}
\end{figure}


In figure~\ref{fig:riblets_bl_Cf}, we plot $C_f$ and drag-reduction percentage $DR\% \equiv (C_{f_\mathrm{smooth}} - C_{f_\mathrm{riblet}})/C_{f_\mathrm{smooth}} \times 100$ in two ways. Firstly, we plot as a function of $x/\delta^*_\mathrm{in}$ to be consistent with the experiments (figure~\ref{fig:riblets_bl_Cf}\textit{a}), and secondly as a function of $Re_\theta$ (figure~\ref{fig:riblets_bl_Cf}\textit{b}). In the former way, $DR\%$ is calculated at matched $Re_x \equiv U_\infty x / \nu$, whereas in the latter way it is calculated at matched $Re_\theta$. In figure~\ref{fig:riblets_bl_Cf}(\textit{a}), we overlay the data points of \cite{choi1997turbulence} (CO97) and \cite{baron1993some} (BQ93) at $x/\delta^*_\mathrm{in}$ locations, where our $Re_{\theta_\mathrm{smooth}}$ is matched with theirs. The experimental data points for $C_f$ are below the DNS counterparts. The values of $C_{f_\mathrm{smooth}}$ and $C_{f_\mathrm{riblet}}$ from the experiments are shifted by about $10\%$ below our DNS curves. This systematic drop is because $C_f = 2/U^+_\infty$, and $U^+_\infty$ is higher in the experiments compared to our DNS (figures~\ref{fig:riblets_bl_stats}\textit{a,c}), owing to the stronger wake profiles in the experiments. Nevertheless, the differences in $DR\%$ between our DNS and experiments is within $1.6\%$ (figure~\ref{fig:riblets_bl_Cf}\textit{a}, inset). We explain this trend by studying the propagation of the deviation in $U^+_\infty$ ($\epsilon_{U^+_\infty}$) to the deviations in $C_f$ ($\epsilon_{C_f}$) and $DR$ ($\epsilon_{DR}$)
\begin{align}
\frac{\epsilon_{C_f}}{C_f} \approx \left( \frac{2}{U^+_\infty} \right) \epsilon_{U^+_\infty}, \quad \epsilon_{DR} \approx \left[ \frac{  2 \Delta U^+  }{{U^+_\infty}^2 \left( 1 + \Delta U^+/U^+_{\infty} \right)^3} \right] \epsilon_{U^+_\infty}. \tag{5.4\textit{a,b}} \label{eq:epsilon_DR}
\end{align}
Equation (5.4\textit{b}) is derived from \cite{garcia2019control}'s semi-empirical relation for $DR$ over riblets (their equation 5). Considering figures~\ref{fig:riblets_bl_stats}(\textit{a,c}), the deviation in $U^+_\infty$ between the experiments and DNS is within $\epsilon_{U^+_\infty} = \pm 2$; with $\Delta U^+ \simeq 1$ and $U^+_{\infty} \simeq 22$, (\ref{eq:epsilon_DR}) yields $\epsilon_{C_f}/{C_f} = \pm 18.2\%$ and $\epsilon_{DR} = \pm 0.7\%$. Consistent with figure~\ref{fig:riblets_bl_Cf}(\textit{a}), $\epsilon_{U^+_\infty}$ significantly propagates to $\epsilon_{C_f}$, but has a small propagation to $\epsilon_{DR}$.

%

Comparing figure~\ref{fig:riblets_bl_Cf}(\textit{a}) with \ref{fig:riblets_bl_Cf}(\textit{b}) signifies the sensitivity of $DR\%$ to the matched Reynolds number definition between $C_{f_\mathrm{smooth}}$ and $C_{f_\mathrm{riblet}}$. When $Re_x$ is matched (figure~\ref{fig:riblets_bl_Cf}\textit{a}), $DR\%$ decreases from $9.6$ to $7.2$ over the range $200 \le x/\delta^*_\mathrm{in} \le 700$, whereas when $Re_\theta$ is matched (figure~\ref{fig:riblets_bl_Cf}\textit{b}), $DR\%$ remains almost constant at $7.5 \pm 0.5$ over the range $600 \le Re_\theta \le 1200$. Overall, despite the inevitable differences between the experiments and DNS, the agreement is quite encouraging. This section indicates a promising potential in $\eta$-grid to afford DNSs of TBLs over complex surfaces at comparable Reynolds numbers with the experiments.

\subsection{Grid saving with Reynolds number}\label{sec:riblet_grid_save}
\begin{figure}
  \centering
  \includegraphics[width=\textwidth,trim={{0.0\textwidth} {0.0\textwidth} {0.0\textwidth} {0.0\textwidth}},clip]{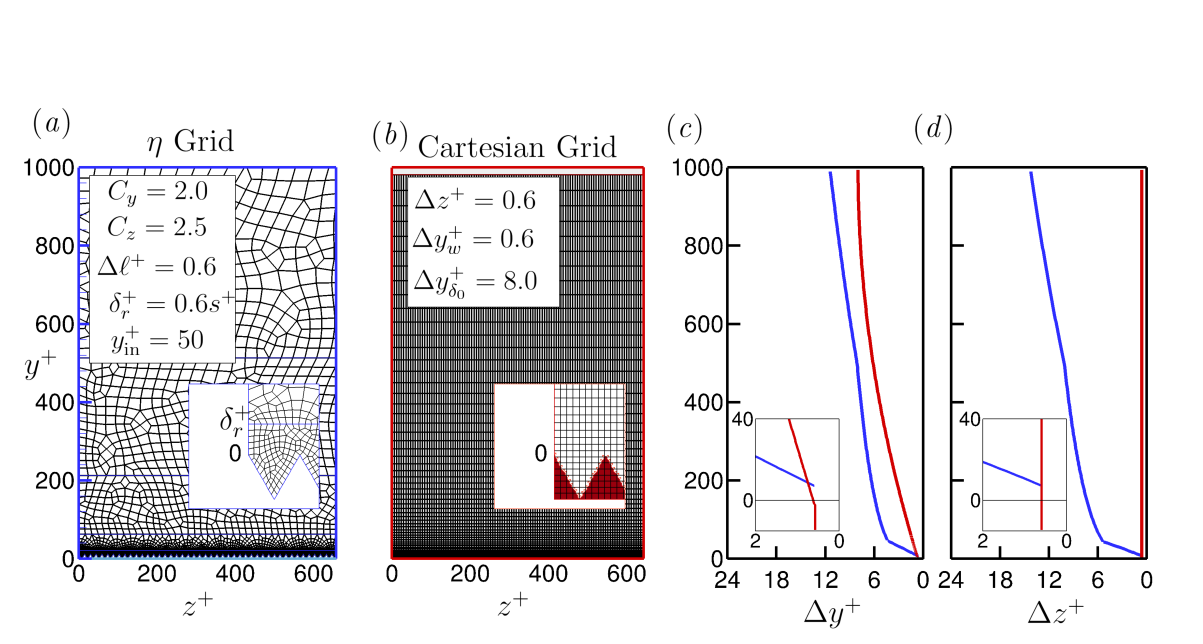}
  \includegraphics[width=.94\textwidth,trim={{0.0\textwidth} {0.08\textwidth} {0.0\textwidth} {0.0\textwidth}},clip]{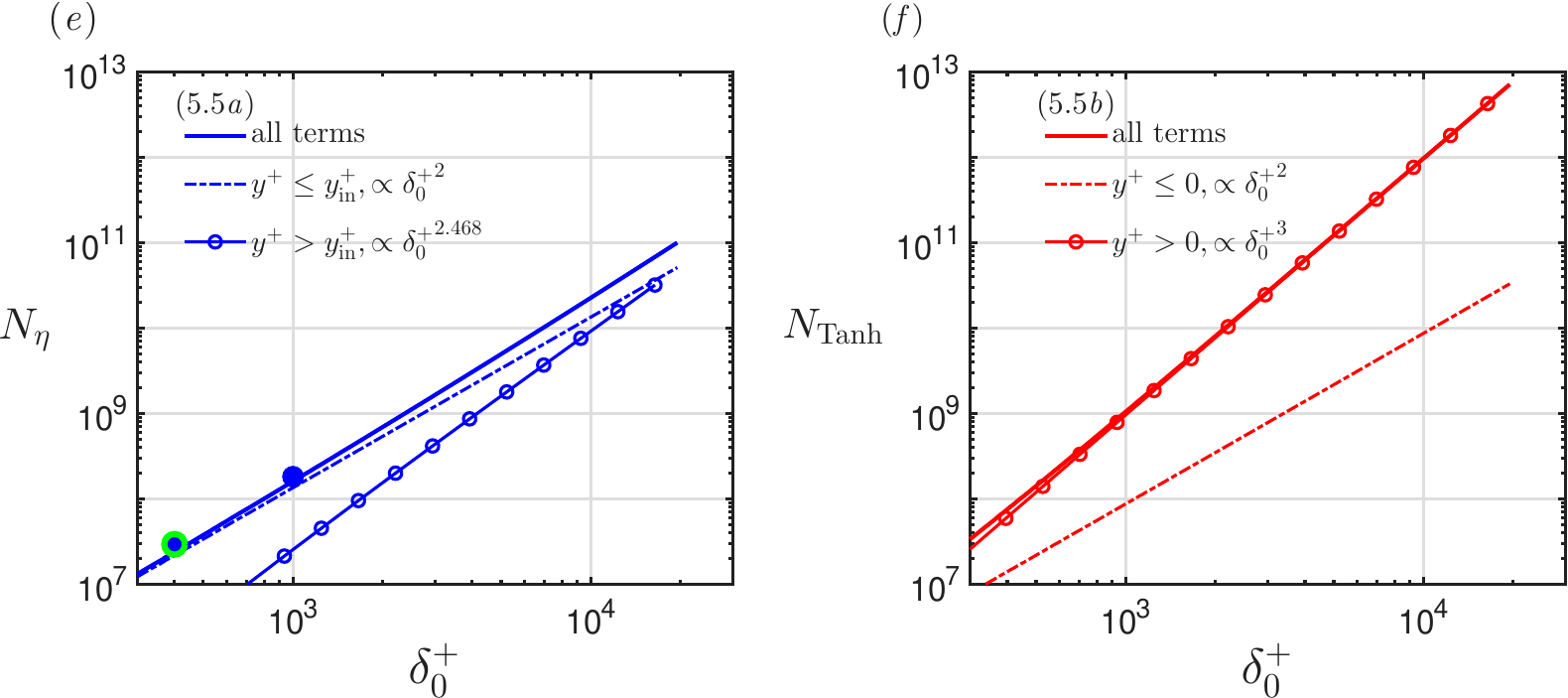} 
  \includegraphics[width=.99\textwidth,trim={{-0.11\textwidth} {0.0\textwidth} {0.0\textwidth} {0.0\textwidth}},clip]{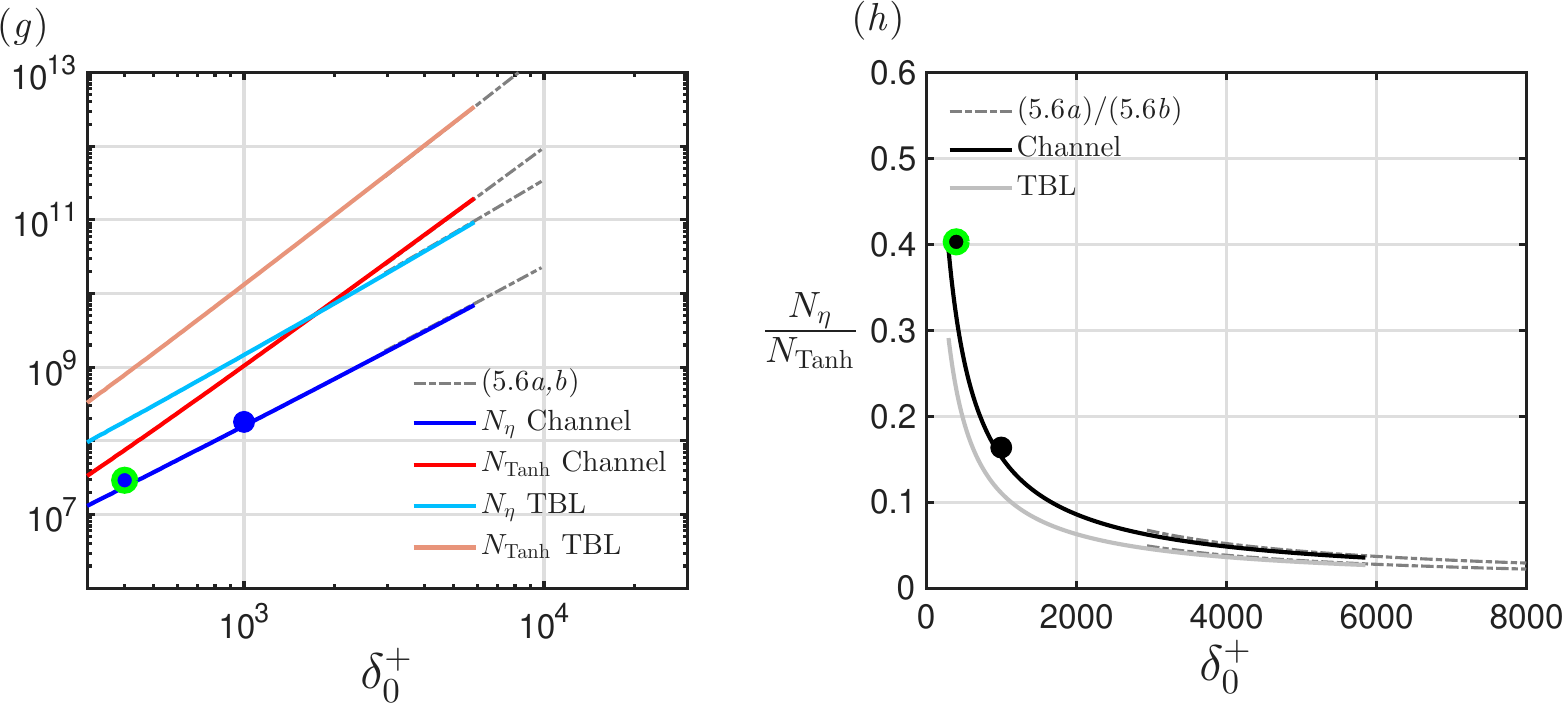}   
  \caption{Grid analysis similar to figures~\ref{fig:Ndof_smooth_channel} and \ref{fig:Ndof_smooth_bl}, but for turbulent open-channel flow and TBL over T615. (\textit{a,b}) Visualise the grid elements, and (\textit{c,d}) plot $\Delta y^+$ and $\Delta z^+$ for turbulent open-channel flow at $\delta^+_0 = 1000$; panel (\textit{a}) and blue curves correspond to $\eta$-grid (5.1\textit{a--c}), and panel (\textit{b}) and red curves correspond to a Cartesian grid with IBM. (\textit{e}) $N_\eta$ versus $\delta^+_0$ for open-channel flow (\ref{eq:N_eta_T615_1}), and its decomposition into the number of grid points up to the inner layer ($y^+ \le y^+_\mathrm{in}$), and beyond. (\textit{f}) $N_\mathrm{Tanh}$ versus $\delta^+_0$ for open-channel flow (\ref{eq:N_IBM_1}), and its decomposition into the number of grid points up to the riblets crest ($y^+ \le  0$), and beyond. (\textit{g}) $N_\eta$ and $N_\mathrm{Tanh}$, and (\textit{h}) their ratio $N_\eta/N_\mathrm{Tanh}$ versus $\delta^+_0$. In (\textit{g,h}), solid lines are from (\ref{eq:N_eta_T615_1}) and (\ref{eq:N_IBM_1}), and gray dashed-dotted lines are from their asymptotic relations (\ref{eq:N_eta_IBM_2}). The bullets at $\delta^+_0 = 400, 1000$ are the actual values of $N_\eta$ from grid generation, following \S\ \ref{sec:grid_generation}, and the bullet with grid outline is the simulated case T615\_S (table~\ref{tab:riblets_channel}, figures~\ref{fig:riblets_setup}\textit{d} and \ref{fig:riblets_channel_stats}\textit{c,d}).}
  \label{fig:Ndof_T615}
\end{figure}

We compare two grids for DNSs of turbulent flows over riblets; $\eta$-grid (5.1\textit{a--c}) (figure~\ref{fig:Ndof_T615}\textit{a}) versus a Cartesian grid with riblets implemented via IBM (figure~\ref{fig:Ndof_T615}\textit{b}). The latter grid is widely utilised by the previous DNS studies of turbulent flows over riblets~\citep{goldstein1995direct,kuwata2022dissimilar,malathi2023riblet,zhdanov2024influence,zhdanov2024net,rowin2025experimental}. The number of grid points with $\eta$-grid $N_\eta$ and a Cartesian grid with hyperbolic-tangent $y$-grid mapping $N_\mathrm{Tanh}$ are
\begin{align}
 N_\eta &= \overbrace{\frac{(L_x L_z/\delta^2_0)}{\Delta x^+ \Delta \ell^+} \left[ \frac{\delta^+_r + \frac{k^+}{2}}{\Delta \ell^+} + \frac{(y^+_\mathrm{in} - \delta^+_r)\ln(C_y/C_z)}{(C_y - C_z)(\kappa y^+_\mathrm{in})^\beta} \right] {\delta^+_0}^2}^{y^+ \le y^+_\mathrm{in} \; \mathrm{(sublayer+inner)}} +
 \overbrace{\begin{cases}
(\ref{eq:N_eta_2}) & \small{\text{Channel}}   \\
(\ref{eq:N_eta_bl_2}) &  \small{\text{TBL}} 
\end{cases}}^{y^+> y^+_\mathrm{in}} \tag{5.5\textit{a}} \label{eq:N_eta_T615_1} \\
N_\mathrm{Tanh} &= \underbrace{\frac{(L_x L_z/\delta^2_0)}{\Delta x^+ \Delta z^+} \left( \frac{k^+}{\Delta y^+_w} \right){\delta^+_0}^2}_{y^+ \le 0 \; \mathrm{(groove)}} +
 \underbrace{\begin{cases}
 (\ref{eq:N_hyp_2})  & \small{\text{Channel}}   \\
 (\ref{eq:N_hyp_bl_2})  &  \small{\text{TBL}} 
\end{cases}}_{y^+ > 0} \tag{5.5\textit{b}} \label{eq:N_IBM_1}
\end{align}
For (\ref{eq:N_eta_T615_1}) and (\ref{eq:N_IBM_1}), we assume that $\delta^+_0 \gg k^+$; this assumption is nearly valid for riblets, given that $\delta^+_0 \gtrsim \mathcal{O}(10^2)$ and $k^+ \sim \mathcal{O}(10)$. We obtain $N_\eta$ (\ref{eq:N_eta_T615_1}) by summing the number of grid points from all blocks of the $\eta$-grid for riblets (figure~\ref{fig:idealised_grid_riblets}). To obtain $N_\mathrm{Tanh}$ (\ref{eq:N_IBM_1}), we consider a grid arrangement following the studies that have employed Cartesian grids with IBM for riblets~\citep{zhdanov2024influence,rowin2025experimental}; $\Delta x^+$ and $\Delta z^+$ are fixed, and $\Delta y^+$ is fixed at $\Delta y^+_w$ from the riblets valley ($y^+ = -k^+$) to crest ($y^+ =0$), and then is expanded following a hyperbolic-tangent mapping up to $y^+ = \delta^+_0 - k^+/2$ (red lines in figures~\ref{fig:Ndof_T615}\textit{c,d}). Considering (5.5\textit{a,b}), the number of grid points over riblets is the number of grid points over the smooth wall (arguments in braces on the right) plus the number of grid points down from the inner layer ($y^+ \le y^+_\mathrm{in}$ in \ref{eq:N_eta_T615_1}), or down from the riblets crest ($y^+ \le 0$ in \ref{eq:N_IBM_1}).

In figure~\ref{fig:Ndof_T615}, we compare $N_\eta$ with $N_\mathrm{Tanh}$ for turbulent flows over T615 ($\alpha = \ang{60}, s^+ = 15$). We showed the accuracy of $\eta$-grid for DNS of turbulent open-channel flow over this riblet geometry at $\delta^+_0 = 400$ (figures~\ref{fig:riblets_channel_stats}\textit{c,d}). In figures~\ref{fig:Ndof_T615}(\textit{a,b}), we report our chosen grid parameters for comparison; we match the grid cell size within the riblet groove between $\eta$-grid ($\Delta \ell^+ = 0.6$) and the Cartesian grid ($\Delta z^+ = \Delta y^+_w = 0.6$). The other grid parameters for $\eta$-grid are chosen based on our extensive study of this grid. For the Cartesian grid, we set $\Delta y^+_{\delta_0} = 8.0$, which is close to the values as set by the previous DNSs over riblets~\citep{endrikat2021influence,zhang2024direct,rowin2025experimental}.



In figures~\ref{fig:Ndof_T615}(\textit{e,f}), respectively we plot $N_\eta$ (\ref{eq:N_eta_T615_1}) and $N_\mathrm{Tanh}$ (\ref{eq:N_IBM_1}) versus $\delta^+_0$ for turbulent open-channel flow. We also decompose $N_\eta$ and $N_\mathrm{Tanh}$ into the partition that resolves the near-riblet region ($\propto {\delta^+_0}^2$), and the one that resolves the log region and beyond. The near-riblet grid covers $y^+ \le y^+_\mathrm{in}$ for $N_\eta$, and $y^+ \le 0$ for $N_\mathrm{Tanh}$; it is of the same order between $N_\eta$ and $N_\mathrm{Tanh}$ (dashed-dotted lines). However, the grid size that resolves the log region and beyond is the major source of disparity between $N_\eta$ and $N_\mathrm{Tanh}$ (lines with circles). This grid size is almost two orders of magnitude smaller for $N_\eta$ compared to $N_\mathrm{Tanh}$. Such significant disparity is due to the constraint of a fixed $\Delta z^+ = 0.6$ for the Cartesian grid with IBM (red line in figure~\ref{fig:Ndof_T615}\textit{d}), whereas with $\eta$-grid, $\Delta z^+_\eta$ expands to $\simeq 15$ by $\delta^+_0 = 1000$ (blue line in figure~\ref{fig:Ndof_T615}\textit{d}). Considering figure~\ref{fig:Ndof_T615}(\textit{e}), up to $\delta^+_0 \simeq 2000$, $N_\eta$ is dominated by the near-wall grid size ($y^+ \le y^+_\mathrm{in}$), and scales $\propto {\delta^+_0}^2$ for channel flow; beyond $\delta^+_0 \simeq 2000$, the grid size above $y^+_\mathrm{in}$ has an increasing contribution to $N_\eta$. On the other hand, for $\delta^+_0 \gtrsim 1000$, more than $90\%$ of $N_\mathrm{Tanh}$ is dominated by the grid size above the riblets crest ($y^+ > 0$), hence $  N_\mathrm{Tanh} \propto {\delta^+_0}^3$. With our chosen grid parameters, we arrive at the following asymptotic relations
\begin{align}
 N_\eta \simeq \begin{cases}
                134.12{\delta^+_0}^2 + 1.35 {\delta^+_0}^{2.468} & \small{\text{Channel}} \\
                351.33{\delta^+_0}^{2.1746} + 3.5 {\delta^+_0}^{2.6746} & \small{\text{TBL}}
               \end{cases}, \;
N_\mathrm{Tanh} \simeq \begin{cases}
                       0.96{\delta^+_0}^3 & \small{\text{Channel}} \\
                       3.73{\delta^+_0}^{3.1746} & \small{\text{TBL}}
                      \end{cases}. \tag{5.6\textit{a,b}} \label{eq:N_eta_IBM_2}               
\end{align}
In figure~\ref{fig:Ndof_T615}(\textit{g}), we plot $N_\eta$ and $N_\mathrm{Tanh}$ for turbulent open-channel flow and TBL, and in figure~\ref{fig:Ndof_T615}(\textit{h}) we plot the ratio $N_\eta/N_\mathrm{Tanh}$. The asymptotic relations (\ref{eq:N_eta_IBM_2}) (gray dashed-dotted lines) are in excellent agreement with the direct relations (5.5\textit{a,b}) (solid lines). Also, the values of $N_\eta$ from the generation of $\eta$-grid following \S\ \ref{sec:grid_generation} (bullets) are in great agreement with (\ref{eq:N_eta_T615_1}), figures~\ref{fig:Ndof_T615}(\textit{e,g}). Figure~\ref{fig:Ndof_T615}(\textit{h}) signifies the enormous grid saving by $\eta$-grid for DNSs of turbulent flows over riblets. By $\delta^+_0 = 2000$, $N_\eta/N_\mathrm{Tanh}$ drops below $0.1$, and by $\delta^+_0 = 6000$, it further drops to $0.04$.    


\section{Conclusions}
We formulated an unstructured grid-generation framework, termed $\eta$-grid, for efficient DNSs of wall-bounded turbulent flows, over smooth and uneven surfaces. Similar to the wall-normal grid mapping by \cite{pirozzoli2021natural}, we increase the grid size proportional to the local Kolmogorov scale $\eta^+$, but we simultaneously increase the wall-normal and spanwise grid sizes ($\Delta y^+_\eta, \Delta z^+_\eta$) based on a revised semi-empirical fit for $\eta^+$. We obtain $\eta^+_\mathrm{fit}$ from processing DNS databases for turbulent channel flows and zero pressure-gradient turbulent boundary layers (ZPG TBLs). For DNSs of turbulent flows over smooth walls, our proposed $\eta$-grid consists of an inner layer with a thickness $y^+_\mathrm{in} \simeq 50$, where the grid resolution is similar to a conventional DNS grid, a uniform $\Delta z^+_\eta \simeq 5 - 6$, but a growing $\Delta y^+_\eta$ from $\Delta y^+_w = 0.3$ to $4$. Beyond $y^+_\mathrm{in}$, the grid sizes expand following $\Delta y^+_\eta = 2.0 \eta^+_\mathrm{fit}, \Delta z^+_\eta = 2.5 \eta^+_\mathrm{fit}$. We extended the $\eta$-grid formulation for application to turbulent flows over riblets, by taking into account the resolution requirements of important flow physics over riblets (secondary flows, Kelvin-Helmholtz rollers). The extended formulation has an additional layer from the riblet surface up to the riblet sublayer $\delta^+_r \simeq 0.6s^+$~\citep{modesti2021dispersive}, where $s^+$ is the riblet spacing, and $\delta^+_r$ is measured from the riblet crest; this layer is well-resolved by square elements of size $\Delta \ell^+ \simeq s^+/30 - s^+/20$. Beyond $y^+  = \delta^+_r$, the formulation follows the smooth-wall $\eta$-grid. This extended formulation is applicable to turbulent flows over roughness, where $\delta^+_r$ becomes the roughness sublayer, and $\Delta \ell^+$ should resolve the smallest roughness wavelength. We implemented $\eta$-grid through a multi-block grid-generation approach with hexahedral elements for Finite Volume Method (FVM), and Spectral Element Method (SEM) solvers.

We tested the accuracy of $\eta$-grid with a SEM solver (SOD2D) and the widely used FVM solver OpenFOAM. We conducted an exhaustive DNS campaign of wall-bounded turbulent flows, including smooth-wall turbulent channel flows up to friction Reynolds number $\delta^+_0 = 1000$, smooth wall ZPG TBLs up to $\delta^+_0 = 737$, as well as turbulent open-channel flow and ZPG TBL over various riblet geometries at $\delta^+_0 \simeq 400$. For all test cases, we generated a fixed $\Delta x^+ \simeq 10 - 15$ in the streamwise direction. We compared the results with the reference DNS and experimental data, as well as finer grid calculations. Accuracy assessments were in terms of skin-friction coefficient, profiles of the mean velocity and r.m.s.\ of velocity fluctuations, and their energy spectrograms. For the smooth and riblet-roughened channel flow cases, compared to the reference DNSs, $\eta$-grid with SOD2D yielded $\lesssim 1\%$ difference, and $\eta$-grid with OpenFOAM yielded $\lesssim 2\%$ difference. Through the finer grid and supporting calculations, we concluded that the larger difference with OpenFOAM is related to its discretisation schemes. For the ZPG TBL cases, with $\eta$-grid and SOD2D we achieved excellent agreement with the reference data up to the end of the log region. We observed differences in the wake region for cases with $Re_\theta \lesssim 2000$ ($\delta^+ \lesssim 730$); this was due to the different inflow conditions and their history effects, as reported in the literature~\citep{schlatter2012turbulent}.

For each test case, we analysed the saving in the number of grid points with $\eta$-grid ($N_\eta$). We compared $N_\eta$ with the number of grid points from a conventional Cartesian grid ($N_\mathrm{Tanh}$), with matched $\Delta x^+$, a fixed $\Delta z^+$ equal to $\Delta z^+_\eta$ at the wall, and a hyperbolic-tangent $y$-grid mapping ($0.3 \le \Delta y^+_\mathrm{Tanh} \le 8$). For turbulent flows over riblets, we compare with a Cartesian grid with riblets implemented via an Immersed Boundary Method (IBM), with $\Delta y^+, \Delta z^+$ close to $\Delta y^+_\eta, \Delta z^+_\eta$ within the riblet groove. For turbulent flows over a smooth wall, $N_\eta/N_\mathrm{Tanh} \propto {\delta^+_0}^{-0.5}$; by $\delta^+_0 = 6000$, $N_\eta/N_\mathrm{Tanh} \simeq 0.11$. For turbulent flows over riblets, $N_\eta/N_\mathrm{Tanh}$ varies from $\propto {\delta^+_0}^{-1.0}$ (for $\delta^+_0 \lesssim 2000$) to $\propto {\delta^+_0}^{-0.5}$ (for $\delta^+_0 \gtrsim 10^4$); by $\delta^+_0 = 6000$, $N_\eta/N_\mathrm{Tanh} \simeq 0.04$ for drag-reducing triangular riblets with tip angle $\ang{60}$ and $s^+ = 15$. We hope that such enormous grid saving encourages the wall-turbulence community to leverage $\eta$-grid, and conduct DNSs at Reynolds numbers that cannot be afforded via Cartesian grids, with today's computational power. 
\\ 

\noindent \textbf{Funding}\\
AR acknowledges funding from the Air Force Office of Scientific Research (AFOSR) under award number FA8655-24-1-7008, monitored by Dr.\ Douglas Smith and Dr.\ Barrett Flake. VK acknowledges his AI4S fellowship within the Generaci\'on D initiative by Red.es, Ministerio para la Transformaci\'on Digital y de la Funci\'on P\'ublica, for talent attraction (C005/24-ED CV1), funded by NextGenerationEU through PRTR. WW acknowledges funding from AFOSR Grant No. FA9550-25-1-0033, monitored by Dr.\ Gregg Abate. OL has been partially supported by a Ramon y Cajal postdoctoral contract (Ref: RYC2018- 025949-I). The authors acknowledge the support given by the Departament de Recerca i Universitats de la Generalitat de Catalunya to the Large-Scale Computational Fluid Dynamics Research Group (Code: 2021 SGR 00902).
We thank EPSRC for the computational time made available on ARCHER2 via the UK Turbulence Consortium (EP/X035484/1), and the UKRI access to the HPC call 2024. We also acknowledge the computational resources provided by Barcelona Supercomputing Center and Red Espa\~nola de Supercomputaci\'on (RES) on MareNostrum V (Nos.\ IM-2025-3-0053, IM-2026-1-0036).\\

\noindent \textbf{Declaration of interests.} The authors report no conflict of interest.

\appendix

\section{Effect of grid aspect ratio with OpenFOAM}\label{sec:dxp_study}
\begin{figure}
  \centering
  \includegraphics[width=\textwidth,trim={{0.0\textwidth} {0.0\textwidth} {0.0\textwidth} {0.0\textwidth}},clip]{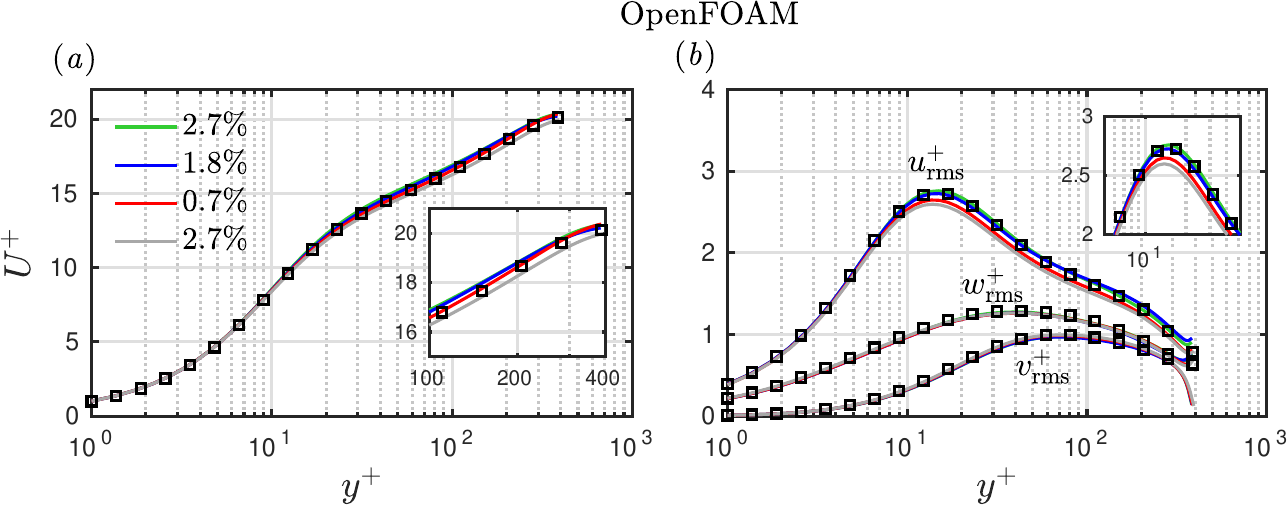}
  \caption{Sensitivity assessment of OpenFOAM to $\Delta x^+$ for turbulent open-channel flow at $\delta^+_0 = 395$ with $\eta$-grid. All cases have identical $yz$-grid, with grid parameters $\Delta y^+_w = 0.3, y^+_\mathrm{in} = 50, C_y = 2.0$ and $C_z = 2.5$, but have different $\Delta x^+$; $\Delta x^+ = 6$ (green), $10$ (blue), $14$ (red) and $18$ (grey). The numbers in (\textit{a}) are $\varepsilon_{C_f}$ relative to the DNS of \cite{moser1999direct}.}
  \label{fig:dxp_study}
\end{figure}

We conducted additional cases with OpenFOAM to assess the sensitivity of statistics to the grid aspect ratio. The test case is an open-channel flow at $\delta^+_0 = 395$ with $\eta$-grid. For all runs, we generate identical $yz$-grids with $\Delta y^+_w = 0.3, y^+_\mathrm{in} = 50, C_y = 2.0$ and $C_z = 2.5$, but we change $\Delta x^+$ from $6$ to $18$. Interestingly, $\varepsilon_{C_f}$ falls below $1\%$ with $\Delta x^+ = 14$, but rises to $2.7\%$ with $\Delta x^+ = 6$ (figure~\ref{fig:dxp_study}\textit{a}).

\bibliography{afosr_refs}
\bibliographystyle{jfm}

\end{document}